\documentclass[longauth]{aa} 
\usepackage{graphicx}
\usepackage{txfonts}
\usepackage{hyperref}
\makeatletter
\renewcommand*\aa@pageof{, page \thepage{} of \pageref*{LastPage}}
\makeatother
\usepackage{orcidlink}
\usepackage{supertabular}
\usepackage{longtable}
\usepackage{CJKutf8}
\usepackage{comment}

\begin{document}

   \title{On the diversity of strongly-interacting Type~IIn supernovae}

    \author{I.~Salmaso\,\orcidlink{0000-0003-1450-0869}
          \inst{1}\fnmsep\thanks{\email{ irene.salmaso@inaf.it}}
          \and
          E.~Cappellaro \inst{1} \and L.~Tartaglia  \inst{2}\fnmsep \inst{1} \and J.~P.~Anderson\,\orcidlink{0000-0003-0227-3451} \inst{3}\fnmsep \inst{4}  \and S.~Benetti  \inst{1} \and M.~Bronikowski \inst{5} \and Y.-Z.~Cai~\begin{CJK*}{UTF8}{gbsn}(蔡永志)\end{CJK*}\,\orcidlink{0000-0002-7714-493X} \inst{6}\fnmsep \inst{7}\fnmsep\inst{8} \and P.~Charalampopoulos\,\orcidlink{0000-0002-0326-6715} \inst{9} \and T.-W.~Chen\,\orcidlink{0000-0002-1066-6098} \inst{10} \and E.~Concepcion \inst{5} \and N.~Elias-Rosa  \inst{1}\fnmsep \inst{11} \and L.~Galbany\,\orcidlink{0000-0002-1296-6887} \inst{11}\fnmsep \inst{12} \and M.~Gromadzki\,\orcidlink{0000-0002-1650-1518} \inst{13} \and C.~P.~Guti\'errez\,\orcidlink{0000-0003-2375-2064} \inst{12}\fnmsep \inst{11} \and E.~Kankare \inst{9} \and P.~Lundqvist \inst{14} \and K.~Matilainen  \inst{15}\fnmsep \inst{16} \and P.~A.~Mazzali \inst{17}\fnmsep \inst{18} \and S.~Moran \inst{16} \and T.~E.~Müller-Bravo\,\orcidlink{0000-0003-3939-7167} \inst{11}\fnmsep \inst{12} \and M.~Nicholl\,\orcidlink{0000-0002-2555-3192} \inst{19} \and A.~Pastorello\,\orcidlink{0000-0002-7259-4624} \inst{1} \and P.~J.~Pessi\,\orcidlink{0000-0002-8041-8559} \inst{14} \and T.~Pessi \inst{20} \and T.~Petrushevska \inst{5} \and G.~Pignata \inst{21} \and A.~Reguitti\,\orcidlink{0000-0003-4254-2724}  \inst{22}\fnmsep \inst{1} \and J.~Sollerman  \inst{14} \and S.~Srivastav\,\orcidlink{0000-0003-4524-6883} \inst{23} \and M.~Stritzinger\,\orcidlink{0000-0002-5571-1833} \inst{24} \and L.~Tomasella\,\orcidlink{0000-0002-3697-2616} \inst{1} \and G.~Valerin  \inst{1}
          }
   \institute{INAF-Osservatorio Astronomico di Padova, Vicolo dell'Osservatorio 5, 35122 Padova, Italy
   \and INAF-Osservatorio Astronomico d’Abruzzo, Via M. Maggini snc, 64100 Teramo, Italy
   \and European Southern Observatory, Alonso de C\'ordova 3107, Casilla 19, Santiago, Chile
    \and Millennium Institute of Astrophysics MAS, Nuncio Monsenor Sotero Sanz 100, Off.
104, Providencia, Santiago, Chile
    \and Center for Astrophysics and Cosmology, University of Nova Gorica, Vipavska 11c, 5270 Ajdov\v{s}\v{c}ina, Slovenia
    \and Yunnan Observatories, Chinese Academy of Sciences, Kunming 650216, P.R. China
    \and International Centre of Supernovae, Yunnan Key Laboratory, Kunming 650216, P.R. China
    \and Key Laboratory for the Structure and Evolution of Celestial Objects, Chinese Academy of Sciences, Kunming 650216, P.R. China
     \and Department of Physics and Astronomy, University of Turku, Vesilinnantie 5, 20500, Finland
     \and Graduate Institute of Astronomy, National Central University, 300 Jhongda Road, 32001 Jhongli, Taiwan
    \and Institute of Space Sciences (ICE, CSIC), Campus UAB, Carrer de Can Magrans s/n, 08193 Barcelona, Spain
    \and Institut d'Estudis Espacials de Catalunya (IEEC), 08860  Castelldefels (Barcelona), Spain
     \and Astronomical Observatory, University of Warsaw, Al. Ujazdowskie 4, 00-478 Warszawa, Poland
     \and The Oskar Klein Centre, Department of Astronomy, Stockholm University, AlbaNova 106 91, Stockholm, Sweden
     \and Nordic Optical Telescope, Aarhus Universitet, Rambla José Ana Fernández Pérez 7, local 5, E-38711 San Antonio, Breña Baja Santa Cruz de Tenerife, Spain
    \and Tuorla Observatory, Department of Physics and Astronomy, 20014 University of Turku, Vesilinnantie 5, Turku, Finland
    \and Astrophysics Research Institute, Liverpool John Moores University, ic2, 146 Brownlow Hill, Liverpool L3 5RF, UK
    \and Max-Planck Institut für Astrophysik, Karl-Schwarzschild-Str. 1, 85741 Garching bei München, Germany
    \and Astrophysics Research Centre, School of Mathematics and Physics, Queens University Belfast, Belfast BT7 1NN, UK
    \and Instituto de Estudios Astrof\'isicos, Facultad de Ingenier\'ia y Ciencias, Universidad Diego Portales, Av. Ej\'ercito Libertador 441, Santiago, Chile
    \and Instituto de Alta Investigación, Universidad de Tarapacá, Casilla 7D, Arica, Chile
    \and INAF-Osservatorio Astronomico di Brera, Via E. Bianchi 46, 23807 Merate (LC), Italy
    \and Astrophysics sub-Department, Department of Physics, University of Oxford, Keble Road, Oxford, OX1 3RH, UK
    \and Department of Physics and Astronomy, Aarhus University, Ny Munkegade 120, 8000 Aarhus C, Denmark
            }

   \date{Received XXXXXXXXXXX; accepted XXXXXXXXXXX}

 
  \abstract
   {Massive stars experience strong mass-loss 
   rates, producing a dense, H-rich circumstellar medium (CSM). After the explosion, the collision and continued interaction of the supernova (SN) ejecta with the CSM power the light curve through the conversion of kinetic energy into radiation. When the interaction is strong, the light curve shows a broad peak and high luminosity lasting for 
   several months. Also the spectral evolution is slower, compared to non-interacting SNe.
   Energetic shocks between the ejecta and the CSM create the ideal conditions for particle acceleration and production of high-energy (HE) neutrinos above 1~TeV.}
   {We study four strongly-interacting Type~IIn SNe: 2021acya, 2021adxl, 2022qml, and 2022wed to highlight their peculiar characteristics, derive the kinetic energy of the explosion and the characteristics of the CSM, infer clues on the possible progenitors and their environment and relate them to the production of HE neutrinos.}
  {We analysed spectro-photometric data of a sample of interacting SNe to determine their common characteristics and derive physical properties (radii and masses) of the CSM and ejecta kinetic energies to compare them to HE neutrino production models.}
   {The SNe analysed in this sample exploded in dwarf, star-forming galaxies and they are consistent with energetic explosions and strong interaction with the surrounding CSM. For SNe~2021acya and 2022wed, we find high CSM masses and mass-loss rates, linking them to very massive progenitors. For SN~2021adxl, the spectral analysis and less extreme CSM mass suggest a stripped-envelope massive star as possible progenitor. SN~2022qml is marginally consistent with being a Type~Ia thermonuclear explosion embedded in a dense CSM. The mass-loss rates for all SNe are consistent with the expulsion of several solar masses of material during eruptive episodes in the last few decades before the explosion. Finally, we find that the SNe in our sample are marginally consistent with HE neutrino production.}
   {}


   \maketitle
%

\section{Introduction}
\label{sec:intro_altre}
The final stages in the lives of massive stars are poorly known. In particular, key processes such as mass-loss mechanisms (e.g., through winds or eruptions) are difficult to characterise (e.g., \citealt{smith_massloss_2014}). 
For example, Wolf-Rayet stars (WR) efficiently lose their external layers because of strong winds driven by radiation pressure \citep{abbott_wr_1982} and when they explode the resulting supernova (SN) lacks signatures of H and in some cases also He (stripped-envelope SNe, SE~SNe, \citealt{clocchiatti_sesne_1997}). On the other hand, luminous blue variables (LBVs) are massive stars that undergo multiple eruptive mass-loss episodes \citep{humphreys_LBV_1994}. For example, \citet{moriya_massloss_2014} find mass-loss rate for SNe~IIn above $10^{-3}\; \rm{M_{\odot}\, yr^{-1}}$, while \citet{dukiya_asassn14il_2024} find an astonishing mass-loss rate $2-7\; \rm{M_{\odot}\, yr^{-1}}$ for the SN~IIn ASASSN-14il.
Commonly, the mass-loss is not steady with time but each episode can shed a significant amount of mass from the progenitor star, perhaps under the influence of a binary companion.
Binarity is an important, perhaps even dominant factor in the production of SE~SNe \citep{sana_binarie_2012}. In fact, massive stars are mostly found in binary systems and the presence of a companion is likely to strip the donor star from its outer H layers after a common envelope (CE) phase \citep{podsiadlowski_ce_1992}. This process can enhance the mass-loss rate even for lower-mass stars and generate a dense CSM around the donor \citep{chevalier_ce_2012}.

The net result of an enormous ($0.01-0.1\;\mathrm{M_{\odot}\,yr^{-1}}$, \citealt{kiewe_massloss_2012}) mass-loss is a massive circumstellar medium (CSM) that is revealed by narrow lines in the spectrum after the explosion (e.g., Type IIn SNe, \citealt{schlegel1990,schlegel1996,filippenko_classification_1997}) and contributes to the luminosity through interaction with the SN ejecta.
The luminosity of interacting SNe is mainly powered by the conversion of kinetic energy into radiation in the shock of the ejecta with the CSM. It is expected that the shock is stronger when a shell ejection occurs shortly before the explosion (less than a few years). In fact, the closer the CSM shell is to the progenitor, the denser it is, and thus the SN ejecta colliding into it give rise to a more intense shock wave. When the CSM has a high density, the interaction can completely mask the internal power source to the extent that the underlying event can even be a non-terminal outburst \citep{vink_massloss_2015}, 
which could be mistaken for a faint core-collapse (CC) SN, or even a thermonuclear explosion \citep{silverman_ptf11kx_2013}.
The density of the CSM and its spatial distribution affect the shape of the light curve \citep{khatami_kasen2023} and the strength of the emission lines, giving rise to asymmetries \citep{andrews_2013L_2017} and peculiar line profiles such as double peaks \citep{andrews_iptf_2018}. Not only that, the mass and density of the CSM can also affect the shape of the light curve \citep{khatami_kasen2023}. Interestingly, \citet{taddia_carnegie_2013} find that there does not appear to be a continuity of properties among SNe~IIn but rather different subtypes, possibly hinting at the presence of different progenitors. 
Although the sample in \citet{taddia_carnegie_2013} was quite small, this trend has been recently observed also with bigger samples of SNe~IIn \citep{hiramatsu_IIn_2024,ransome_IIn_2024}.

Objects powered by strong interaction are rare. However, some extraordinary objects have been found in the past and present very well-sampled multiband light curves and spectra.
A well-known SN that showed huge interaction 
in brightness and duration of the light curve
is SN~2010jl \citep{smith_2010jl_2012,fransson_2010jl_2014,ofek_2010jl_2014}, a luminous Type~IIn SN with a bright ($\sim -20$~mag at peak) light curve that lasted more than 1000$\,\rm{days}$ and spectra dominated by H Balmer lines with symmetric, electron-scattering driven profiles. These characteristics are interpreted as signs of interaction with a massive ($>3\;\mathrm{M_{\odot}}$), H-rich CSM, possibly produced by an LBV progenitor \citep{fransson_2010jl_2014}. 
LBVs and hypergiant stars have been proposed as progenitors of several SNe~IIn \citep{gal-yam_2005gl_2007,gal-yam_2005gl_2009}.
Another example is SN~2013L \citep{andrews_2013L_2017,taddia_2013L_2020}, which also 
had an observed light curve lasting 1500$\,\rm{days}$ and peaking at around $-19$~mag. The spectra are also slow-evolving and dominated by Balmer lines but the profile of the H$\alpha$ line is significantly asymmetric, with a blue shoulder that is interpreted as due to the emergence of the shock wave not fully hidden by electron scattering \citep{taddia_2013L_2020}. We will use these two SNe as reference throughout the paper.

Interaction can occur also for a limited time range, temporarily increasing the luminosity and changing the shape of the emission lines. This is the case, for example, of SN~1998S \citep{fassia_98S_2000,fassia_98s_2001}, whose broad Balmer emission lines disappeared and reappeared again during the spectroscopic evolution due to the presence of at least two distinct CSM shells \citep{fassia_98s_2001}. 

Interaction is a powerful phenomenon that was proposed to power many SNe, among which, the superluminous SN class (SLSNe, \citealt{smith_2006gy_2007,gal-yam_SLSN_2012}), so as to explain why they are so luminous and with such slow evolution. The mechanism powering SLSNe is still debated, but the slow rise seems to favour a central engine (particularly, a magnetar, \citealt{kasen_magnetar_2010}) in a massive ejecta with long diffusion time. 
However, we caution that \citet{hiramatsu_IIn_2024} showed that there is no clear transition between SNe~IIn and SLSNe and that the arbitrary cut should be removed.

In some cases, the CSM distribution may be strongly asymmetric. If it is also particularly close to the progenitor, the ejecta can quickly and completely engulf it. Narrow lines are then no longer visible in the spectra and the broad lines only show P-Cygni profiles typical of an expanding photosphere, but interaction continues within the inner ejecta, providing energy to the light curve. In these cases, it is difficult to determine whether the luminosity is fully due to interaction or to a central engine such as a magnetar, but the extreme energy and duration of these events seem to point toward a combination of the two phenomena \citep{Kangas2022,pessi_hidden_2023,salmaso_faa_2023}. 

From a multimessenger point of view, interacting SNe~IIn are particularly interesting because the shocked regions may provide a favourable environment for particle acceleration. These accelerated particles can decay into high-energy (HE) neutrinos, which, in principle, may be observed by neutrino detectors.
In the years, there have been some tentative associations of interacting SNe with HE neutrino events. The first one was SN~2011fh, associated with a cascade event detected one day after the optical light curve peak. More recently, the Type~Ibn SN~2023uqf has been found inside the errorbox of neutrino IC-231004A \citep{reusch_2023uqf_neutrino_2023,stein_2023uqf_neutrino_2023} and more or less in time coincidence with the neutrino detection. However, given the cosmic SN rate, there is the possibility that this is a random association \citep{petropoulou_iin_neutrini_2017}, also because the neutrino errorbox is usually a couple of square degrees in size.
Moreover, the current understanding is that even with state-of-the-art neutrino detectors a SN would need to explode within a few Mpc to produce a detectable flux \citep{valtonen_icecube_2023}. However, HE neutrino production depends on the strength of the interaction and the efficiency of the acceleration process. Therefore, it is important to provide empirical constraints on whether interacting SNe can produce HE neutrinos, as well as to characterise their energetics and ejecta/CSM physical conditions.

In this paper, we focus on SNe that show clear narrow lines in their spectra and are thus classified as SNe~IIn. We present a sample of four interacting SNe that displayed a high luminosity and slow spectro-photometric evolution.
The analysis of the follow-up spectro-photometric data is used to derive global physical parameters such as kinetic energy, mass-loss rate, and mass of the CSM and to investigate their similarities and differences, as well as to explore the potential production of HE neutrinos.
The paper is structured as follows: we give an overview of the objects and of their host galaxies in Sect.~\ref{sec:sne}. In Sect.~\ref{sec:bol_altre} we derive and analyse the bolometric light curves. In Sect.~\ref{sec:pitik} we derive the CSM mass and radius and mass-loss rate and confront the parameters with theoretical predictions for neutrino production. Finally, we summarise our work and present our conclusions in Sect.~\ref{sec:disc_altre}.

\section{The sample}
\label{sec:sne}
Our goal is the selection of strongly-interacting SNe among all the SNe classified as IIn. An example of such objects found in the literature is SN~2010jl. These transients are the most appealing candidates for emitting HE neutrinos, although other less powerful objects cannot be ruled out. The SNe that correspond to our criteria in the literature are quite rare and often with little data, especially at very late phases. In this work, we build a sample of well-observed strongly-interacting SNe to analyse their properties. The criteria we use to identify the new transients may be useful in the future to enlarge this sample.

The sample was built selecting newly-discovered transients announced through services such as AstroNotes\footnote{\url{https://www.wis-tns.org/astronotes}} and Astronomer's Telegrams\footnote{\url{https://astronomerstelegram.org}} that met our selection criteria. We evaluated both the classification spectrum and the light curve and selected SNe classified as Type~IIn, with a relatively fast rising time 
among the luminous SNe ($t_{rise}<40$~days), and high but not extreme luminosity ($L_{peak}\sim10^{43}\;\mathrm{erg\,s^{-1}}$). These criteria tend to exclude SLSNe,  which are at least one order of magnitude more luminous ($L_{peak}\gtrsim10^{44}\;\mathrm{erg\,s^{-1}}$ or $<-21$~mag) and tend to have a longer rise time to peak ($t_{rise}\gtrsim70$~d) \citep{lunnan_slsne_2015}. 
Despite the claims in \citet{hiramatsu_IIn_2024}, the debate on the mechanism powering SLSNe is not fully settled, therefore, we decided to exclude them altogether from the sample, although this does not mean that they cannot be powered by interaction or produce HE neutrinos. Also, a preference was given to nearby events (typically within redshift $z\sim0.1$, so that a more complete follow-up could be ensured).
Following our procedure, we selected 11 targets in three years, which are listed in Tab.~\ref{tab:tutte}, together with their coordinates, redshift, distance modulus, and galactic absorption. We also report some notable parameters derived from the pseudo-bolometric $g,r,i$ light curve (for details on the computation, see Sect.~\ref{sec:bol_altre}), namely: the rise time $t_{rise}$, defined as the time between the last non-detection and the first peak; the peak luminosity $L_{peak}$; the integrated luminosity in the first $\sim$200~days $E_{200}$. 
These parameters will be used as criteria to ensure that the sample is not polluted by objects that show interaction at very late times or that underwent brief bursts of interaction over short periods of time.
The case of SN~2021adxl is a bit peculiar since its poorly-constrained rise time is $<91$~days, but given the light curve shape and luminosity, very similar to SN~2010jl \citep{fransson_2010jl_2014}, it is likely that the explosion date was shortly before the peak. Therefore, we elected to keep it in the sample.

\begin{table*}
\caption{SNe we selected for follow up.}
    \centering
    \resizebox{\textwidth}{!}{\begin{tabular}{cccccccccc}
    \hline\hline
       SN & $\alpha$  & $\delta$  & Redshift & Distance& $A_V$\tablefootmark{*} & $t_{rise}$ & $t_{rise,min}$ & $L_{peak}$ & $E_{200}$ \\
       & (hh:mm:ss) & ($^{\circ}:':''$) & & modulus (mag) & (mag) & (days) & (days) & $(\times 10^{43}\; erg\,s^{-1})$ & $(\times 10^{49}\; erg)$\\\hline
        SN~2021crx & 13:21:53.976	& +08:36:20.34	  & $0.067$\tablefootmark{a} & 37.36 & 0.069 & $<49$ & 34 & 0.2 & 0.5\\
        SN~2021gci & 18:34:40.186	& +22:53:42.02  & $0.084$\tablefootmark{b} & 37.69 & 0.401 & $<34$ & 25 & 0.3 & 1.0\\
        SN~2021kwj & 18:19:49.510 &	+56:10:01.20  & $0.025$\tablefootmark{b}& 35.12 & 0.110 & $<32$ & 23 & 0.1 & 0.3\\
        SN~2021qim & 12:42:45.210	& +73:08:20.33  & $0.031$\tablefootmark{b} & 35.65 & 0.069 & $<26$ & 24 & 0.2 & 0.02\\
        SN~2021acya & 04:02:13.760 &	-28:23:29.72  & $0.06203$\tablefootmark{a} & 37.10  & 0.037 & $<48$ & 33 & 3.0 & 10\\
        SN~2021adxl & 11:48:06.940 &	-12:38:41.71  & $0.01790$\tablefootmark{a} & 34.51 & 0.080 & $<91$ & 8 & 1.7 & 9 \\
        SN~2022iaz & 12:30:31.315 &	-19:04:42.13  & $0.067$\tablefootmark{b} & 37.39 & 0.111 & $<16$ & 5 & 1.0 & 0.6\\
        SN~2022owx & 14:24:36.230	& +04:33:30.49  & $0.026$\tablefootmark{a} & 35.35 & 0.083 & $<30$ & 27 & 0.5 & 0.3\\
        SN~2022qml & 22:29:45.502	& +13:38:24.11  & $0.0473$\tablefootmark{a} & 36.60 & 0.167 & $<37$ & 34 & 0.9 & 4 \\
        SN~2022wed & 07:24:15.497	& +19:04:52.71	  & $0.116$\tablefootmark{a}& 38.56 & 0.148 & $<22$ & 17 & 0.8 (first peak) & 7\\
        & & & & & & & & 1.4 (second peak) & \\
        SN2023awp & 15:30:01.536	& +12:59:15.15  & $0.014$\tablefootmark{a} & 34.79 & 0.105 & $<18$ & 14 & 0.1 & 0.2\\\hline
    \end{tabular}}
    \tablefoot{$t_{rise}$ is the time between the last non-detection and the first peak, while $t_{rise,min}$ is the time between the first detection and the first peak. $L_{peak}$ is the peak luminosity and $E_{200}$ is the integrated luminosity in the first 200~days since discovery.\\
    \tablefoottext{a}{Heliocentric redshift was measured by averaging the position of the narrow emission lines in our observed spectra.}\\
    \tablefoottext{b}{Retrieved from the Transient Name Server (TNS).}\\
   \tablefoottext{*}{Galactic extinction along the line of sight was retrieved from the NASA/IPAC Extragalactic Database (NED).}}
    \label{tab:tutte}
    \end{table*}

\begin{figure}[!htb]
    \centering
    \includegraphics[width=\columnwidth]{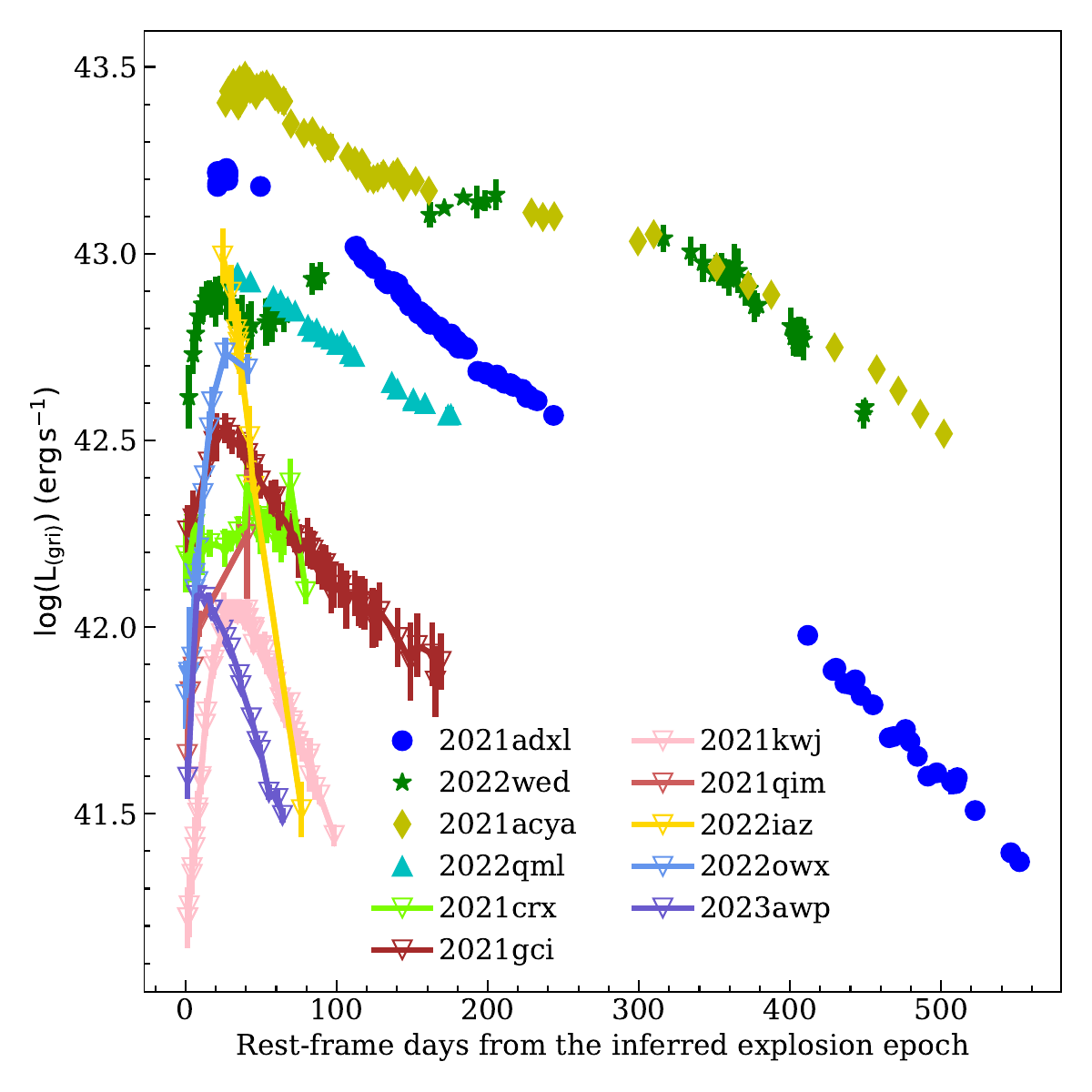}
    \caption{Pseudo-bolometric $g,r,i$ light curves of all the SNe we followed with our program. Empty reverse triangles indicate the SNe that, after further analysis, were excluded from the final sample. All the phases are corrected for time dilation (this is true throughout the paper).}
    \label{fig:lc_escluse}
\end{figure}

As we will detail in the following, it turned out that not all SNe we selected with our criteria proved to evolve as expected and 7 out of 11 candidates were excluded from more detailed analysis for not showing signs of strong interaction. Their light curve was generally dimmer than the four strongly-interacting SNe that remained in the sample and faded faster. This can clearly be seen in Fig.~\ref{fig:lc_escluse}, where we show their pseudo-bolometric light curve built using only optical filters (see Sect.~\ref{sec:bol_altre} for the computation).

\begin{figure}[!htb]
    \centering
    \includegraphics[width=\columnwidth]{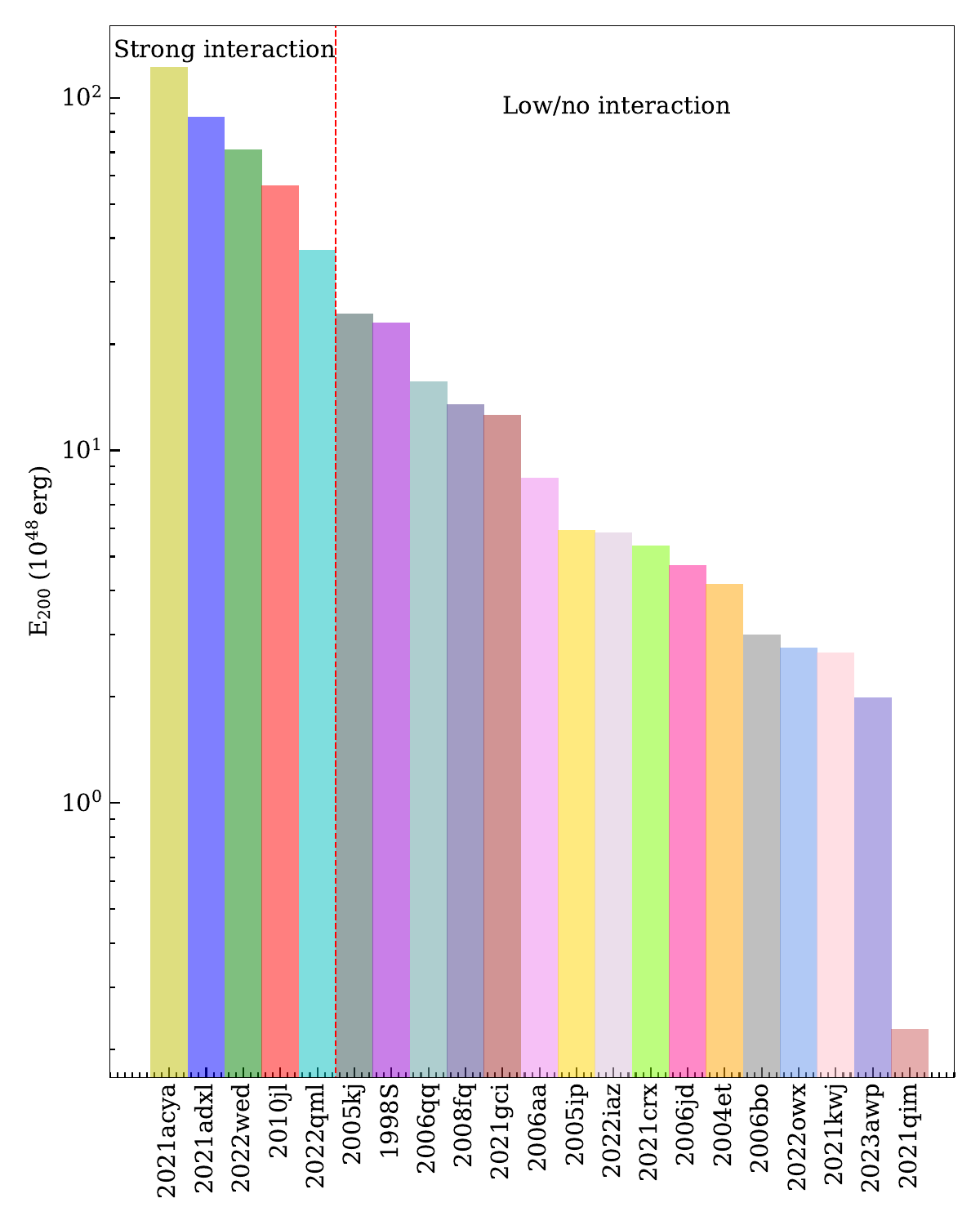}
    \caption{Integrated luminosity in the first 200~days for each SN in our sample and some objects 
   from literature for comparison. The dashed vertical line divides the strongly-interacting sub-sample from the more mildly-interacting SNe.}
    \label{fig:hist_lum}
\end{figure}

To discriminate the transients that display strong interaction, we used the measured luminosity integrated over the first 200$\,\rm{days}$ and imposed that this parameter $E_{200} \geq 3\times10^{49}\;\mathrm{erg}$. This criterion 
is determined a-posteriori to identify the objects with prolonged, strong interaction. In fact, with longer integration times, the contamination arises from objects undergoing prolonged but mild interaction 
(\citealp[e.g., SN~2005ip,][$E_{200} \leq 1\times10^{49}\;\mathrm{erg}$]{taddia_carnegie_2013}). Conversely, with shorter integration times there is significant contamination from objects experiencing strong interaction for a limited period (e.g., SN~1998S). Our parameter is a compromise between the two cases and ensures that all the selected SNe are strongly-interacting. 
The excluded SNe had a lower energy display in the first 200~days compared to the selected ones. This can be seen from Fig.~\ref{fig:hist_lum}, where we also add the values calculated for the strongly-interacting SN~2010jl, the mildly-interacting SN~1998S, and the normal Type~IIP SN~2004et \citep{sahu_04et_2006} as reference for different amounts of interaction. 
We also add the seven SNe~IIn in the sample by \citet{taddia_carnegie_2013}: SNe~2005ip, 2005kj, 2006aa, 2006bo, 6006jd, 2006qq, and 2008fq. These prove to be polluters, since, although they show signs of interaction in the spectra, they are less luminous and their light curve has a shorter duration, thus implying the presence of less interaction.
An intermediate case is that of SN~2021gci, which has significantly more energy than the others, and with a more luminous first peak, albeit still one order of magnitude below the selected transients. On the other hand, SNe~2022iaz and 2022owx have more luminous peaks but faded too fast, providing a small total amount of energy.

This exercise denotes that a single spectrum and the early light curve is not enough to discern between strongly-interacting and normal interacting SNe, thus causing the sample to be contaminated with objects that have to be rejected later. The very minimum information required to identify a strongly-interacting SN includes at least photometric coverage of 2/3 months and a couple of good resolution spectra with enough signal-to-noise (S/N) at different epochs to check the evolution. Ideally, one should have a well-sampled light curve for the first 200~days to properly distinguish between strongly-interacting SNe and SNe with low interaction based on the energy input.

The four SNe selected here that fulfill our criteria are then 2021acya, 2021adxl, 2022qml, and 2022wed. In the following, we will describe each object in more detail. We note 
that a paper presenting detailed observations for one of the targets in our list, SN~2021adxl, was recently published \citep{brennan_adxl_2024}. We will refer to their observations and results in the relevant sections.

\begin{figure}
    \centering
    \includegraphics[width=\columnwidth]{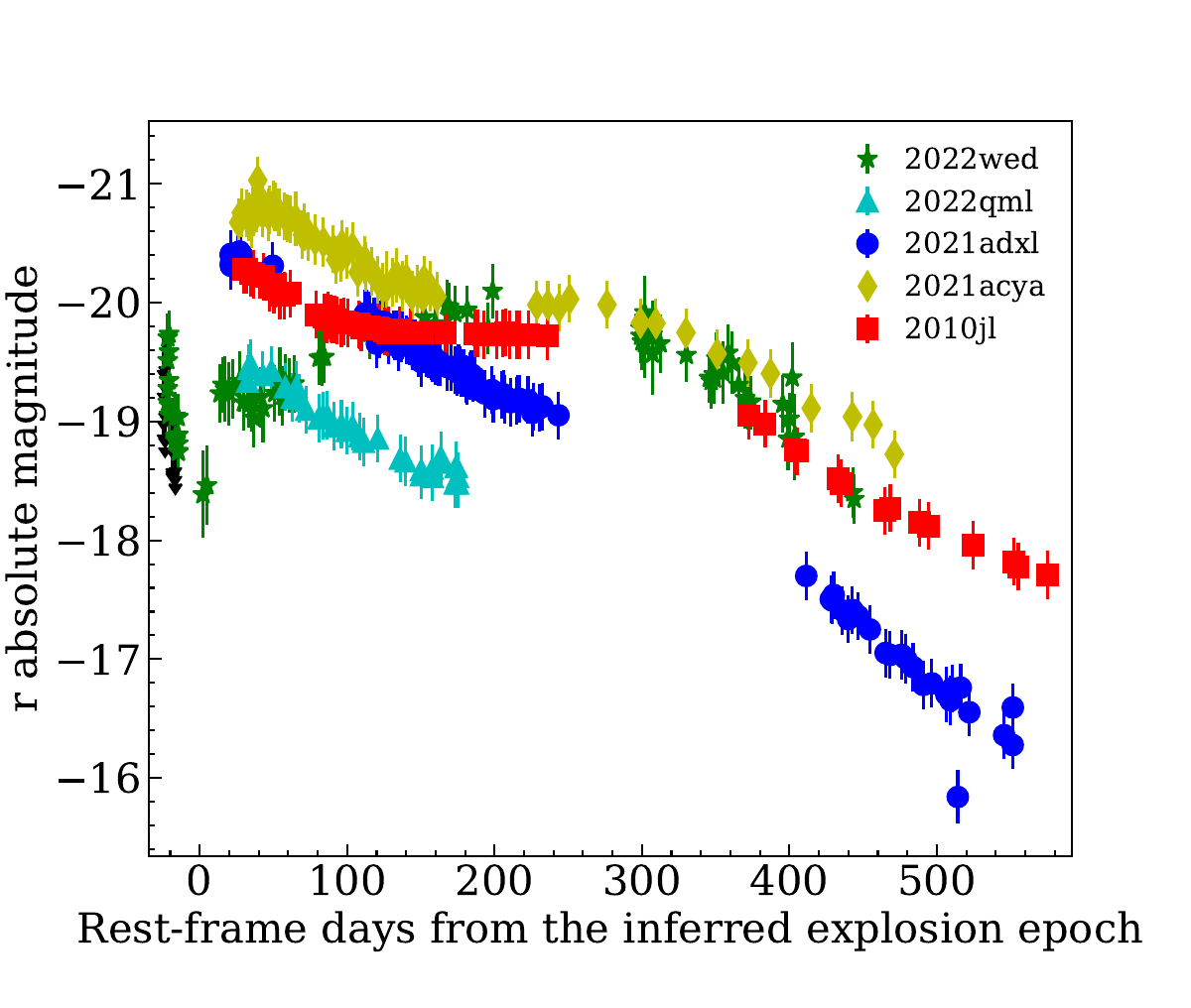}
    \caption{Absolute magnitude $r$-band light curves for the SNe~IIn in our sample as well as the well-studied Type~IIn SN~2010jl \citep{smith_2010jl_2012,fransson_2010jl_2014} as comparison. Black arrows indicate upper limits. All photometry is K-corrected and the phases are corrected for time dilation.}
    \label{fig:r-abs}
\end{figure}

In Fig.~\ref{fig:r-abs}, we show the $r$-band absolute light curves for the four SNe in our sample along with SN~2010jl. This SN will be our reference throughout the whole analysis by virtue of its spectro-photometric characteristics, which perfectly fall within our parameters. Details on the observations and data reduction are given in Sects.~\ref{sec:obs} and \ref{sec:redu}. As the explosion date, we take the middle epoch between the last non-detection and the discovery date for all SNe but for SN~2021adxl (cf. Sect.~\ref{subsec:adxl}). To compute the absolute magnitude, we corrected the apparent magnitudes for Galactic extinction and distance (see Sect.~\ref{subsec:reddening} for details). The transients span a range of 2~mag in absolute magnitude at maximum light and show different light curve shapes, although all of them are in general more long-lasting than ordinary Type~II SNe, which fade on faster timescales ($\sim $2~mag in 100 days).

\subsection{SN~2021acya}
\label{subsec:acya}
SN~2021acya was discovered on $30^{\rm{th}}$ October 2021 (MJD 59518.029) in the $orange$ band by the Asteroid Terrestrial-impact Last Alert System (ATLAS, \citealt{tonry_atlas_2018}) at a magnitude of 18.124, while the last non-detection was only two days prior \citep{tonry_disc_acya_2021}. As the explosion epoch, then, we took MJD $59517\pm1$. It was then classified as SN~IIn on $25^{\rm{th}}$ November 2021 \citep{ragosta_class_acya}.

The rise to the peak is relatively slow, as it reaches the brightest absolute magnitude of $-20.3\pm0.1$ in the $r$ band on MJD~$59537\pm1$, 20$\,\rm{days}$ after the first detection. A slow decline follows the peak, which flattens into a plateau at $\sim$160$\,\rm{days}$ lasting for $\sim150\,\rm{days}$. A linear luminosity decline then resumes and lasts up to 480$\,\rm{days}$ after the explosion, when the SN is finally lost. The shape of the late-time $r$-band light curve from the starting of the plateau matches well that of SN~2010jl (Fig.~\ref{fig:r-abs}) but it is slightly brighter at all phases, probably indicating a stronger interaction in the case of SN~2021acya.

\begin{figure*}
    \centering
    \includegraphics[width=\textwidth]{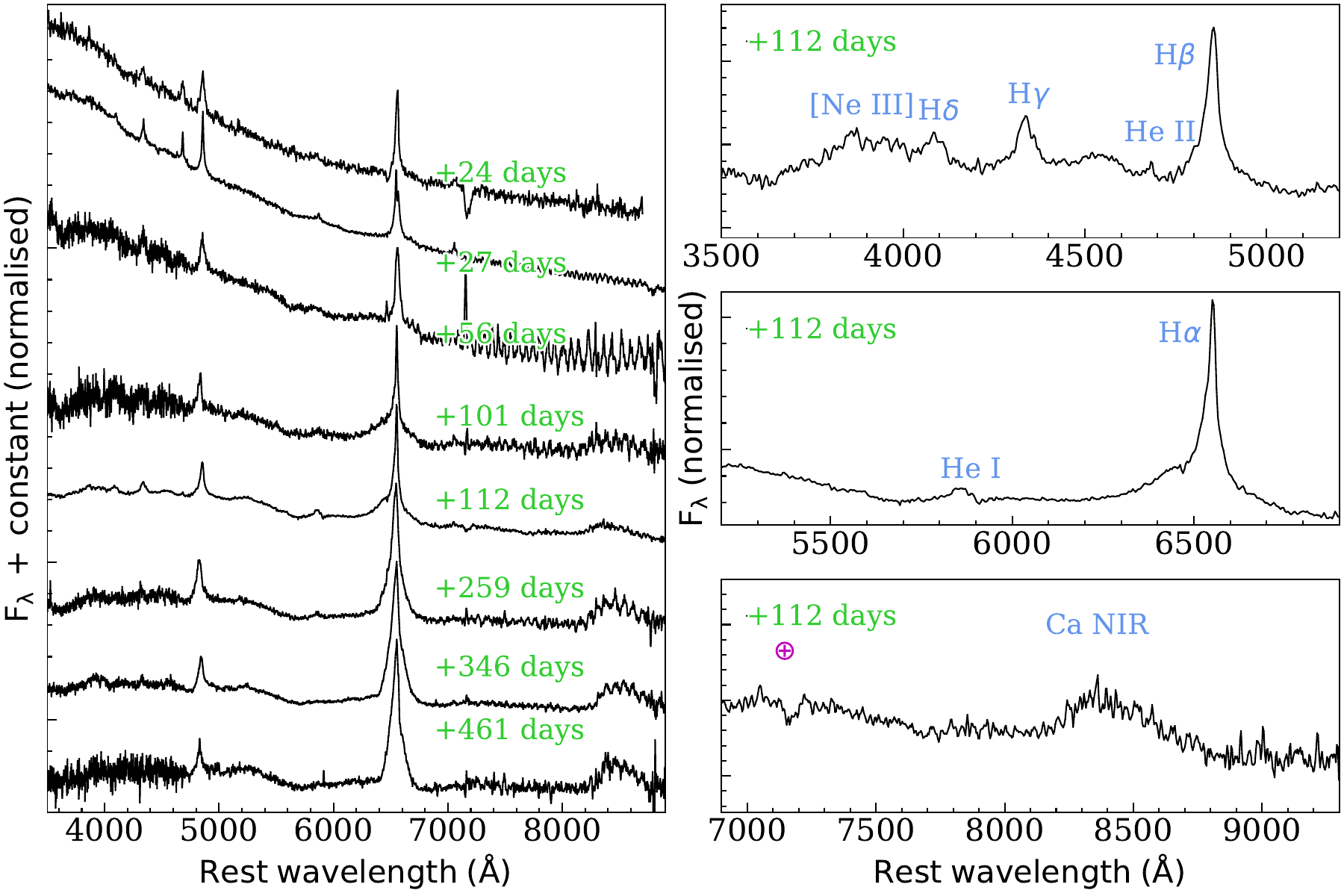}
    \caption{Spectral evolution of SN~2021acya through the most significant spectra. Numbers close to the spectra indicate the phase from explosion corrected for time dilation. All spectra are scaled with respect to the H$\alpha$ and arbitrarily shifted for better visualisation. The spectra are not corrected for extinction.}
    \label{fig:spec_evol_acya}
\end{figure*}

The spectral evolution of SN~2021acya is shown in Fig.~\ref{fig:spec_evol_acya}. At first glance, the spectra seem to show a very slow evolution, which is typical of long interacting SNe. At a closer look, however, there are significant changes in the continuum shape and the width of the emission lines.
The first spectrum of SN~2021acya at +24$\,\rm{days}$ shows a hot (10000~K) continuum and the only prominent features are Balmer emission lines. The line profile is almost symmetrical, with an electron scattering profile. Therefore, the width of the lines cannot be used to trace the bulk motion of the gas. The spectrum at +27$\,\rm{days}$ is slightly cooler and we can detect emission from \ion{He}{I}~$\lambda5876$. Afterwards, the continuum cools down to 8000~K and, from phase +56$\,\rm{days}$, a bump starts to emerge in the bluer part of the spectrum that becomes more and more evident. This feature is usually attributed to a plethora of Fe lines, such as \ion{Fe}{II}, that show up when the medium is heated by a strong shock \citep{chevalier_iin_shock_1994, mazzali_1998bw_2001}. At +101$\,\rm{days}$ a broad emission from the \ion{Ca}{II}~NIR triplet $\lambda\lambda\lambda 8498,8542,8662$ appears, and its intensity increases as the evolution proceeds. On the spectrum at +112$\,\rm{days}$ we tentatively identify [\ion{Ne}{III}]~$\lambda 3869$ but the low S/N makes it difficult to see the feature in other spectra. \ion{He}{I} is visible until +259$\,\rm{days}$, while it is not detected in the spectrum at +346$\,\rm{days}$. The following spectra are all similar and dominated by H$\alpha$, H$\beta$, and Ca~NIR.

\subsection{SN~2021adxl}
\label{subsec:adxl}

SN~2021adxl was discovered on $3^{\rm{rd}}$ November 2021 (MJD 59521.540) by \citet{fremling_disc_adxl_2021} with a magnitude of 14.41 in the $r$ band but its last non-detection was on MJD 59431.690, almost three months before the discovery. It was then classified as SN~IIn on $2^{\rm{nd}}$ February 2022 \citep{de_class_adxl_2022}. To aid in the comparison, we adopted as explosion date a day closer to the first observation considering the similarity of the bolometric light curve with SN~2010jl, which shows a fast ($\sim20$~days) rise to the peak. In fact, the peak magnitude is only slightly higher while the initial decline is almost identical for the two SNe and also the colour evolution is similar (cf. Sect.~\ref{subsec:reddening}). For these reasons, we arbitrarily chose MJD $59500^{+22}_{-78}$ as the explosion epoch, to match the peak to that of SN~2010jl.
Given this, the luminosity in the $r$ band peaks at $-20.4\pm0.2$~mag on MJD~$59529\pm4$, 29 days after our estimated explosion date.

\begin{figure*}
    \centering    \includegraphics[width=\textwidth]{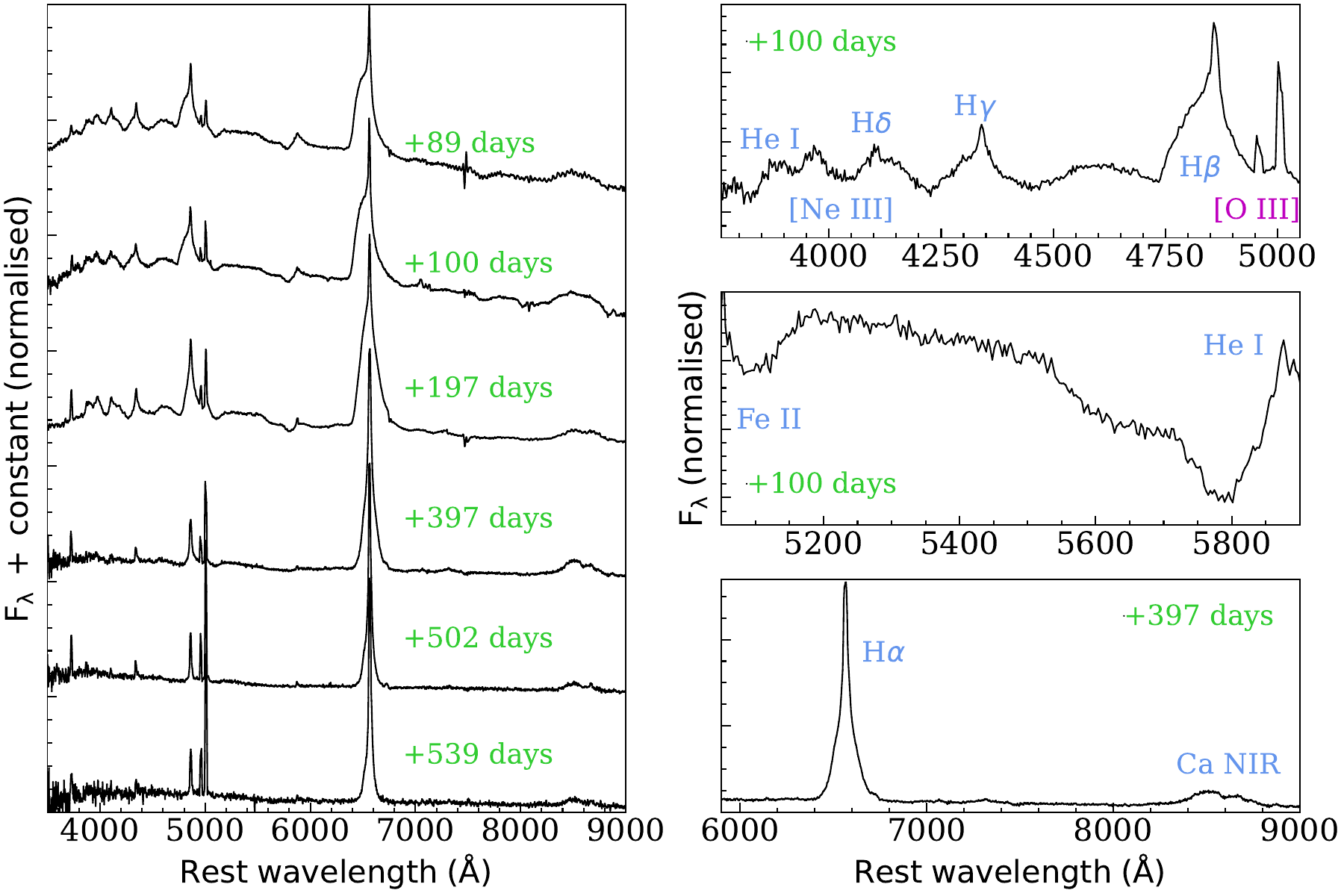}
    \caption{Spectral evolution of SN~2021adxl through the most significant spectra. Numbers close to the spectra indicate the phase from explosion corrected for time dilation. All spectra are scaled with respect to the H$\alpha$ and arbitrarily shifted for better visualisation. The spectra are not corrected for extinction.}
    \label{fig:spec_evol_adxl}
\end{figure*}

The spectral evolution of SN~2021adxl is shown in Fig.~\ref{fig:spec_evol_adxl}. The earliest spectrum of SN~2021adxl was taken at +89$\,\rm{days}$ and the continuum is already almost flat. The main emission lines are those of the Balmer series and \ion{He}{I}~$\lambda5876$, the latter showing a distinct P-Cygni profile, as well as a small bump from \ion{Ca}{II}~NIR. The blue bump due to the \ion{Fe}{II} forest, in this case, is less accentuated. We also clearly see the absorption component of the P-Cygni profile of \ion{Fe}{II}~$\lambda 5169$.
The presence of broad He and Fe P-Cygni profiles when they are absent in H in a SN~IIn is unusual and will be further discussed in Sect.~\ref{subsec:multifit}. 
There are also narrow emissions from [\ion{O}{III}]~$\lambda\lambda4959,5007$ due to the background starburst region. Notably, the H$\alpha$ and H$\beta$ lines are highly asymmetric. The profile has a flat top and electron scattering wings on the sides but with a distinct blue shoulder that was observed in SN~2013L \citep{taddia_2013L_2020}. This composite profile is attributed to the shock front being exposed, giving the boxy profile, and to electron scattering forming the wings \citep{taddia_2013L_2020}. The later spectra are almost identical until phase +197$\,\rm{days}$, when the Balmer lines have nearly symmetrical profiles with electron scattering wings. At this point, the P-Cygni profiles on \ion{He}{I} and \ion{Fe}{II} have also disappeared. The evolution then proceeds with the shrinking of the broad emission lines until the last spectrum at +~539$\,\rm{days}$, which is dominated by a faint, broad H$\alpha$ emission. We do not detect high ionisation lines, probably because the spectral resolution is not very high and the intensity of these features is low. However, \citet{brennan_adxl_2024} identify [\ion{Ne}{V}]~$\lambda 3346$, [\ion{O}{III}]~$\lambda 4363$, [\ion{Ne}{V}]~$\lambda 3346$, [\ion{Ca}{V}]~$\lambda 6086$, [\ion{Fe}{VII}]~$\lambda 6087$, and [\ion{Fe}{X}]~$\lambda 6365$ in their spectrum at +480~days.

\begin{figure}
\centering
    \includegraphics[width=\columnwidth]{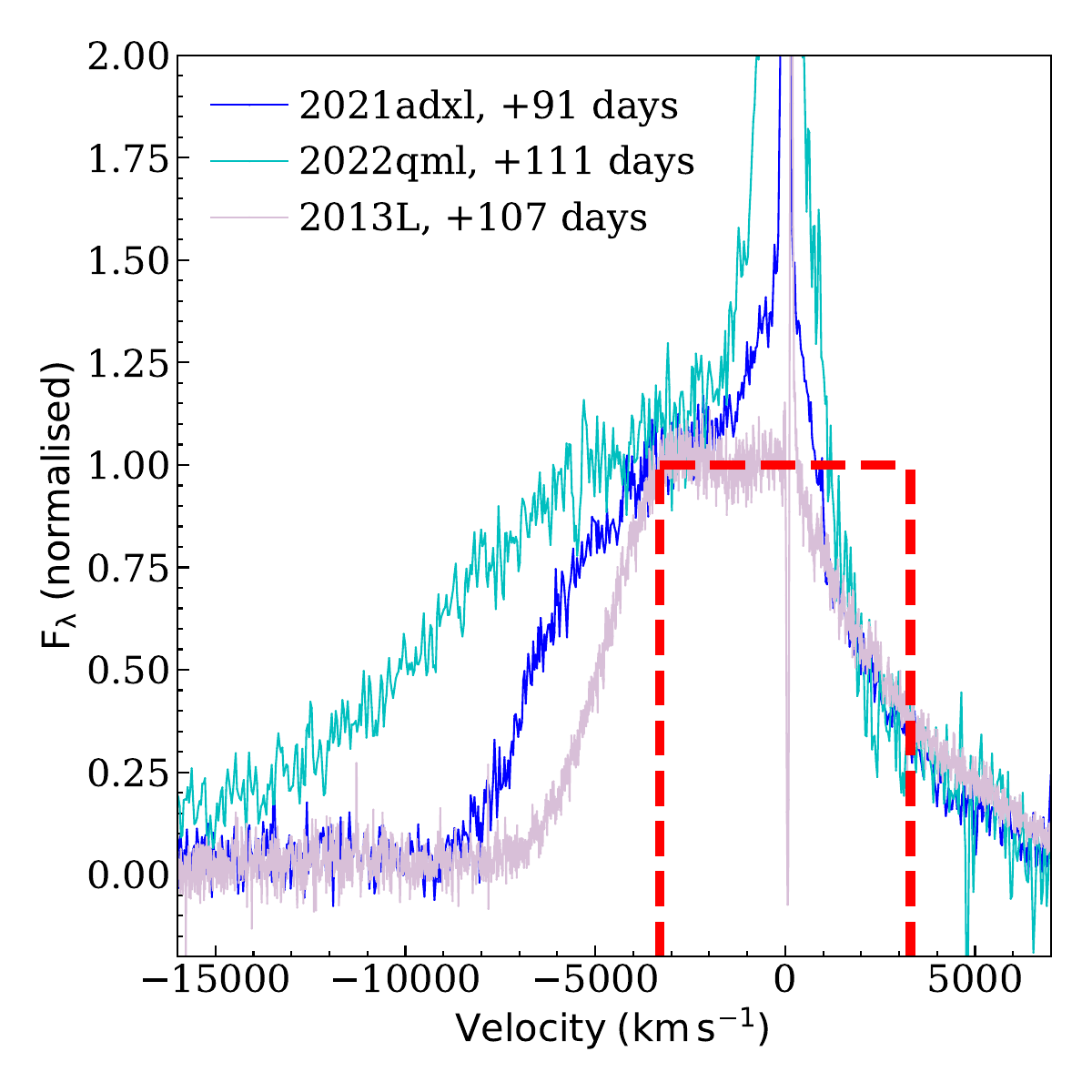}
    \caption{Spectral comparison between SNe~2021adxl, 2022qml, and 2013L from \cite{taddia_2013L_2020} (zoom on the H$\alpha$ line). The red, dashed line highlights the boxy region of the line profile due to the shock. All spectra are redshift-corrected, continuum-subtracted, and rescaled for better visualisation.}
    \label{fig:confronto_adxl_13l}
\end{figure}

In Fig.~\ref{fig:confronto_adxl_13l}, we show a zoom-in on the H$\alpha$ region on the spectrum of SN~2021adxl at +91$\,\rm{days}$ compared to a spectrum of SN~2013L \citep{taddia_2013L_2020} at a similar phase, which showed a peculiar blue shoulder on the broad H$\alpha$ emission. \citet{taddia_2013L_2020} explain the profile as the combination of a boxy profile originating in the shocked shell with the extended wings caused by electron scattering, where the red side of the boxy component would be lost due to occultation of the receding shocked shell by the inner ejecta with high optical depth. To show this, a red, dashed box is added to the plot, roughly indicating the profile of a pure shocked shell with a velocity of $3300\;\rm{km\,s^{-1}}$, devoid of electron scattering. On the blue side, the box is matched to the blue shoulder, while its reflection on the red side shows the missing flux. For such strong occultation, the line-emitting region must be located just above the opaque ejecta (the similar profile of SN~2022qml will be commented in the next section).

\subsection{SN~2022qml}
\label{subsec:qml}

The SN was discovered on $2^{\rm{nd}}$ August 2022 (MJD~59794.030) at a magnitude 18.137 in the $cyan$ band \citep{tonry_disc_qml_2022} and classified as SN~IIn on $27^{\rm{th}}$ August 2022 \citep{gutierrez_class_qml_2022}. The peak is bright, at $-19.46\pm0.1$~mag in the $r$ band on MJD~$59800\pm1$, and showing a linearly declining light curve after that. The rise was very rapid, given the last non-detection only one day before the discovery. We adopt a best estimate of the explosion epoch as MJD~$59793\pm1$.

\begin{figure*}
    \centering
    \includegraphics[width=\textwidth]{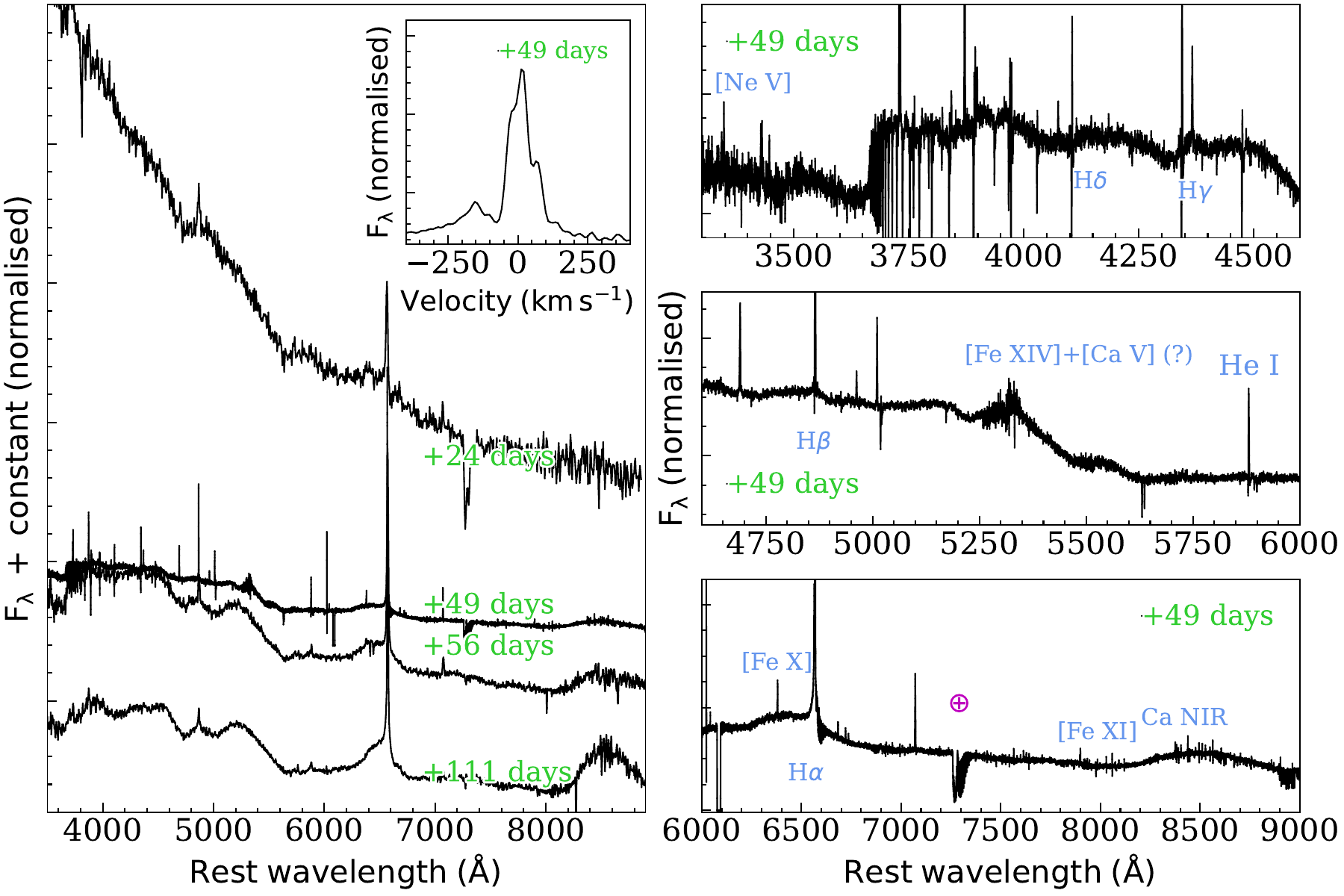}
    \caption{Spectral evolution of SN~2022qml through the most significant spectra. Numbers close to the spectra indicate the phase from explosion corrected for time dilation. All spectra are scaled with respect to the H$\alpha$ and arbitrarily shifted for better visualisation. The spectra are not corrected for extinction. We also plot the narrow H$\alpha$ in velocity space to show the P-Cygni profile due to the wind that formed the CSM.}
    \label{fig:spec_evol_qml}
\end{figure*}

The spectral evolution of SN~2022qml is shown in Fig.~\ref{fig:spec_evol_qml}. The spectrum at +24$\,\rm{days}$ has a blue continuum, implying a high temperature, and the only notable feature is the narrow H$\alpha$ on top of a very broad but shallow emission. The spectrum at +49$\,\rm{days}$ starts to show a distinctive blue bump due to a forest of \ion{Fe}{II} lines and a broad, asymmetric H$\alpha$ with a boxy profile smoothed by electron scattering wings. We also identify the narrow line of [\ion{Fe}{X}]~$\lambda6375$ and, possibly, [\ion{Ne}{V}]~$\lambda3426$, [\ion{Fe}{XI}]~$\lambda7892$ and a broader emission that could be due to a blend of coronal [\ion{Fe}{IV}]~$\lambda5303$ with [\ion{Ca}{V}]~$\lambda5309$. The absorption in the P-Cygni profile of the narrow H$\alpha$ is also clearly visible, which allows for measuring the velocity of the progenitor wind
\footnote{We refer to it as wind for simplicity but the CSM could have been produced also by eruption or binary interaction. This is true throughout the paper.}
($\sim100\;\mathrm{km\,s^{-1}}$). This feature is not identifiable in the other spectra of this object due to their lower resolution. At +56$\,\rm{days}$ the only prominent features are Balmer lines, [\ion{Fe}{X}]~$\lambda6375$, and \ion{He}{I}~$\lambda5876$. After that, the high ionisation lines disappear while the spectra remain dominated by the H$\alpha$, until phase +111$\,\rm{days}$, when the \ion{Ca}{II}~NIR $\lambda\lambda\lambda 8498,8542,8662$ emerges with a broad, symmetric emission. A zoom on the H$\alpha$ region of this spectrum is also plotted in Fig.~\ref{fig:confronto_adxl_13l}. The shape of the blue shoulder is similar to SNe~2021adxl and 2013L but it is broader, indicating a higher shock velocity.

The strong blue bump of SN~2022qml is reminiscent of a feature seen in Type~Ia-CSM SNe (e.g., SN~2002ic, \citealt{kotak_2002ic_2004}) but also in Type~Ic SNe (e.g., SN~1998bw \citealt{galama_1998bw_1998,kulkarni_1998bw_1998}). As we mentioned, the presence of the blue bump is usually attributed to a forest of \ion{Fe}{II} lines. In the case of SN~2022qml, the strong bump, which is the most extreme in our sample, suggests a high Fe abundance in the ejecta. A massive star that produces Fe must also show high abundance of O and Ca, as SN~1998bw did. However, SN~2022qml spectra do not show any \ion{O}{I} features and only a small bump likely due to Ca~NIR. Also, the shape of the light curve and its faster decline, compared to the other SNe in our sample, is suggestive of a different origin for this SN. It fact, it is possible that it was not a SN~IIn but rather a Type~Ia-CSM SN. This would be in line with what found by \citet{leloudas_iacsm_2015}, who showed that simulated thermonuclear SN spectra were consistently misclassified as SNe~IIn once the underlying SN flux was 
a fraction $\sim 0.2-0.3$ or below with respect to the continuum. If this is the case, we probably missed the emergence of Si lines because the SN ejecta were still embedded in the CSM cocoon when the SN faded.
    
\subsection{SN~2022wed}
\label{subsec:wed}

SN~2022wed was discovered on $21^{\rm{st}}$ September 2022 (MJD~59843.954) \citep{fremling_disc_wed_2022} at a magnitude 20.46 in the $r$ band, while the last non-detection was on MJD~59839.484. As explosion epoch we chose MJD~$59841\pm2$. It was then classified as SN~IIn on $27^{\rm{st}}$ February 2023 \citep{hiramatsu_class_wed_2023}.

The $r$ light curve shows a first peak at $-19.20\pm0.03$~mag on MJD~$59863\pm2$, 22~days after the explosion. After a short decline lasting around 40$\,\rm{days}$, the light curve shows a second, broader and brighter peak at $-19.68\pm0.01$~mag +230$\,\rm{days}$ after the explosion followed by a very slow decline up to +460$\,\rm{days}$.

\begin{figure*}
    \centering
    \includegraphics[width=\textwidth]{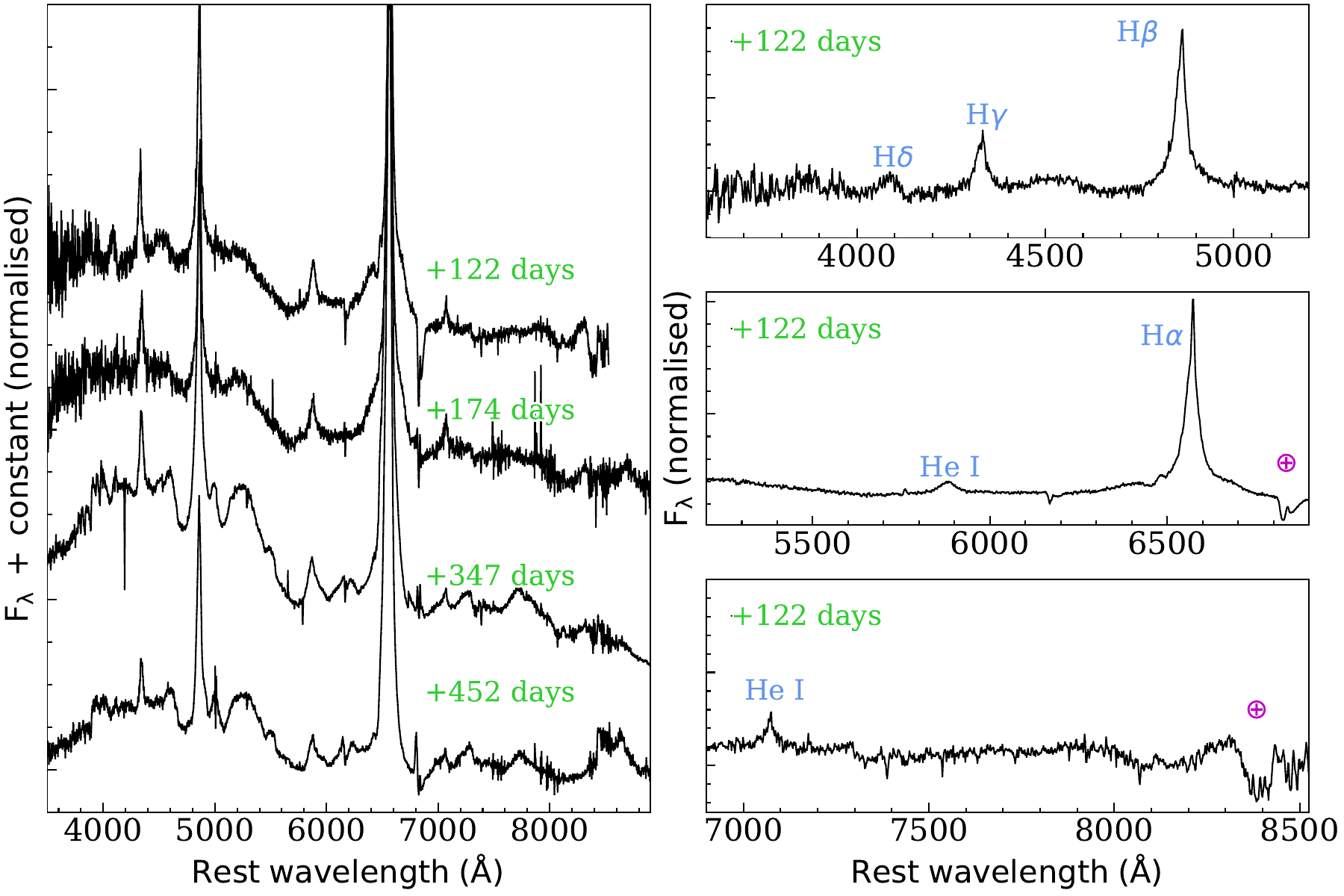}
    \caption{Spectral evolution of SN~2022wed. Numbers close to the spectra indicate the phase from explosion corrected for time dilation. All spectra are scaled with respect to the H$\alpha$ and arbitrarily shifted for better visualisation. The spectra are not corrected for extinction.}
    \label{fig:spec_evol_wed}
\end{figure*}

The spectral evolution of SN~2022wed is shown in Fig.~\ref{fig:spec_evol_wed}. The first spectrum was taken at +122$\,\rm{days}$ and shows an already cool (6000~K) continuum with the distinctive blue bump of \ion{Fe}{II} which, however, is less pronounced than for SN~2022qml. The main emission lines are the Balmer series and \ion{He}{I} at $\lambda5876$ and $\lambda7065$~\AA. The H$\alpha$ profile in this SN is symmetric with electron scattering wings. The spectrum at +347$\,\rm{days}$ shows a small bump at the position of [\ion{Ca}{II}]~$\lambda\lambda 7291,7324$. The H$\alpha$ feature has shrunk considerably and the Lorentzian wings are almost invisible. The last spectrum shares the same features, but H$\alpha$ is narrower and more symmetric, while the blue bump seems to decrease slightly. Moreover, the Ca~NIR triplet $\lambda\lambda\lambda 8498,8542,8662$ finally appears.

\subsection{Host galaxies}
\label{sec:hosts}

\begin{table*}[htbp]
 \caption{Main information for the hosts of our SN sample.}
 \label{tab:hosts}
    \centering
    \begin{tabular}{ccccc}
    \hline\hline
  SN & Host & Absolute mag (NUV) &Absolute mag (\textit{r}) & alternative designation\\\hline
  2021acya & GALEXASC~J040213.69-282329.8 & $-15.3$ & $-17.5$ & $\cdots$ \\
  2021adxl & WISEA~J114806.88-123841.3 & $-16.8$ & $-17.5$ & GALEXASC~J114806.87-123843.3,\\
  & & & & UVQS~J114806.88-123841.3\\
  2022qml & SDSS~J222945.52+133823.5 & $\cdots$ & $-15.9$ & GALEXASC~J222945.55+133823.7,\\
 & & & &  GALEXMSC~J222945.49+133821.1\\
  2022wed & unknown & $\cdots$ &  $-16.6$& $\cdots$\\\hline
    \end{tabular}
    \tablefoot{ The absolute $r$-band magnitudes are measured on our images while those in the NUV are retrieved from NED.}
\end{table*}

\begin{figure}
    \centering
    \includegraphics[width=0.5\columnwidth]{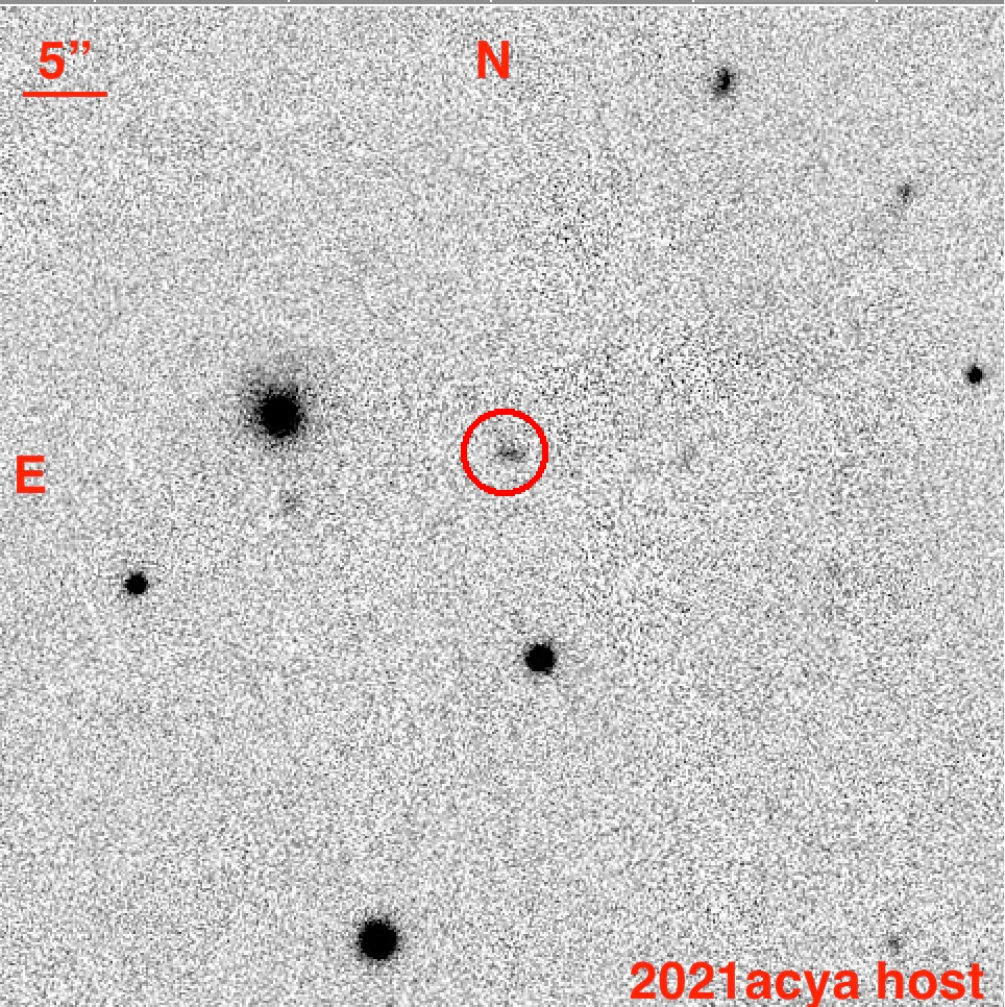}\includegraphics[width=0.5\columnwidth]{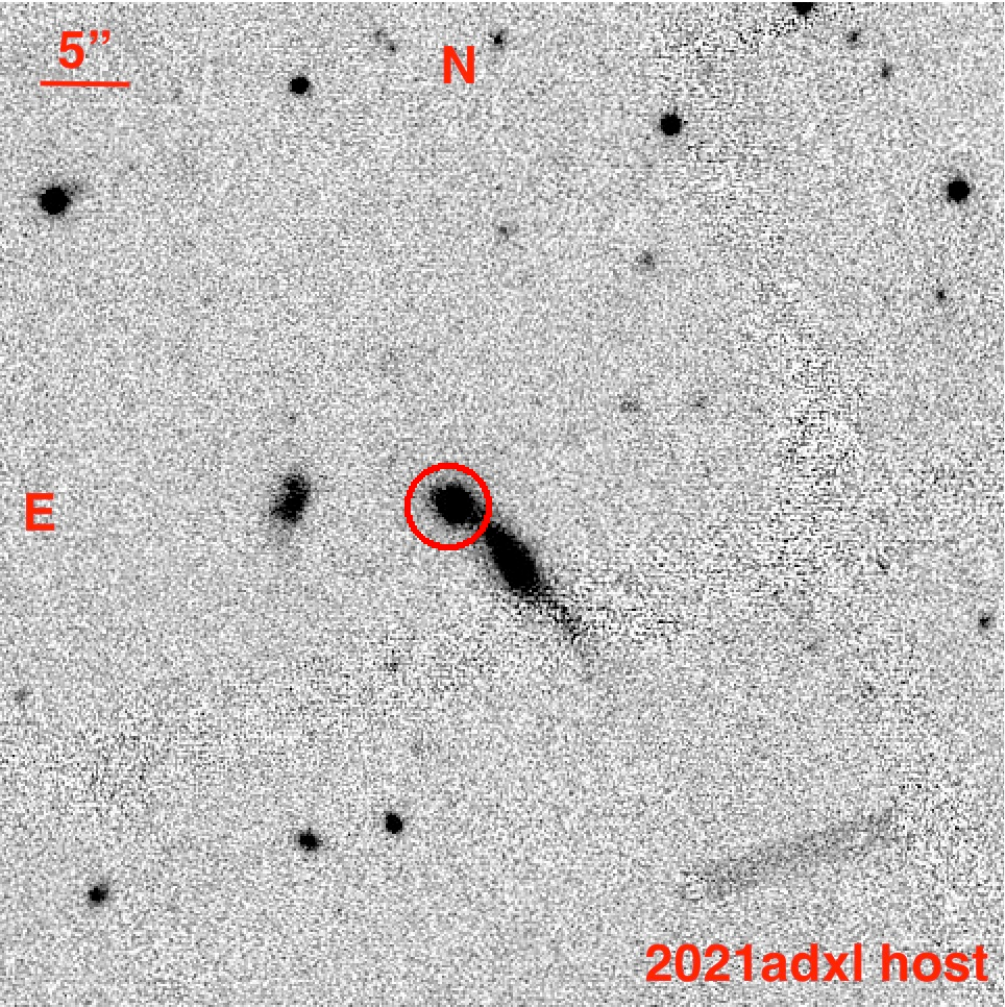}
    \includegraphics[width=0.5\columnwidth]{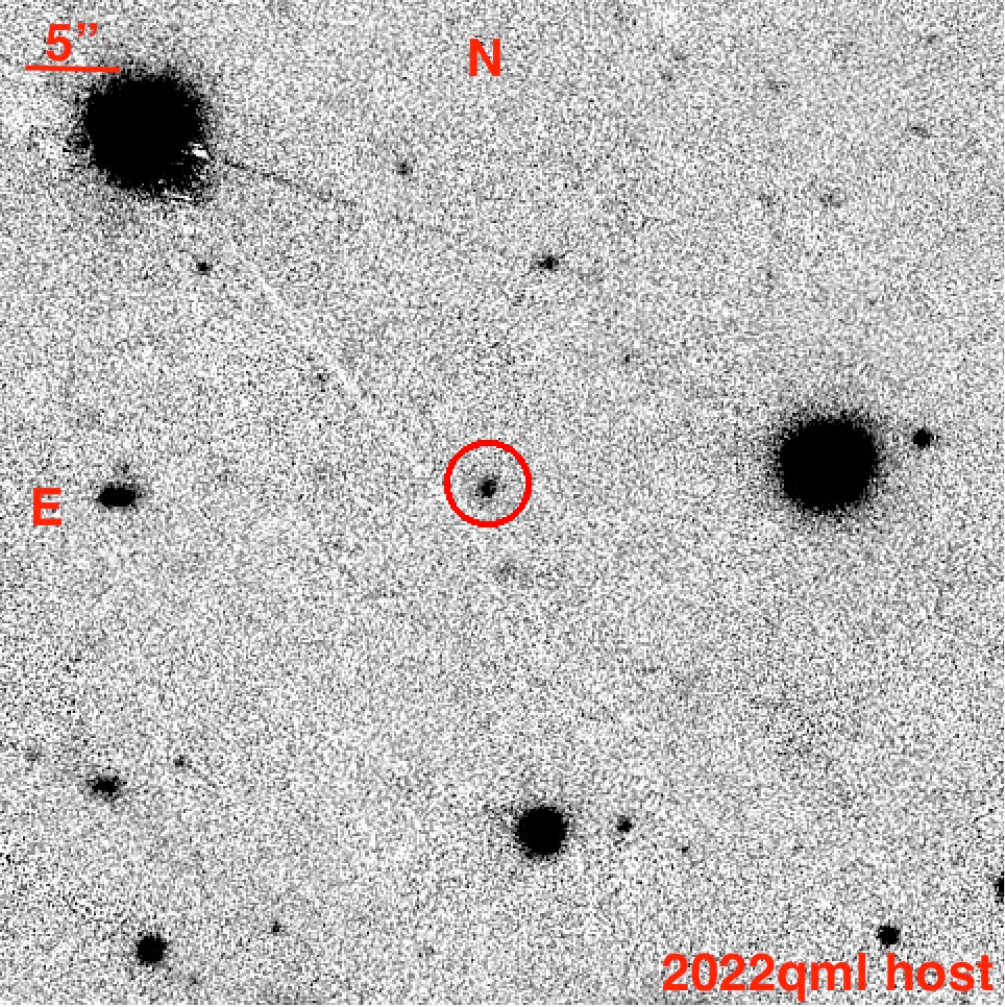}\includegraphics[width=0.5\columnwidth]{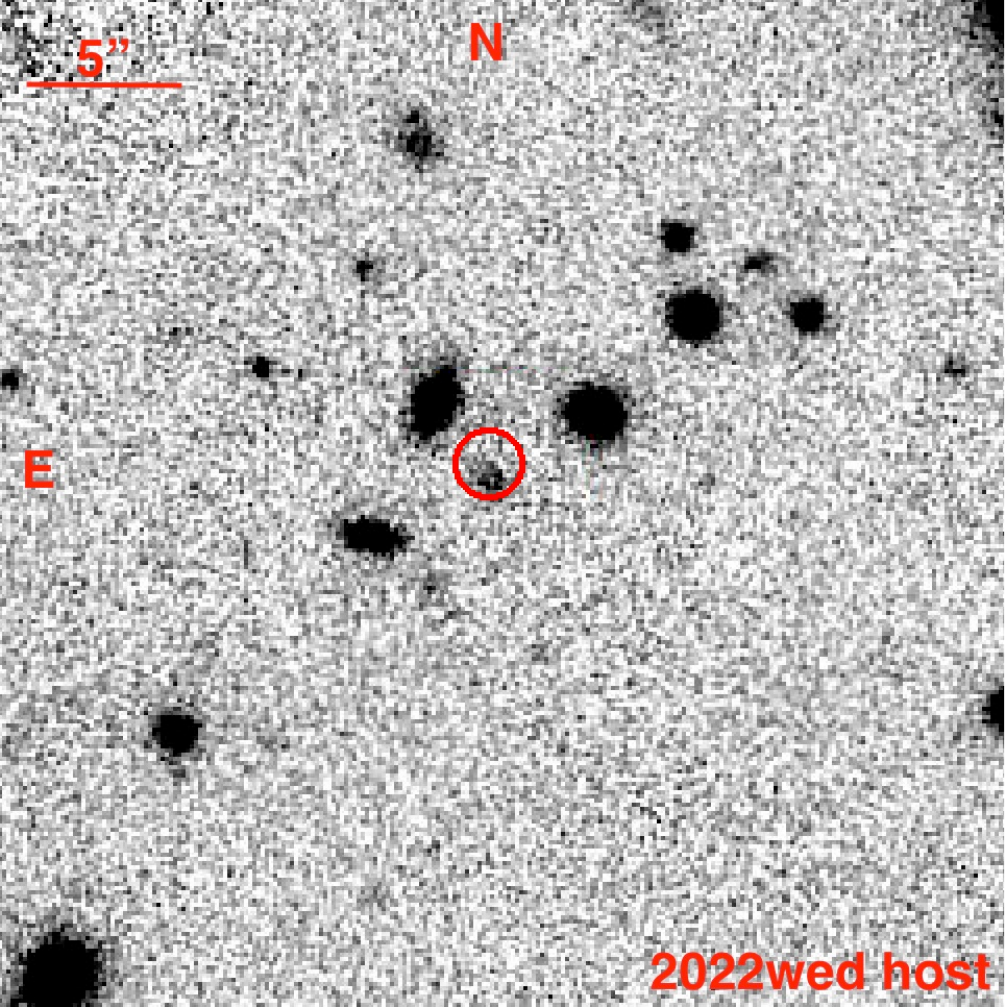}
    \caption{Pre-explosion $r$-band images of the host galaxies of our SNe~IIn retrieved from PS1. Red circles indicate the SNe position. \textit{Top left:} SN~2021acya. \textit{Top right:} SN~2021adxl. \textit{Bottom left:} SN~2022qml. \textit{Bottom right:} SN~2022wed.}
    \label{fig:host}
\end{figure}

The main characteristics of the host galaxies of the SNe in our sample are summarised in Table~\ref{tab:hosts}. The SN host galaxies as retrieved from the Panoramic Survey Telescope and Rapid Response System (Pan-STARRS, PS1, \citealt{ps1}) in the $r$ band are shown in Fig.~\ref{fig:host}.
Interestingly, all known hosts are small, probably dwarf galaxies.

The case of SN~2022wed deserves some discussion, as the SN is close to three IR sources listed on NED (WISEA J072415.72+190456.3, WISEA J072415.05+190455.7, and WISEA J072415.95+190449.4) whose redshifts are unknown. Given the angular separation and assuming these galaxies have the same redshift of the SN, the radial distances from the centre of each galaxy would be 11, 15, and 17~kpc, respectively, which allows for the SN to belong to any of them. However, in the pre-explosion image produced by stacking PS1 observations between MJD $55182-56638$, a faint source can be seen at the location of the SN (see Fig.~\ref{fig:host}, the bottom right panel). Its position is slightly offset from the SN, 0.01 arcsec in right ascension and 0.8 arcsec in declination, which, assuming it is at the same redshift of the SN, gives a radial distance of 1.7~kpc. Its Full-Width-at-Half-Maximum (FWHM) is about 1.3~arcsec, comparable with a point-like source in the image we have. We apply a Point-Spread-Function (PSF) fit to the source and find an apparent magnitude $r=22.4\pm0.1$~mag, which, if we assume the same redshift of the SN, translates to an absolute magnitude of $-16$~mag, similar to that of the host galaxies of the other SNe in the sample.

Motivated by the presence of narrow [\ion{O}{III}] and by the archival image of the host of SN~2021adxl, which shows a bright spot in correspondence with the SN position (see Fig.~\ref{fig:host}, the panel on the upper right), we attempted to use the line ratios of [\ion{N}{II}]~$\lambda6583$/H$\alpha$, [\ion{O}{III}]~$\lambda5007$/H$\beta$, and [\ion{O}{II}]~$\lambda3727$/[\ion{O}{III}]~$\lambda5007$ to derive an estimate of the oxygen abundance (which is assumed to trace the metallicity). We performed the measurements on the spectrum at +539$\,\rm{days}$, since it is the one where the SN~contribution is smaller and the narrow lines due to the \ion{H}{II} region are more evident. Unfortunately, we are only able to measure an upper limit for [\ion{N}{II}]~$\lambda6583$ because the line cannot be resolved from the strong and broad H$\alpha$ but the metallicity is consistent with a solar/subsolar composition according to the N2-versus-O3 calibrator. Based on a similar analysis but with spectra of better resolution, \cite{brennan_adxl_2024} report a subsolar (0.1~Z$_{\odot}$) metallicity. Their relative flux measurements are consistent with ours but for H$\alpha$, which is three times lower than ours. This is not surprising considering that there is probably some residual contamination from the SN. For [\ion{N}{II}] they have an actual measure 100 times smaller than our conservative limit. The metallicity measured by \cite{brennan_adxl_2024} is lower than the ones reported for SNe~2010jl and 2013L, both around 0.3~Z$_{\odot}$ \citep{stoll2011,taddia_2013L_2020}. This is consistent with regions of high star formation in low-mass galaxies such as the host of SN~2021adxl \citep{yates_metal_2012}.
For the other SNe, the hosts are too faint to properly distinguish the structure of the galaxy but, interestingly, the hosts of SNe~2021acya and 2022qml are also UV bright 
(as per the magnitudes reported in NED), which is indicative of a high specific star formation rate (SFR).

\begin{figure}
    \centering
    \includegraphics[width=\columnwidth]{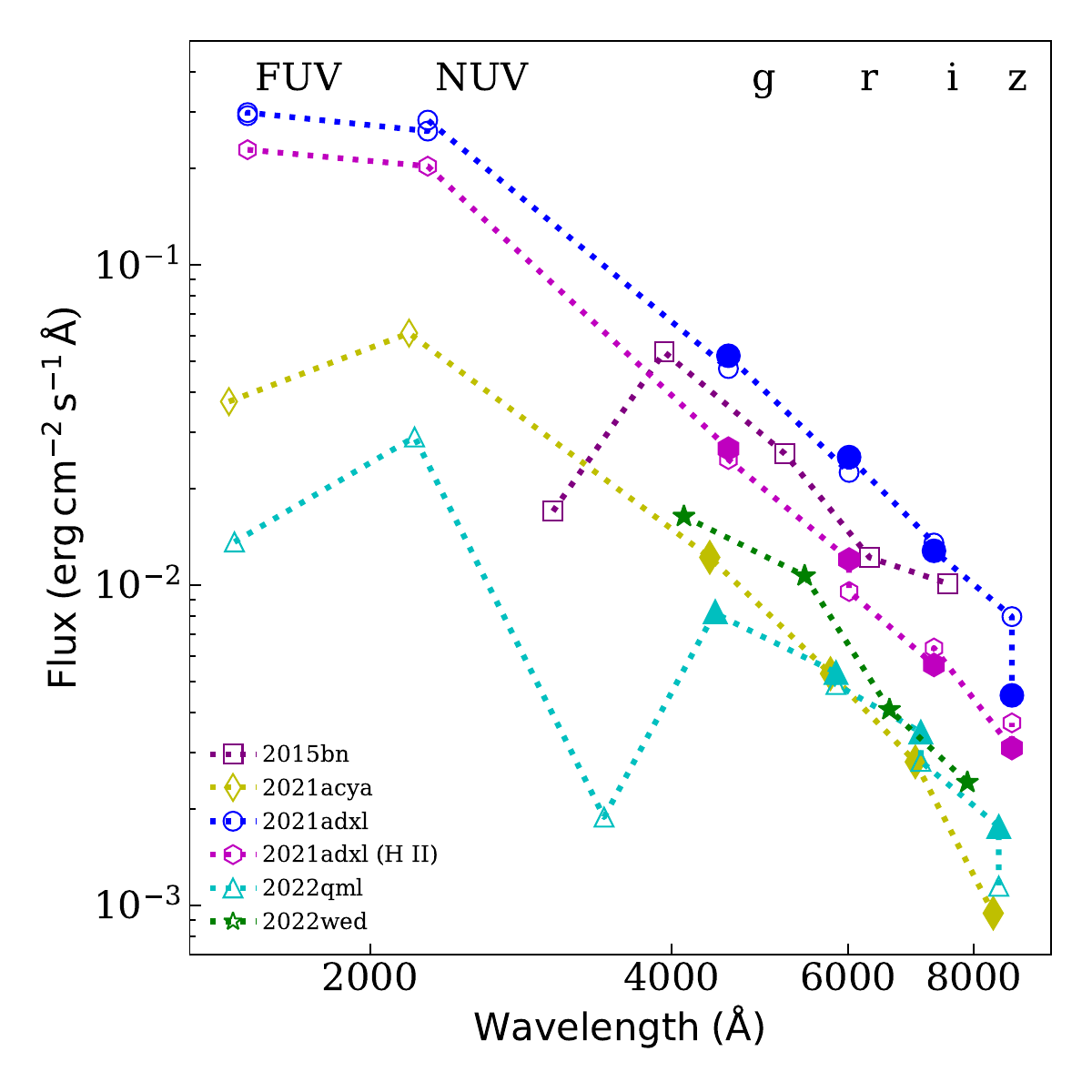}
    \caption{SED of the host galaxies. Larger, filled points represent our measurements on the PS1 template images, while empty points are from NED (or \citealt{brennan_adxl_2024}, in the case of SN~2021adxl). The fluxes are corrected for the luminosity distance and they are shown at the rest frame wavelength position.}
    \label{fig:sed_host}
\end{figure}

We attempted a measure of the spectral energy distributions (SEDs) of the hosts of our SNe in the template images from PS1 (see Sect.~\ref{subsec:phot}). On each image, we used the code \texttt{SExtractor} \citep{bertin_sextractor_1996} to obtain the Kron magnitude of the source.
For SN~2021adxl, we repeated the measurement twice, first fitting the whole galaxy and then just the \ion{H}{II} region. We show the fluxes as a function of wavelength in Fig.~\ref{fig:sed_host}, the measurements retrieved from NED and, in the case of SN~2021adxl, from \cite{brennan_adxl_2024} are also reported. Also, for comparison, we added the SED of the host galaxy of SN~2015bn \citep{nicholl_2015bn_2016}, a superluminous (SL) Type-I SN, which was chosen because its host has one of the most complete SEDs. From Fig.~\ref{fig:sed_host}, it appears that all hosts are faint and that their SEDs are quite blue. In particular, there is no significant difference between the SED of the whole host of SN~2021adxl and the \ion{H}{II} region alone aside from the total flux. The shape of the SED of our SN hosts matches well that of the host of SN~2015bn. This is interesting because the hosts of SLSNe at low redshift are often blue dwarf galaxies with high star formation \citep{lunnan_hosts_2014,schulze_hosts_2018}.

In general, dwarf galaxies in the local Universe produce more stars per unit mass than massive galaxies and, furthermore, they seem to have a top-heavy initial mass function (IMF) that allows them to produce a higher fraction of massive stars than with standard Saltpeter IMF \citep{dabringhauser_imf_2012,marks_imf_2012}. 
Interestingly, SNe~IIn in general do not tend to follow the star formation as traced by the H$\alpha$ emission \citep{habergham_host_2014}, possibly indicating multiple progenitor scenarios \citep{ransome_host_2022}. However, luminous SNe~IIn do occur more frequently in metal-poor environments with young stellar populations \citep{moriya_environment_IIN_2023}. This is consistent with the notion that strongly interacting SNe are massive stars from the higher end of the IMF and appear with higher frequency in dwarf galaxies.

\subsection{Analysis on the line profile and emission}
\label{subsec:multifit}
During the interaction between the SN ejecta and the CSM, four main 
regions should be considered: from the outside, \textit{i}) the unshocked CSM, \textit{ii}) the CSM that was shocked by a forward shock (FS), \textit{iii}) the SN ejecta shocked by the reverse shock (RS), and \textit{iv}) the unshocked SN ejecta that are expanding fast \citep{chevalier_iin_shock_1994}. 
If the CSM is dense 
($\geq 3\times 10^{-15}\; \rm{g\,cm^{-3}}$), a cool dense shell (CDS) forms between the FS and RS \citep{chevalier_cds_1979c_1985}. An opaque CDS has the same effect of an expanding photosphere \citep{chugai_taue_2001}.
In the case of strong interaction, the heated CSM dominates the optical emission. Since emissions from the CSM can remain bright for months or years, it is possible that the ejecta thermal energy fades before the CSM becomes transparent and thus one may never see the characteristic P-Cygni profiles indicative of an expanding photosphere \citep{Smith_interaction_handbook_2017}. In general, in SNe~IIn one may expect a narrow ($v\sim100\;\rm{km\,s^{-1}}$) component due to the unshocked CSM and a broader ($v\sim5000-10000\;\rm{km\,s^{-1}}$) component which, depending on the optical depth of the CSM can originate either from the ionised SN ejecta or from electron-scattering in the CSM \citep{huang_csm_2018}. Furthermore, if the CSM is asymmetric \citep{stritzinger_05ip_06jd_2012}, or if dust is present \citep{fox_dust_2011}, they can affect the shape of the line profiles, thus making the theoretical interpretation more difficult.

\begin{figure}
\centering
	\includegraphics[width=\columnwidth]{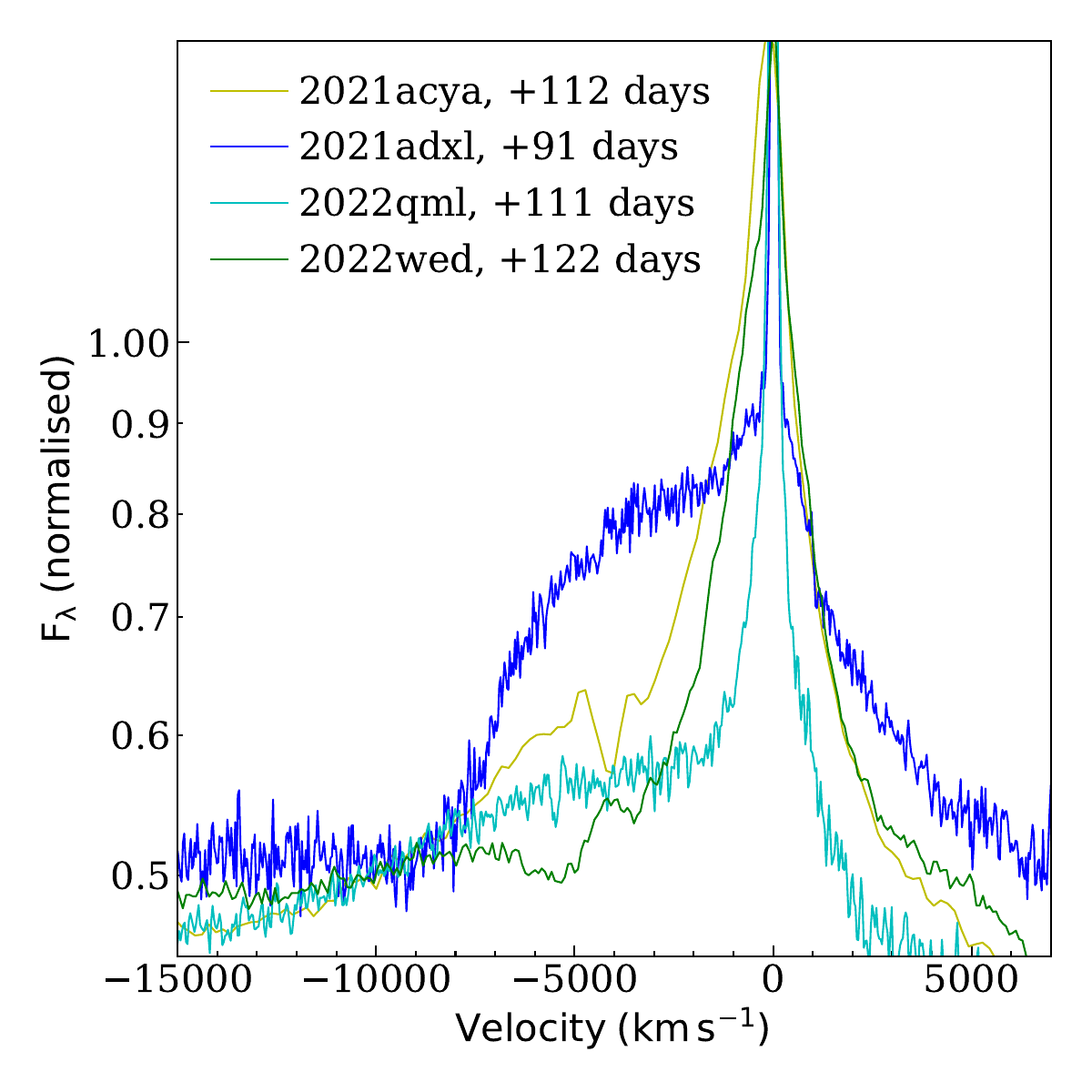}
    \caption{Zoom on the H$\alpha$ region of medium-resolution spectra at $\sim100\,\rm{days}$. All spectra are normalised with respect to the H$\alpha$ and shown in logarithmic scale for better visualisation.}
    \label{fig:highres_halpha}
\end{figure}

In Fig.~\ref{fig:highres_halpha}, we plot a zoom on the H$\alpha$ region of spectra taken around 100$\,\rm{days}$ from the explosion for all the SNe in our sample. SNe~2021adxl and 2022qml have broad, blue, asymmetric profiles with a narrow emission centered at rest wavelength. On the other hand, SNe~2021acya and 2022wed have a more symmetric broad line (Tables and are available online only, see the Data availability section). In the case of SN~2021acya, there is a small dip that may be a narrow P-Cygni absorption but it is probably an instrumental artifact, since it is not visible at other epochs.

Given the diversity in shape of the H$\alpha$ across our sample, it is difficult to directly compare them in terms of FWHM, line position, and total flux. Opting to apply the same method to all of them while being conscious that the procedure will not be optimal, we performed a multi-component fit for the H$\alpha$ lines, including the contribution of a broad Gaussian and a narrow Lorentzian function. We chose these rather simple profiles because they can ensure a more direct comparison among all the SNe in the sample, even if they might not provide the most perfect fit, especially for complicated profiles such as those of SNe~2021adxl and 2022qml. 
In some spectra, adding a second, intermediate-width Gaussian component 
representing the post-shock gas \citep{Smith_interaction_handbook_2017} allowed for a better fit. This is a relatively faint feature that was ignored for the rest of the analysis. For each component, we fit the position, peak intensity, and width. Given the asymmetric profile of the line in some objects, the position of the peak of the broad component was allowed to vary with respect to the narrow one. All the measurements were performed on reddening-corrected spectra.

\begin{figure*}[htbp]
    \centering
    \includegraphics[width=\textwidth]{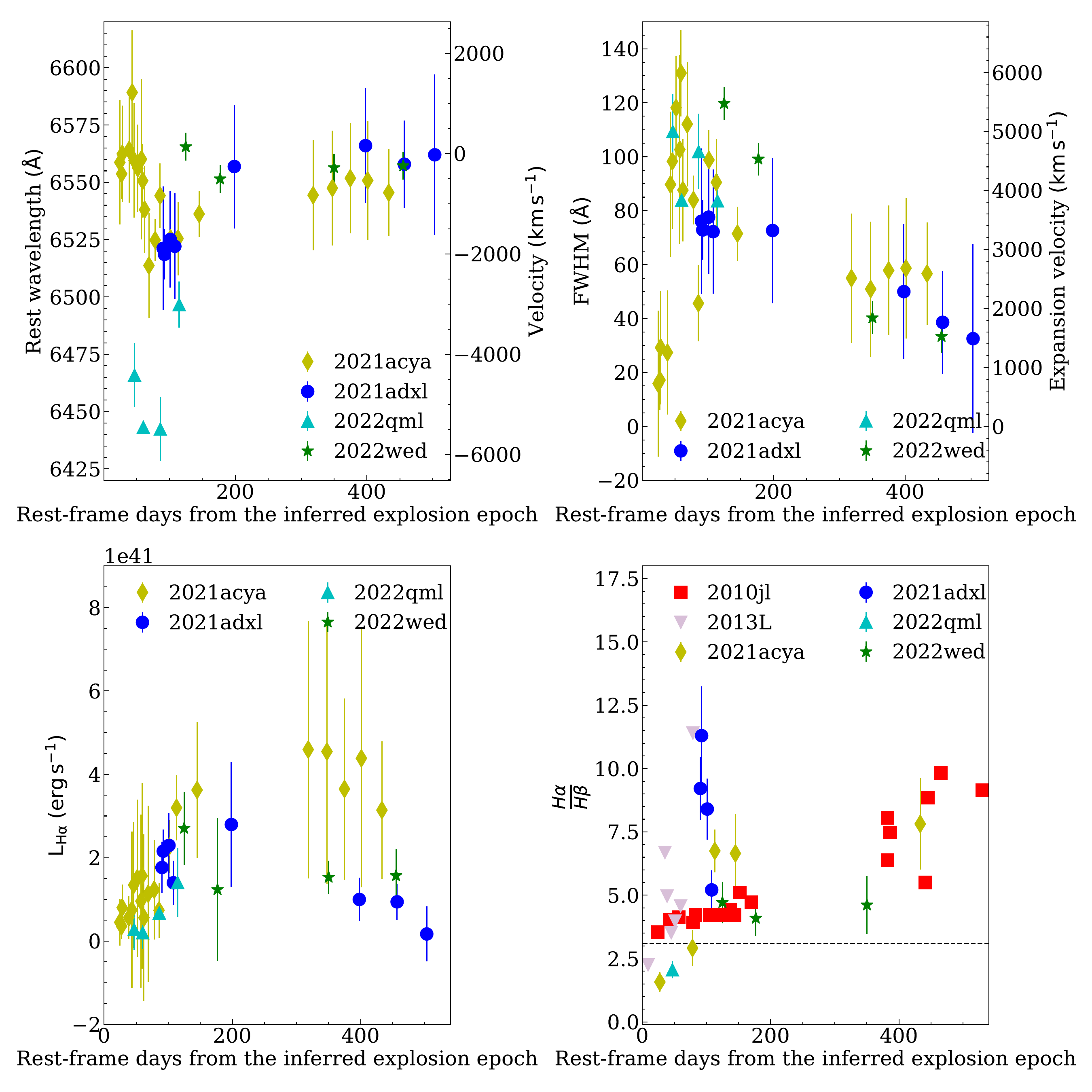}
    \caption{Results of the multifit on the broad H$\alpha$ and H$\beta$ for the SNe in the sample. \textit{Top left:} Central wavelength of the H$\alpha$ line. \textit{Top right: }FWHM of the H$\alpha$ line. \textit{Bottom left: } H$\alpha$ luminosity. \textit{Bottom right: }Ratio of $\frac{H\alpha}{H\beta}$ from the values measured for our spectra. The dashed line indicates the division between recombination and collision, which is considered to be 3.1 as in \citet{Osterbrock2006} (case B). Data from SN~2010jl taken from \cite{fransson_2010jl_2014} and from SN~2013L taken from \cite{taddia_2013L_2020} are also added, given their spectral similarity to the SNe in the sample.}
    \label{fig:param_HalphaHbeta}
\end{figure*}

For all the SNe in the sample, position, FWHM, and intensity of the narrow component are constant within the errors (calculated by summing in quadrature the root mean square and the resolution). 
Also, the FWHM of the narrow component is not resolved in the fit with the exception of the spectrum of SN~2022qml at +49$\,\rm{days}$, which yields a FWHM of 12.9~\AA.
The average position of the narrow line was taken as the rest frame reference for each transient.

In the upper left plot of Fig.\ref{fig:param_HalphaHbeta}, the rest-frame position of the centre of the broad H$\alpha$ component is shown. All our SNe show evolution in the rest-frame position. In SN~2021acya, in particular, the position of the line centre is initially at zero velocity with respect to the reference, then rapidly shifts to the blue with a maximum offset of about $2000\mathrm{\;km\,s^{-1}}$ that later reduces to $\sim1000\mathrm{\;km\,s^{-1}}$ at 300$\,\rm{days}$. SN~2021adxl appears to show the same late-time behaviour, while for SN~2022wed the variation is less extreme. SN~2022qml is the most extreme, with a broad peak significantly shifted towards the blue, especially at early epochs. This is the effect of the blue shoulder, which has peculiar prominence in this SN. In fact, while the blue shoulder is fairly common, such extreme shifts are more rarely observed (see, e.g., SNe~1997cy, \citealt{turatto_1997cy_2000}, and 2007rt, \citealt{trundle_2007rt_2009}).

The shift of the broad peak could have different explanations.
One possibility is that the red wing of the H$\alpha$ is obscured by dust, which forms after the explosion but could also be already present in the CSM \citep{lucy_dust_87A_1989}. However, this does not explain why the position of the peak shifts back again to the rest frame.
The blueward emission could be due to a mechanism by which the photons passing through the shock front multiple times acquire energy that generates bumps in the bluer part of the emission line \citep{ishii_Lint_2024}. 
Our favoured interpretation, however, is that the shift is due to occultation of the receding line-forming region: if the H$\alpha$ is forming in a region close to the photosphere just above a CDS, this would imply a deficit in the red-ward flux, which depends heavily on the density distribution \citep{dessart_blueshift_2005}. 
This is depicted in model A of \citet{dessart_interazione_2016}, with a massive SN ejecta ramming into a dense CSM. During the luminous phase of the light curve, the photosphere corresponds to the CDS and the emitting region is moving outward. 
This last scenario naturally explains the progressive shift to more symmetrical lines as the optical depth decreases and the part of the line-forming region that is receding along our line of sight is revealed. 
The mechanism of the line formation is hence the same for all the sample of SNe, while the optical depth of the electron scattering region varies.
This same argument has been used also to explain the blueshift often observed in emission lines of SNe~II \citep{anderson_blueshift_ii_2014}. However, in this case we do not observe a correlation with the light curve peak luminosity.

In the upper right panel of Fig.~\ref{fig:param_HalphaHbeta}, the FWHM of the broad component is shown. All SNe in our sample start with very broad ($\gtrsim 80$ \AA) FWHMs ($\gtrsim 4000 \;\rm{km\,s^{-1}}$) that then shrink to $\sim40-60$ \AA~ ($2000-3000 \;\rm{km\,s^{-1}}$) around 300$\,\rm{days}$. SN~2021acya, however, starts with a lower FWHM and then broadens significantly. Considering the gaps in the spectral follow-up, it is possible that all SNe underwent the same evolution. A viable explanation for this behaviour is that at early epochs the photosphere is not (yet) hot enough to show a significant electron scattering effect, which becomes dominant later on.

In the bottom left plot of Fig.~\ref{fig:param_HalphaHbeta}, the flux of the broad H$\alpha$ is shown. A progressive increase in the flux is observed in all SNe, with a peak between 200 and 300$\,\rm{days}$ after the explosion. This is followed by a decrease in the case of SNe~2021adxl and 2022wed, while it appears to remain constant (albeit within very large errorbars) for SN~2021acya. A high H$\alpha$ luminosity is often correlated to strong interaction \citep{chugai_Halpha_1991}. If this is true also for our SNe, the strength of interaction increases with time and peaks after the light curve peak. The epoch at which the H$\alpha$ flux starts to decrease is likely related to the extent and density profile of the interacting CSM. 
A more extended CSM would sustain the interaction for a longer time, while a rapidly declining density would drop the strength of interaction and thus the luminosity. In this respect, the decrease for SN~2022wed happens earlier than for the others, around the maximum of the second peak. Possibly, this denotes a rapid change in the CSM density for this particular object. On the other hand, for SNe~2021acya and 2021adxl the change happens later on, perhaps indicating a less steep decline in the CSM density. Also, note that the total H$\alpha$ luminosity is higher for SN~2021acya compared to SN~2021adxl, possibly, due to a higher CSM mass for the former.

Finally, the ratio of $\frac{H\alpha}{H\beta}$ for both the narrow and the broad components is calculated. For the narrow line, the ratio is constant within the errors and close to the 3.1 limit value for the case B hydrogen recombination \citep{Osterbrock2006}. For the narrow lines, this indicates that 
H is excited by radiation. The broad lines ratio is shown in the bottom right panel of Fig.~\ref{fig:param_HalphaHbeta}, where we consider only ratios with an error below 25\%. The values are significantly higher than for the narrow lines and the ratio changes over time. This happens when electrons are pushed to higher 
energy levels through collisions, in regions of high density and optical depth, confirming that all SNe in our sample have a strong shock component in their light emission \citep{chevalier_iin_shock_1994}. This is also in agreement with what is found in the literature, for example, for SNe~2010jl and 2013L \citep{fransson_2010jl_2014,taddia_2013L_2020}.

\begin{figure}[htbp]
\centering
    \includegraphics[width=\columnwidth]{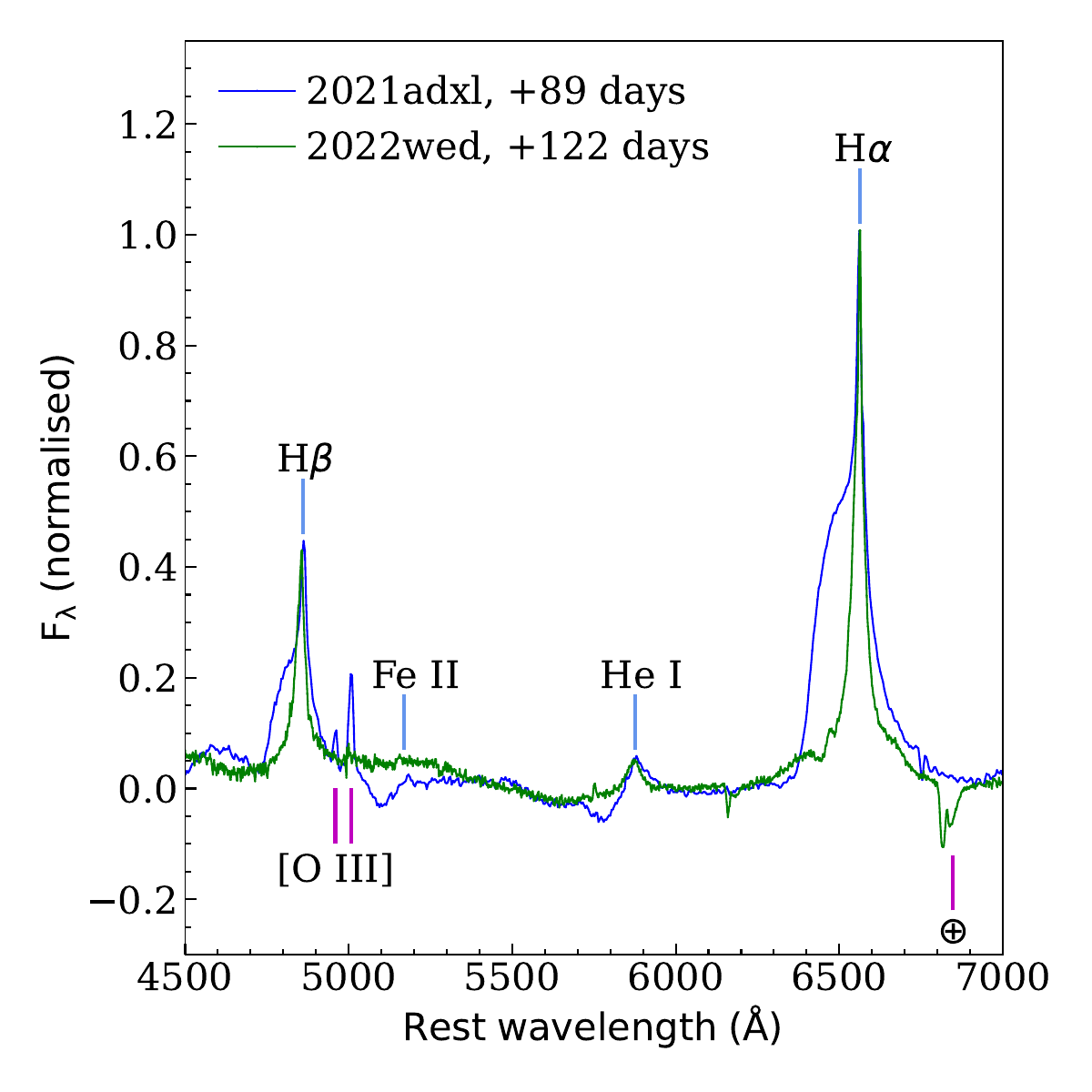}
    \caption{Spectral comparison between SNe~2021adxl and 2022wed. All spectra are redshift-corrected, continuum-subtracted, and rescaled for better visualisation. We can also clearly see the P-Cygni profiles of \ion{He}{I}~$\lambda5876$ and \ion{Fe}{II}~$\lambda 5169$ for SN~2021adxl.}
    \label{fig:confronto_adxl_wed}
\end{figure}

 \citet{brennan_adxl_2024} pointed to a possible two-component absorption feature in the \ion{He}{i} P-Cygni profile of SN~2021adxl, which they explain as either a blend with the \ion{Na}{I}~D $\lambda\lambda5890,5896$ or the effect of an asymmetric explosion.
In Fig.~\ref{fig:confronto_adxl_wed}, a comparison of the spectrum of SN~2021adxl at +91$\,\rm{days}$ with the one of SN~2022wed at +136$\,\rm{days}$ is shown. The 
starting position of the blue bump excess matches the one of SN~2022wed and the presence of Fe in this SN is confirmed by \ion{Fe}{II}~$\lambda5169$ in multiple spectra. This shows that the possible high-velocity \ion{He}{i} component mentioned by \cite{brennan_adxl_2024} is rather better understood as the emergence of the blue bump due to \ion{Fe}{ii} emission (cf. Sect.~\ref{subsec:indice}). Instead, we agree on the interpretation of the lower velocity component as \ion{He}{i} at $\sim 3000\,\mathrm{km\,s^{-1}}$. This feature will be discussed later on.

\begin{figure}[htbp]
    \centering
    \includegraphics[width=\columnwidth]{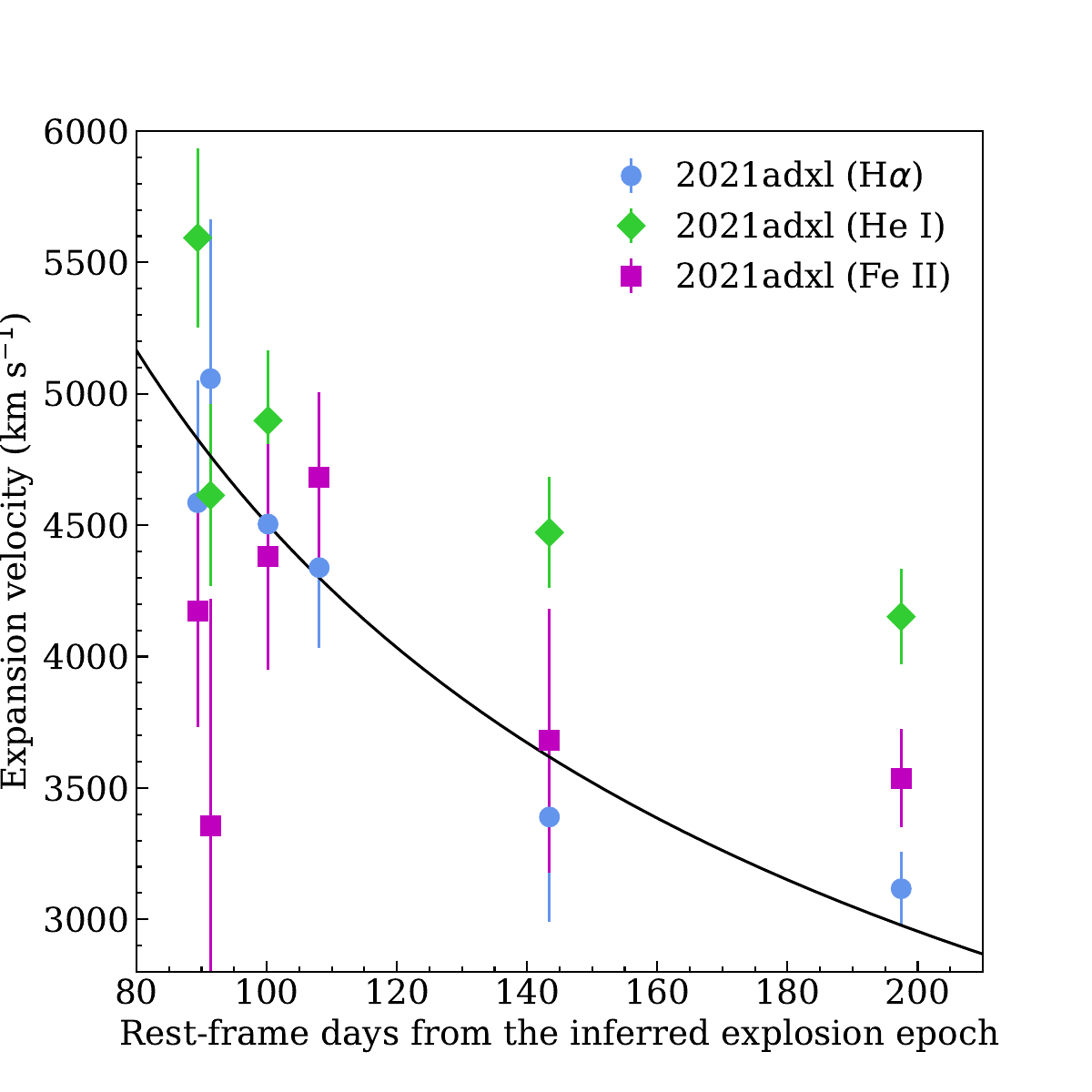}
    \caption{Photospheric expansion velocity of SN~2021adxl from the \ion{He}{I} P-Cygni absorption (green diamonds) and \ion{Fe}{II} (blue squares) along with the shock velocity derived from the blue shoulder of the broad H$\alpha$ (magenta reverse triangles). The black line represents the best exponential decline fit of the H$\alpha$ shock velocity based on our measurements.}
    \label{fig:v_adxl}
\end{figure}

When the shocked region is exposed, it is possible to measure the shock velocity $v_{sh}$ from the blue shoulder of the H$\alpha$ line. The CDS is confined between FS and RS and it is optically thick but radially thin. Photons emitted from such a structure will give a boxy profile to the line \citep{dessart_slsneiin_2015} and the blueward limit of the emission is a direct measure of $v_{sh}$. Even in the case of a composite profile with electron scattering wings, it is still possible to recover the shock velocity as long as the blue shoulder is visible. As shock velocity, we took the Doppler shift 
measured on the left corner of the blue shoulder compared to the position of the narrow emission. The measurements were performed on the spectra where the boxy component is visible and are shown in Fig.~\ref{fig:v_adxl}, while the highest and lowest values are reported in Table~\ref{tab:par_mis}.
The measurements have a significant scatter due to the uncertainty of the blue shoulder position, but it is still possible to fit them to a declining power-law, which gives $v_{sh} \propto t^{-0.61}$. The same was done for SN~2022qml, finding $v_{sh} \propto t^{-0.23}$, more similar to the velocity evolution \cite{taddia_2013L_2020} find for SN~2013L. The $v_{sh}$ measured on the spectra are reported in Table~\ref{tab:par_mis}, along with other parameters inferred from the light curves and spectra.

In Fig.~\ref{fig:v_adxl}, the measurements on the expansion velocity performed on the P-Cygni absorption of \ion{He}{I} and \ion{Fe}{II} for SN~2021adxl are also added. Considering the large errorbars, they have a slightly higher velocity and a similar trend to the shock velocity measured on the H$\alpha$ blue shoulder at all epochs. The P-Cygni profile is indicative of a somewhat extended gas shell above a photosphere and thus of a continuum, while H$\alpha$ shows no absorption components. This could imply the presence of an inner He layer, on top of which there is an optically thin H layer that allows for the formation of the boxy profile on the H$\alpha$ line. Our P-Cygni measurements, in fact, are comparable with the expansion velocity measured, for example, for SN~2004et \citep{sahu_04et_2006}, which was a normal SN~II. A possible scenario is that all H comes from the shocked CSM, while He and Fe are part of the actual SN ejecta. 
Considering the mass-loss (for which we will calculate the rate in Sect.~\ref{sec:pitik}), it is possible that SN~2021adxl is actually a SE SN \citep{woosley_SE_1995} and the narrow H line comes from the fraction that is unshocked.

\begin{figure}[htbp]
    \centering
    \includegraphics[width=\columnwidth]{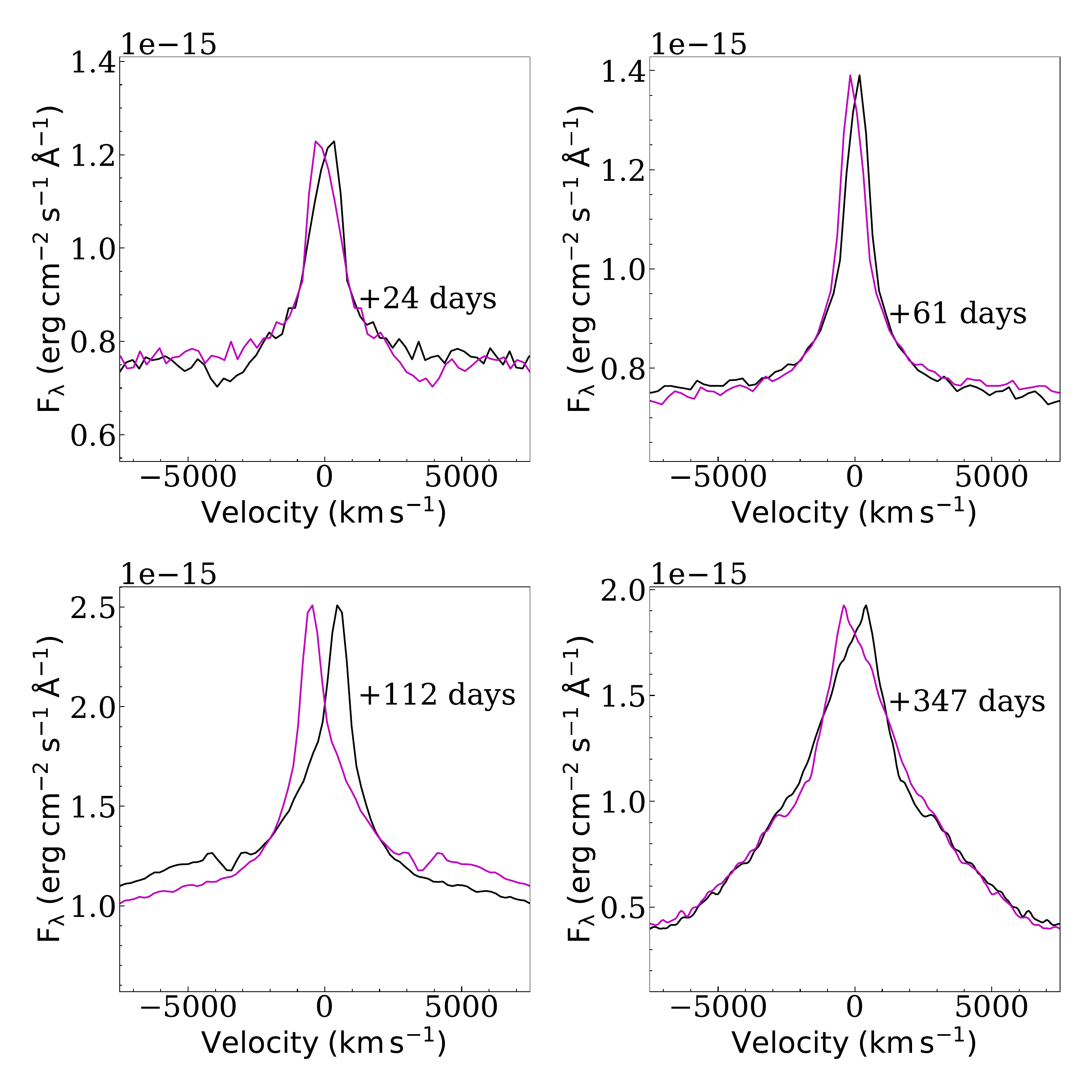}
    \caption{Zoom on the H$\alpha$ profile of selected spectra of SN~2021acya. Magenta lines represent the reflected profile. The profiles have a velocity shift of 150 (+24$\,\rm{days}$), 100 (+61$\,\rm{days}$), 500 (+112$\,\rm{days}$), and 300~$\mathrm{km\,s^{-1}}$ (+347$\,\rm{days}$).}
    \label{fig:fransson28}
\end{figure}

Estimating the shock velocity is very tricky when the line profile does not clearly show a blue shoulder, as it is the case for SNe~2021acya and 2022wed. Following \cite{fransson_2010jl_2014}, we show in Fig.~\ref{fig:fransson28} a zoom of the H$\alpha$ spectral region of SN~2021acya, with the blue side folded over the red one. The lines were shifted so that the broad wings could match, introducing a velocity shift of the narrow peaks. Initially, the shift is low, about $100\mathrm{\,km\,s^{-1}}$, and grows to a maximum of $500\mathrm{\,km\,s^{-1}}$ at +119$\,\rm{days}$. These values are similar to those determined by \cite{fransson_2010jl_2014}, who interpreted the shift as the sign of the bulk velocity of the scattering medium. In this context, the first measurements correspond to the wind velocity and the subsequent acceleration is possibly due to the shock catching up with the wind.

Because of the higher optical depth in the scattering region, the shocked gas remains hidden in the spectra of these SNe, thus precluding a direct measurement of the shock velocity. Therefore, given the similarities between the spectra of SNe~2021acya and 2022wed with 2010jl, we adopted the same value as \cite{fransson_2010jl_2014}, $3000\mathrm{\,km\,s^{-1}}$ at +320$\,\rm{days}$. To estimate its evolution, we took the trend of Eq.~9 of \cite{taddia_2013L_2020} and fit it to this point. As a second estimate, we also used the same trend of SN~2021adxl fitted to this point. Finally, we also fit the formula of \cite{taddia_2013L_2020} to SNe~2021adxl and 2022qml. This gives us an estimation of how the shock velocity could realistically vary. Both estimates were used for the calculations in Sect.~\ref{sec:pitik}.

\begin{figure}[htbp]
    \centering
    \includegraphics[width=\columnwidth]{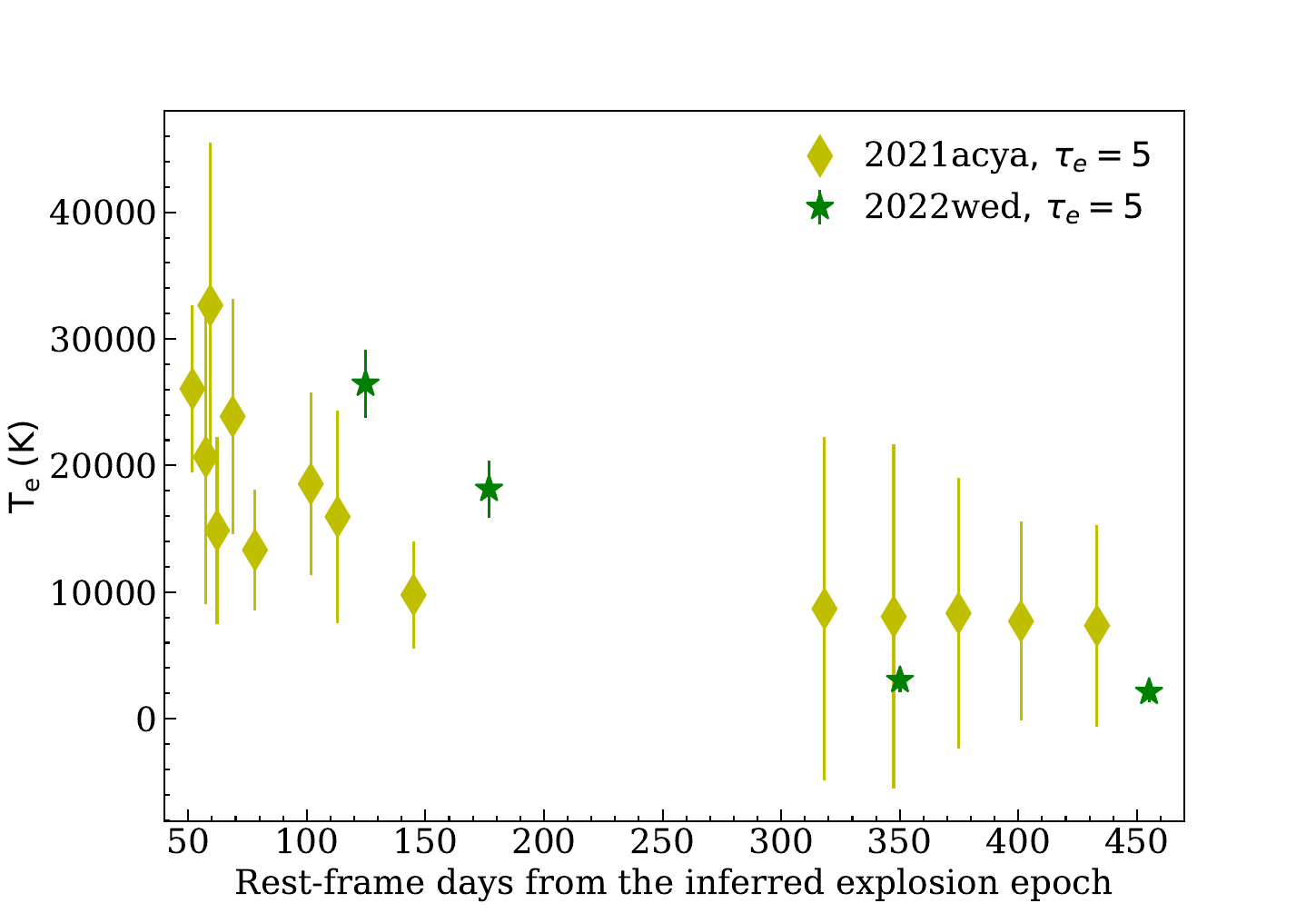}
    \caption{Electron-scattering temperature for SNe~2021acya and 2022wed as inferred from the broad H$\alpha$ FWHM, assuming an optical depth $\tau_e=5$.}
    \label{fig:tescat}
\end{figure}

When the line profile is fully attributed to electron scattering, such as for SNe~2021acya and 2022wed, it is possible to calculate the electron scattering temperature $T_e$ from the FWHM of the broad emission using the equation from \cite{fransson_2010jl_2014}: $\mathrm{FWHM}\approx 647\tau_e(T_e/10^4\,\mathrm{K})^{\frac{1}{2}}\mathrm{\;km\,s^{-1}}$, where $\tau_e$ is the optical depth, and inverting it to extrapolate the values of $T_e$ for each spectrum of SNe~2021acya and 2022wed. A symmetric Lorentzian profile implies $\mathrm{\tau_e}\geq 1$, and typical values found in the literature range around 4 -- 5 \citep{chugai_taue_2001}. For our calculations, $\tau_e=5$ was adopted, thus giving temperatures in the range $6.0\times10^3 - 2.6\times10^4$~K. This is in line with what is derived for SN~2010jl \citep{fransson_2010jl_2014}. The ranges of $T_e$ are reported in Table~\ref{tab:par_mis} and its evolution is shown in Fig.~\ref{fig:tescat}, where only the fits with errors $<25\%$ were considered. The $T_e$ evolution is better followed for SN~2021acya. It shows a rapid increase to $T_e \sim 30000\,$K with the emergence of the shock followed by a similarly rapid decrease and a long tail at a roughly constant $T_e \sim 8000\,$K. The few measurements for SN~2022wed show a similar trend but with a higher temperature during the rapidly declining phase. This is coherent with the observed luminosity evolution that suggests a delayed phase of enhanced interaction.

\subsection{Blue bump excess}
\label{subsec:indice}

\begin{figure}[htbp]
    \centering
    \includegraphics[width=\columnwidth]{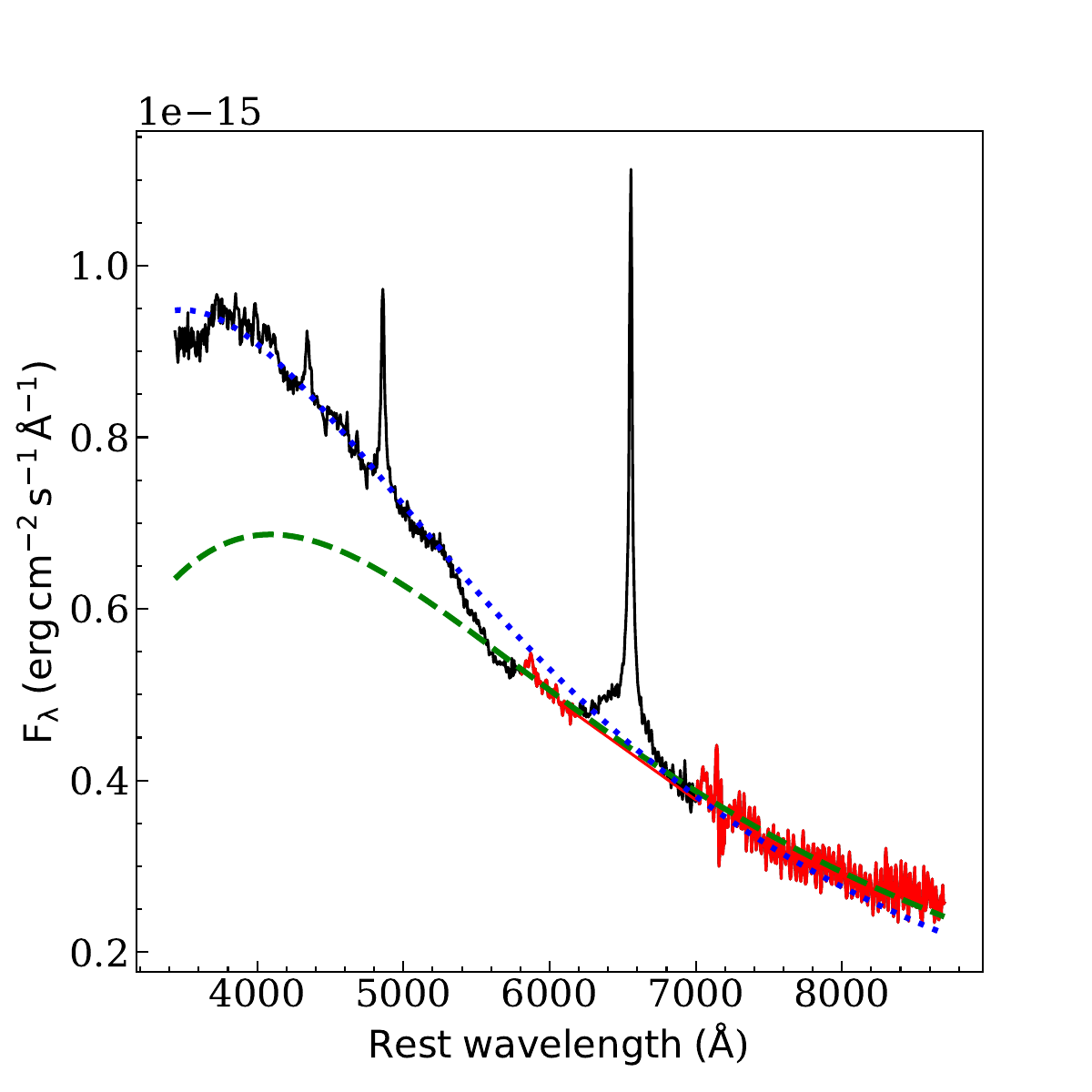}
    \caption{Spectrum of SN~2021acya at +61$\,\rm{days}$ (black). The green dashed line is the estimated BB flux from the fit on the red part of the spectrum, while the blue dotted line is the estimated BB flux from the fit on the whole spectrum. Notice that the blue BB does not fit well the spectrum between $5000-6000\;$\AA.}
    \label{fig:croci}
\end{figure}

As we already noticed, a peculiar feature in the spectra of many interacting SNe is a blue bump bluewards from $5500$~\AA. This was first observed in SN~1988Z \citep{turatto_1988Z_1993} and it is typically attributed to a forest of \ion{Fe}{II} lines (see, e.g., \citealt{foley_2006jc_2007,smith_2011hw_2012}). The relative strength of this feature varies depending on the objects considered and the phases.
In general, the spectrum of interacting SNe is a combination of a hot continuum and emission lines, some isolated and others, such as those contributing to the blue bump, blended.
To estimate the relative contribution of the two components, we fit a black body (BB) function in the red part on the redshift-corrected spectra ($\lambda5800$~\AA\textcolor{white}{,} onward), avoiding the most prominent emission lines (in particular, H$\alpha$). 
However, we caution that the BB fits fails after roughly 200$\,\rm{days}$ (100$\,\rm{days}$ for SN~2022qml).
An example of the fit region and the estimated BB is shown in Fig.~\ref{fig:croci}. Here, the black line is the observed spectrum of SN~2021acya at +61$\,\rm{days}$, while the red bands show the location of the fitting region and the green line the fitted BB function. A blue line representing the BB fit to the whole spectrum is also added to show that the fit is less optimal in this case due to the blue bump.
The blue bump excess is defined as the ratio between the difference in flux between the observed flux in the blue (avoiding the most prominent emission lines, in particular, H$\beta$) and the flux below the BB fitting on the redder part of the spectrum at the same wavelengths, all divided by the total BB flux.

\begin{figure}[htbp]
    \centering
    \includegraphics[width=\columnwidth]{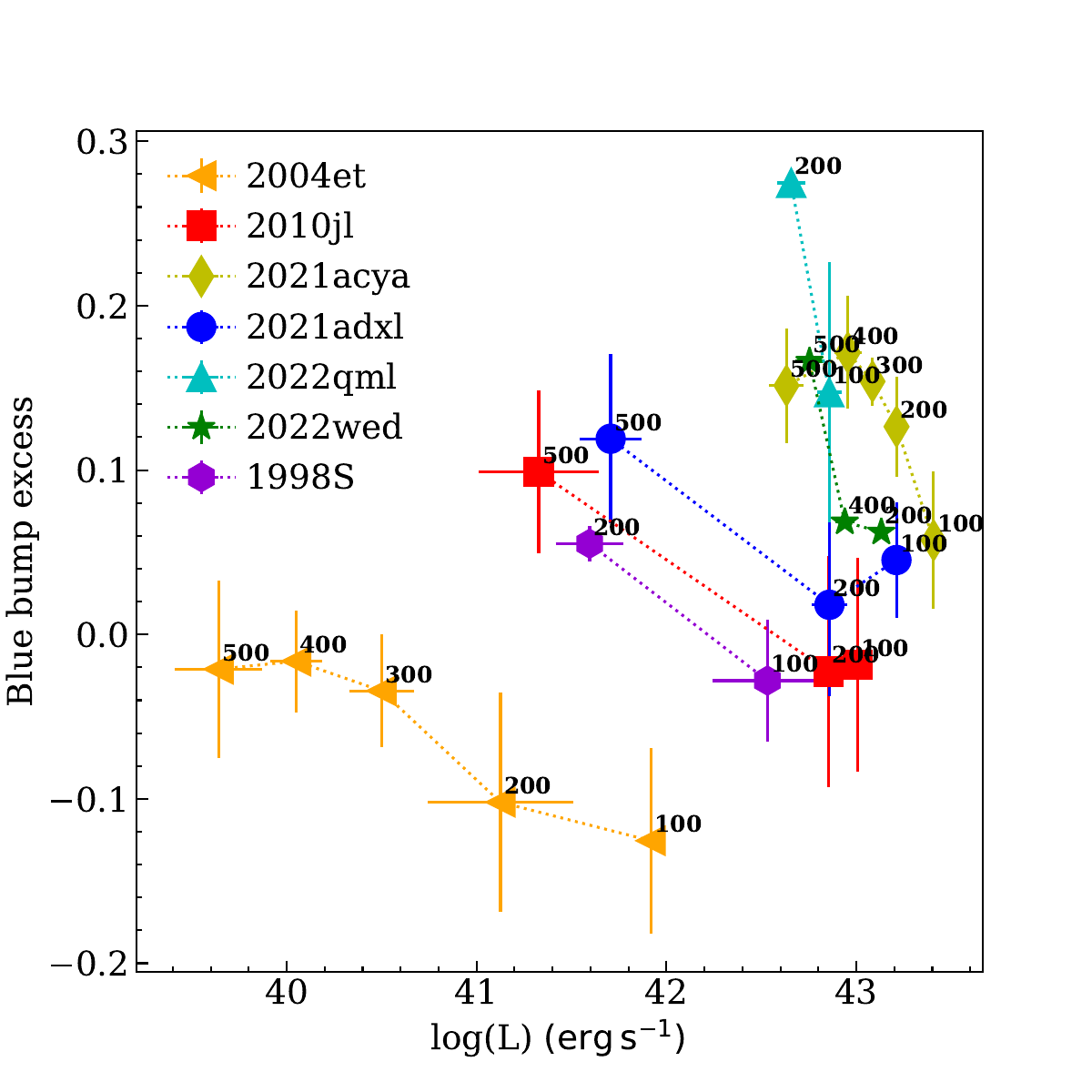}
    \caption{Blue bump excess as a function of the \textit{g,r,i} luminosity. Numbers close to the points indicate the average phase.}
    \label{fig:blue_lum}
\end{figure}

Considering the gaps in the observations and the different phases of the SNe in our sample, a mean blue bump excess was computed every 100$\,\rm{days}$ since the explosion to aid in the comparison. It is shown, compared to the mean luminosity at the same phases derived from the \textit{g,r,i} pseudo-bolometric light curve (see Sect.~\ref{sec:bol_altre} for details on the computation), in Fig.~\ref{fig:blue_lum}, where three reference objects are also added to represent the effect of different amounts of interaction: the strongly-interacting SN~2010jl, the mildly interacting SN~1998S, and the Type~IIP SN~2004et, where moderate interaction only started after $\sim800$$\,\rm{days}$ \citep{kotak_04et_dust_2009}.
The evolution with time is significant. The excess is small (in some cases, even negative, meaning that the fitted BB exceeds the measured flux, probably due to line blanketing) at early phases, and tends to grow after roughly 100$\,\rm{days}$, while the luminosity decreases. 
SNe~2010jl and 2021adxl seem to show milder blue bump excesses at early phases than SNe~2021acya and 2022wed, while SN~2022qml is the most extreme, with a stark increase in the blue bump excess after 100$\,\rm{days}$. SN~1998S exhibits a similar behaviour, albeit with lower absolute values of the blue bump excess.
 SN~2004et, on the other hand, shows an always negative excess. This is in line with what we would expect from a non-interacting SN\footnote{Indeed SN~2004et underwent late rebrightening (phase $\sim 800$$\,\rm{days}$) due to interaction \citep{kotak_04et_dust_2009}. However, while almost all SNe are expected to undergo interaction, sooner or later, since all massive stars suffer from mass-loss episodes, the interaction happening at such late phases is orders of magnitude less strong than for strongly-interacting SNe, since the density of the CSM is lower.}.
 
The H${\alpha}$ luminosity can be used as a tracer of the strength interaction \citep{chugai_Halpha_1991} to verify whether it correlates with the strength of the blue bump excess, however, there appears to be none.
The time at which the interaction component is dominant with respect to the BB continuum is probably due to a combination of CSM density, strength of interaction, and possibly the relative Fe abundance in the ejecta with respect to the other elements. A detailed spectral modelling may help to disentangle the origin of the feature in different objects, but it appears that the presence of the blue bump excess alone is not indicative of strong interaction. A combination of parameters such as high bolometric and H${\alpha}$ luminosity is also required.
However, there seems to be a mild correlation with the CSM optical depth that works as follows: interacting SNe with a strong blue bump excess also show a boxy H$\alpha$ line profile. In turn, the boxy profile is exposed when the electron scattering optical depth of the heated CSM is smaller.

\section{Bolometric light curves}
\label{sec:bol_altre}

\begin{figure}[htbp]
	\includegraphics[width=\columnwidth]{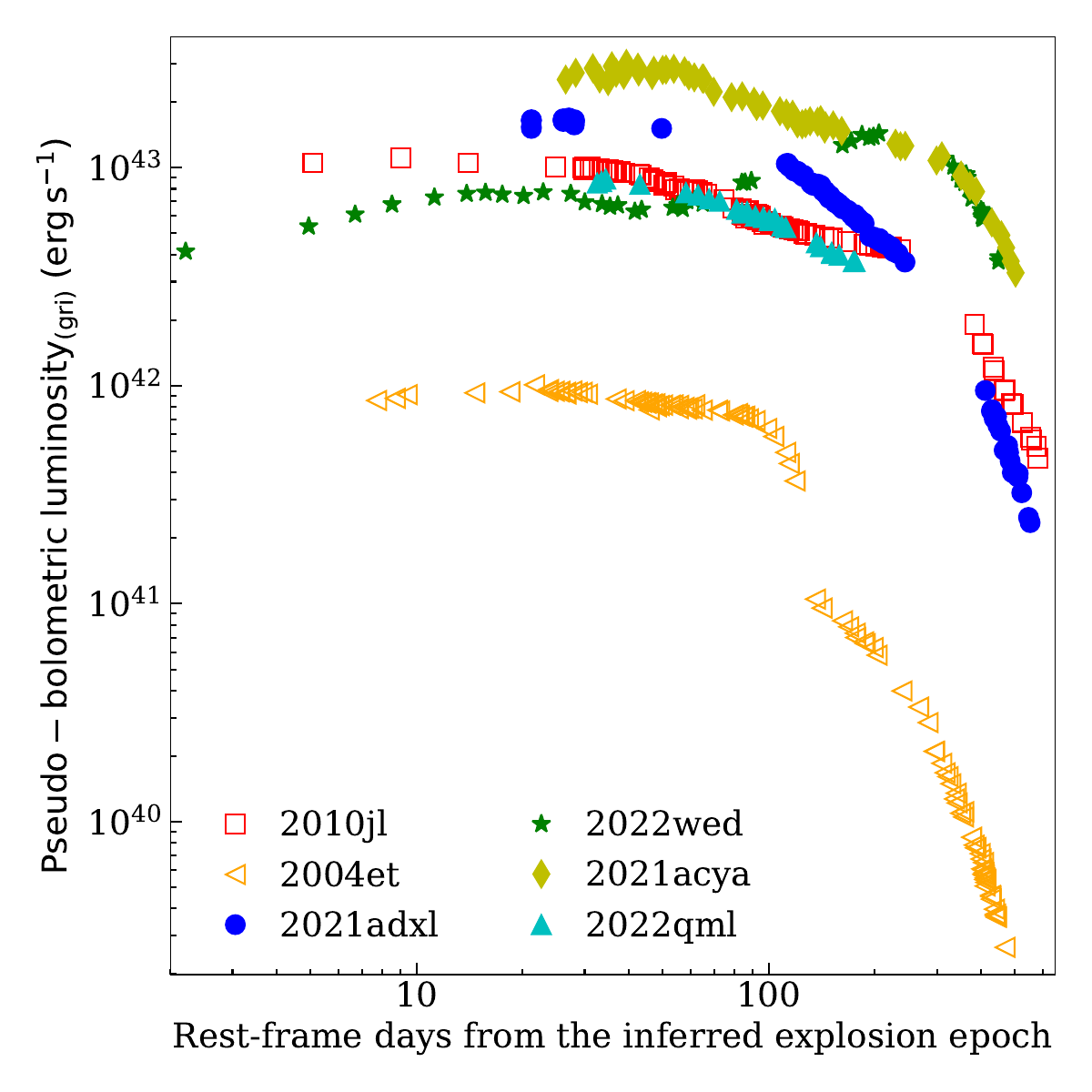}
    \caption{Pseudo-bolometric light curves computed using only optical filters. We also add the bolometric light curves, calculated in the same way, of SNe~2010jl and 2004et for comparison.}
    \label{fig:bolom_altre_log}
\end{figure}

The SNe in our sample have different coverage in terms of wavelength and phases. To ensure a more objective comparison among them, we computed a pseudo-bolometric light curve for all SNe integrating the flux in the \textit{g,r,i} filters. The magnitudes are corrected for extinction and converted to flux densities using photometric zero points\footnote{\url{http://svo2.cab.inta-csic.es/theory/fps/}}. The flux is then integrated in the sampled spectral region using a trapezoidal rule and assuming zero flux below/above the limit defined by the filter equivalent width of the bluer/redder filter, respectively. The measured flux is converted into luminosity using the adopted distance modulus. The results for our four SNe is shown in Fig.~\ref{fig:bolom_altre_log}, where the pseudo-bolometric light curves of SNe~2004et and 2010jl, calculated in the same way, are also included for comparison. A logarithmic scale is used to emphasise the change of slope between the early and late phases.

SN~2021acya is the most energetic SN in our sample, with an extremely high $E_{200}$ (Table~\ref{tab:tutte}). After the very bright peak, SN~2021acya shows a slow decline followed by a plateau (better seen in Fig.~\ref{fig:lc_escluse}) that lasts from +150~days until +300~days. The decline rate both before and after the plateau is similar to that of SN~2010jl.
SN~2021adxl shares some similarities with SN~2010jl, in particular, the break in the light curve around +400$\,\rm{days}$. However, the decline rate is much higher and, in fact, SN~2021adxl is about 1.5~mag fainter than SN~2010jl at $\sim500$$\,\rm{days}$.
SN~2022qml has the shortest light curve, since it is observed for less than 200$\,\rm{days}$, after which the SN is lost because of conjunction with the Sun. When it re-emerged from behind the Sun, the SN was no longer detected. The decline rate is also similar to that of SN~2010jl.
Finally, SN~2022wed has the most peculiar light curve shape, with its two peaks, the second much brighter than the first.  
This is probably due to a CSM arranged in two shells with different mass and density. It also does not show a break at 400$\,\rm{days}$, however, the decline rate after the second peak is almost identical to that of SN~2021acya after the plateau.

 All the SNe in our sample are bright and long lasting like SN~2010jl and in some cases also the decline 
rate is similar. The prototype of Type~II plateau SN~2004et, on the other hand, is completely different, since it is $1-2$ orders of magnitude fainter at all phases and also its plateau phase is considerably shorter. This gives a measure of the significant energy contribution from interaction for the selected SNe.

\begin{figure}
    \centering
    \includegraphics[width=\columnwidth]{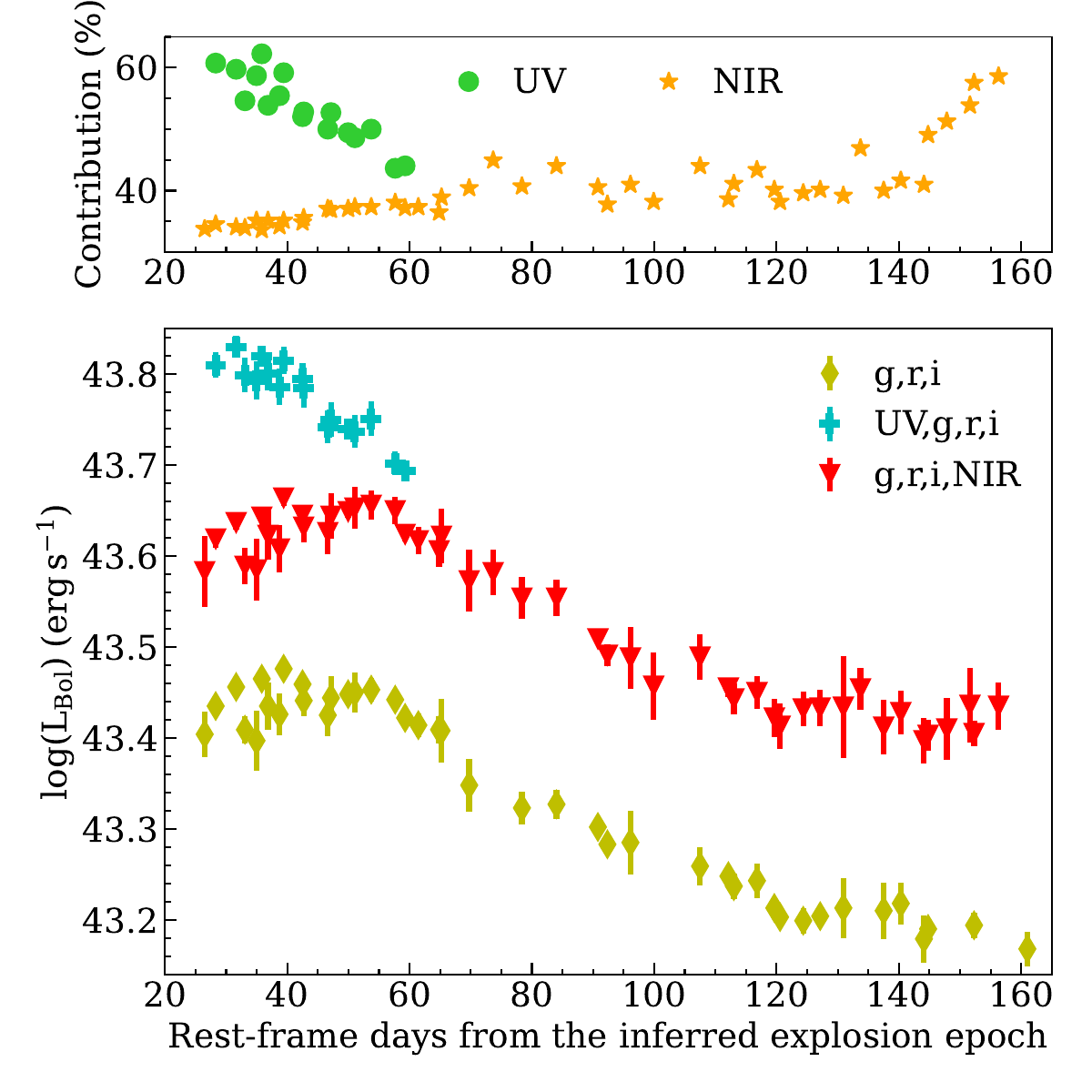}
    \caption{UV and NIR contribution for SN~2021acya. The upper panel shows the differences between the luminosities computed using UV and NIR observations with respect to the bolometric light curve computed taking into consideration only optical bands.}
    \label{fig:excess_acya}
\end{figure}

Along with the direct comparison, we tried to gauge the true bolometric luminosity taking into account UV and NIR bands. In this respect, SN~2021acya has the best photometric coverage. To estimate the contribution of UV and NIR photometry, in Fig.~\ref{fig:excess_acya} the bolometric light curve computed with the contribution from the UV or NIR flux is compared to the one computed using only \textit{g,r,i}. The contributions are calculated as the ratio between the UV or NIR flux and the total flux. While the NIR contribution at early phases is small (around 20\%) and rises up to 40\% at later times, the UV contribution is significant, up to 60\%, but then it decreases rapidly. A similar UV excess was also measured in SN~2010jl and other interacting SNe \citep{fransson_2010jl_2014}. In SN~2010jl, the early NIR light curves followed the decline of the optical ones, while after $\sim200$$\,\rm{days}$ they flattened in $J$ and $H$ and even increased the flux in $K$. This was attributed to pre-existing dust in the CSM that was re-heated by the optical emission \citep{fransson_2010jl_2014}. For SN~2010jl, the NIR bands contribute 20 to 50\% of the bolometric flux in the phase range $200-600$$\,\rm{days}$. This is in line with the NIR contribution that we measure $100-200$$\,\rm{days}$ after the maximum for SN~2021acya.

The same check was also performed on SNe~2021adxl (for which we have a sequence of \textit{u,z,J,H,K} observations), 2022qml (for which we have a sequence of $u$ observations), and 2022wed ($u$ observations). The $u$ band contribution is consistent in all SNe, adding alone between $\sim5-10$\% of the total flux at early phases. The NIR observations of SN~2021adxl, on the other hand, behave like in SN~2021acya, giving a flux $\sim25-35$\% higher. Also, the NIR contribution appears to increase with time.  

The considerable flux difference that is found when adding the NIR contribution is in line also with \citet{martinez_bol_2022}, where they show that NIR observations are fundamental to properly reconstruct the bolometric light curve of SNe~II.
To compute the total luminosity we will use in the next section, the contribution of UV and NIR bands, when available, was added to the optical ones.

In Table~\ref{tab:slope} the measured decline rates in the log-log scale for SNe~2021acya, 2021adxl, and 2022qml are reported (the measurement was not performed on SN~2022wed since its light curve shape is very different from the others). \cite{fransson_2010jl_2014} interpret the break as the time when the FS emerges at the outer edge of the dense CSM, while the specific slope is a sign of the CSM density. In particular, a steeper slope is indicative of a smaller density, since the radiation diffuses earlier. 

\begin{table}[htbp]
\caption{Slopes of the bolometric light curve.}
    \label{tab:slope}
    \centering
    \begin{tabular}{ccc}
    \hline\hline
         SN & Slope (before break) & Slope (after break) \\
          & $\mathrm{erg\,s^{-1}}$/100$\,\rm{days}$ & $\mathrm{erg\,s^{-1}}$/320$\,\rm{days}$\\
         \hline
         2021acya & $-0.739$ & $-2.23$ \\
         2021adxl & $-1.327$ & $-4.39$ \\
         2022qml & $-0.705$ & $\cdots$\\
         2010jl & $-0.536$\tablefootmark{*} & $-3.39$\tablefootmark{*}\\
         \hline
    \end{tabular}
    \tablefoot{Slopes are measured from peak to $\sim400$$\,\rm{days}$ after the explosion and from $\sim400$$\,\rm{days}$ after the explosion onward.\\
    \tablefoottext{*}{From \citet{fransson_2010jl_2014}.}}
\end{table}

In \citet{khatami_kasen2023}, the authors explore the 
diversity of SN light curves when interaction with the CSM is involved. The key parameter in their model is the break-out parameter $\xi$:
\begin{equation}
    \xi=\tau_0 \beta_0 \eta^{-\alpha} \sim t_{esc}/t_{sh}
\end{equation}
where $t_{esc}$ and $t_{sh}$ are, respectively, the timescale for photons to escape ahead of the shock and the dynamical timescale of the shock, $\eta=M_{CSM}/M_{ej}$ is the ratio of CSM to ejecta mass, with $M_{CSM}$ and $M_{ej}$ the masses of CSM and ejecta, respectively, $\beta_0=v_{ej}/c$ is the ejecta velocity with respect to the speed of light, $\tau_0=\frac{\kappa M_{CSM}}{4\pi R^2_{CSM}}$ is the characteristic optical depth of the CSM, with $R_{CSM}$ the radius of the CSM and $\kappa$ the opacity, and $\alpha=1/2$ if $\eta>1$, while $\alpha=1/(n-3)$ if $\eta<1$, where \textit{n} is the power-law exponent of the density profile of the ejecta.
Depending on the values of $\xi$ and $\eta$, \citet{khatami_kasen2023} identify four different scenarios: \textit{i}) edge-breakout with light CSM ($\xi\gg1,\eta\ll1$), \textit{ii}) edge-breakout with heavy CSM ($\xi\gg1,\eta\gtrsim1$), \textit{iii}) interior-breakout with light CSM ($\xi\lesssim1,\eta\ll1$), and \textit{iv}) interior-breakout with heavy CSM ($\xi\lesssim1,\eta\gtrsim1$). Different combinations of $\xi$ and $\eta$ generate different light curves (\citealp[their Fig.~3]{khatami_kasen2023}).
Judging by the light curve shape of our targets, they are well represented by case \textit{iv}, with a CSM comparable to or even more massive than the ejecta mass colliding with it and a dynamical timescale comparable to or longer than the escape time (although SN~2022wed, with its long second peak, could also be part of case \textit{ii}, with an escape timescale longer than the dynamical one).

\section{Mass-loss rate and mass of the CSM}
\label{sec:pitik}

For the four SNe in our sample, the high luminosity, slow decline and narrow emission lines in the spectra are all indicative of CSM/ejecta interaction as source of the luminosity. In fact, their luminosity is at all phases $1-2$ orders of magnitude brighter than typical CC SNe. In turn, if the luminosity is attributed to radioactive material, it would require a $^{56}$Ni mass of the order of $1\;\rm{M_{\odot}}$ or more.

In the contest of CSM/ejecta interaction, it is possible to have an estimate of the mass-loss following the approach of \cite{ishii_Lint_2024}, in particular their Eq.~15, which is derived in \cite{kokubo_kiss15s_2019}. The formula is: 
\begin{equation}
\begin{split}
    \dot{M}=  0.04 &\,M_{\odot}\,yr^{-1}\left(\frac{L_{tot}}{8\times 10^{43}\,erg\,s^{-1}}\right)  \left(\frac{v_w}{40\,km\,s^{-1}}\right) \\& \cdot \left(\frac{\epsilon}{0.3}\right)^{-1} 
    \left(\frac{v_{sh}}{4400\,km\,s^{-1}}\right)^{-3},
    \end{split}
\end{equation}
\noindent where $L_{tot}$ is the total bolometric luminosity, $v_{sh}$ is the shock velocity, $\epsilon$ is the radiation conversion efficiency, which is assumed to be 0.3 as in \citet{kokubo_kiss15s_2019}, and $v_w$ is the wind velocity of the CSM. The wind velocity of SN~2022qml is inferred from the position of the narrow H$\alpha$ P-Cygni absorption in the high-resolution spectrum at +~51$\,\rm{days}$, which gives $\mathrm{100\;km\,s^{-1}}$. For the other SNe the resolution is too poor to properly identify this feature but an upper limit is measured from the FWHM of the narrow line, which is reported in Table~\ref{tab:par_mis}. Upper limits are consistent with the value adopted as reference by \citet{ishii_Lint_2024}, that is $\mathrm{\leq 47.2\,km\,s^{-1}}$. We remind that LBVs have been proposed as progenitors of strongly-interacting SNe~IIn. On the one hand, LBVs have faster winds, of the order of $\mathrm{200 \,km\,s^{-1}}$ \citep{humphreys_lbv_1988} and on the other hand they are known to emit a huge amount of mass through short eruptive episodes (e.g., Eta Carinae has a calculated mass-loss rate of $0.075\;\rm{M_{\odot}\, yr^{-1}}$ \citealt{andriesse_etacar_1979}). For the sake of the calculation, we assumed $v_w=\mathrm{100\;km\,s^{-1}}$ for all SNe, while the shock velocity $v_{sh}$ comes from the estimates in Sect.~\ref{subsec:multifit}.

\begin{table*}[htbp]
 \caption{Parameters derived from measurements on the light curve and spectra of the SNe in our sample.}
    \label{tab:par_mis}
    \centering
    \begin{tabular}{ccccccc}
    \hline\hline
       SN & $t_{rise}$ & $v_w$ & $v_{sh}$ & $T_e$  & $E_{200}$\tablefootmark{\#} & $E_{rad}$ \\
        & (days) & $\mathrm{(km\,s^{-1})}$ & $\rm{(km\,s^{-1})}$ & $\rm{(\times 10^4\;K)}$ & $\rm{(\times 10^{49}\; erg)}$ & $\rm{(\times 10^{51}\; erg)}$\\\hline
        2021acya & $<48$ & $<600$\tablefootmark{a} & -- & $1.5-0.6$\tablefootmark{e} & 15\tablefootmark{\#} & 0.76 \\
        2021adxl & $<91$ & $<200$\tablefootmark{a} & $5000-2000$\tablefootmark{c} & -- & 9 & 0.31 \\
        2022qml & $<37$ & $100$\tablefootmark{b} & $10000-7000$\tablefootmark{d} & -- & 4 & 0.08 \\
        2022wed & $<27$ & $<300$\tablefootmark{a} & -- & $2.6-1.0$\tablefootmark{f} & 7 & 0.39\\\hline
    \end{tabular}
    \tablefoot{\tablefoottext{\#}{The value of $E_{200}$ for SN~2021acya differs from the one reported in Table~\ref{tab:tutte} since the $E_{200}$ reported here are calculated on the full bolometric and not on the pseudo-bolometric $g,r,i$ light curve as done previously and SN~2021acya has a huge UV contribution in the first 60~days.}\\
\tablefoottext{a}{Upper limit measured on the FWHM of the narrow H$\alpha$}.\\
\tablefoottext{b}{Measured from the H$\alpha$ narrow P-Cygni absorption.}\\
\tablefoottext{c}{Measured on spectra from phase +89$\,\rm{days}$ to phase +197$\,\rm{days}$.}\\
\tablefoottext{d}{Measured on spectra from phase +43$\,\rm{days}$ to phase +111$\,\rm{days}$.}\\
\tablefoottext{e}{Measured on spectra from phase +61$\,\rm{days}$ to phase +461$\,\rm{days}$.}\\
\tablefoottext{f}{Measured on spectra from phase +122$\,\rm{days}$ to phase +452$\,\rm{days}$.}
}
\end{table*}

From the shock velocity and the duration of the interaction phase, it is possible to calculate the radius of the CSM and, from the wind velocity, the duration of the mass-loss phase. Then, the mass of the shocked CSM $M_{CSM}$ is derived by integrating the mass-loss rate and considering the duration of the interaction phase as seen from the light curve. These parameters for each SN are reported in Table~\ref{tab:csm}, along with the energy calculated from the integration of the total luminosity, and the ranges are calculated using one or the other $v_{sh}$ estimate described in Sect.~\ref{subsec:multifit}.
Several factors can affect our computation: as mentioned in Sect.~\ref{sec:bol_altre}, the bolometric luminosity is likely underestimated because of the of lack UV and NIR observations for most of the SNe in the sample. This means that the mass-loss rates and hence the CSM masses are lower limits and the SNe could have shed even more mass. 
On the other hand, a steady wind velocity of $100\;\rm{km\,s^{-1}}$ was assumed. This value was measured only on the spectrum of SN~2022qml and it may be different for the other SNe. However, given the upper limit from the narrow FWHM reported Table~\ref{tab:par_mis}, the actual value for the other SNe is likely of the same order of magnitude.
The conversion efficiency is also an unknown factor but it cannot affect the estimate of the mass loss by more than a factor 2.
The parameter that most affects the calculation is the shock velocity, both because of the difficulty to estimate it and because of its weight in the equation. While we have a good estimate of the shock velocity for SNe~2021adxl and 2022qml, the evolution of $v_{sh}$ for SNe~2021acya and 2022wed is a mystery due to their higher opacity. However, it cannot be much higher than our estimate, otherwise a blue shoulder would have been visible in the spectra. If, on the other hand, the values were lower than our estimate, this would imply an even larger mass-loss rate than what was derived.

\begin{table*}
\caption{Parameters derived from our calculations based on \cite{ishii_Lint_2024}.}
    \label{tab:csm}
    \centering
    \begin{tabular}{cccccc}
    \hline\hline
    SN & $\dot{M}$ & $t_{\dot{M}}$ & $M_{CSM}$ & $R_{CSM}$ &  $E_K$\\
     & $(M_{\odot}\,yr^{-1})$ & \textit{(yr)} & $(M_{\odot})$ & $(10^{16}\,cm)$ &  ($\times10^{51}\;erg$)\\
    \hline
    2021acya & $0.06-0.8$ & $34-37$ & $18-19$ & $1.1-1.2$ &  2.53 \\
    2021adxl & $0.01-0.4$ & $24-39$ & $4.6-6.4$ & $0.8-1.2$ &  1.04 \\
    2022qml & $0.008-0.02$ & $31-33$ & $0.5-0.6$ & $0.9-1.0$ & 0.28 \\
    2022wed & $0.005-0.4$ & $32-34$ & $14-15$ & $1.0-1.1$ & 1.29 \\
    \hline
    \end{tabular}
    \tablefoot{ The ranges are calculated using one or the other $v_{sh}$ estimate described in Sect.~\ref{subsec:multifit}.}
\end{table*}

Even with all the caveats due to the assumptions described above, the values of $\dot{M}$ and $M_{CSM}$ are much higher than those expected for a steady wind and more in line with eruptive processes \citep{matsumoto_eruzioni_2022}, in agreement with an LBV progenitor (for reference, \citealt{fassia_98s_2001} find for SN~1998S a mass-loss rate of $2\times10^{-5}\;M_{\odot}\,yr^{-1}$, two orders of magnitude lower than our values, which are instead more in line with the eruptive episodes in Eta Carinae).
To produce a CSM mass of $5-20\;\rm{M_{\odot}}$ requires a huge progenitor, the formation of which could be challenging for stellar evolution models.
The only SN in our sample with a considerably smaller $M_{CSM}$ is SN~2022qml.
This, together with the different spectra and shorter light curve duration, points toward a possible thermonuclear explosion whose progenitor was embedded in a H-rich CSM.

\begin{figure}
    \centering
    \includegraphics[width=\columnwidth]{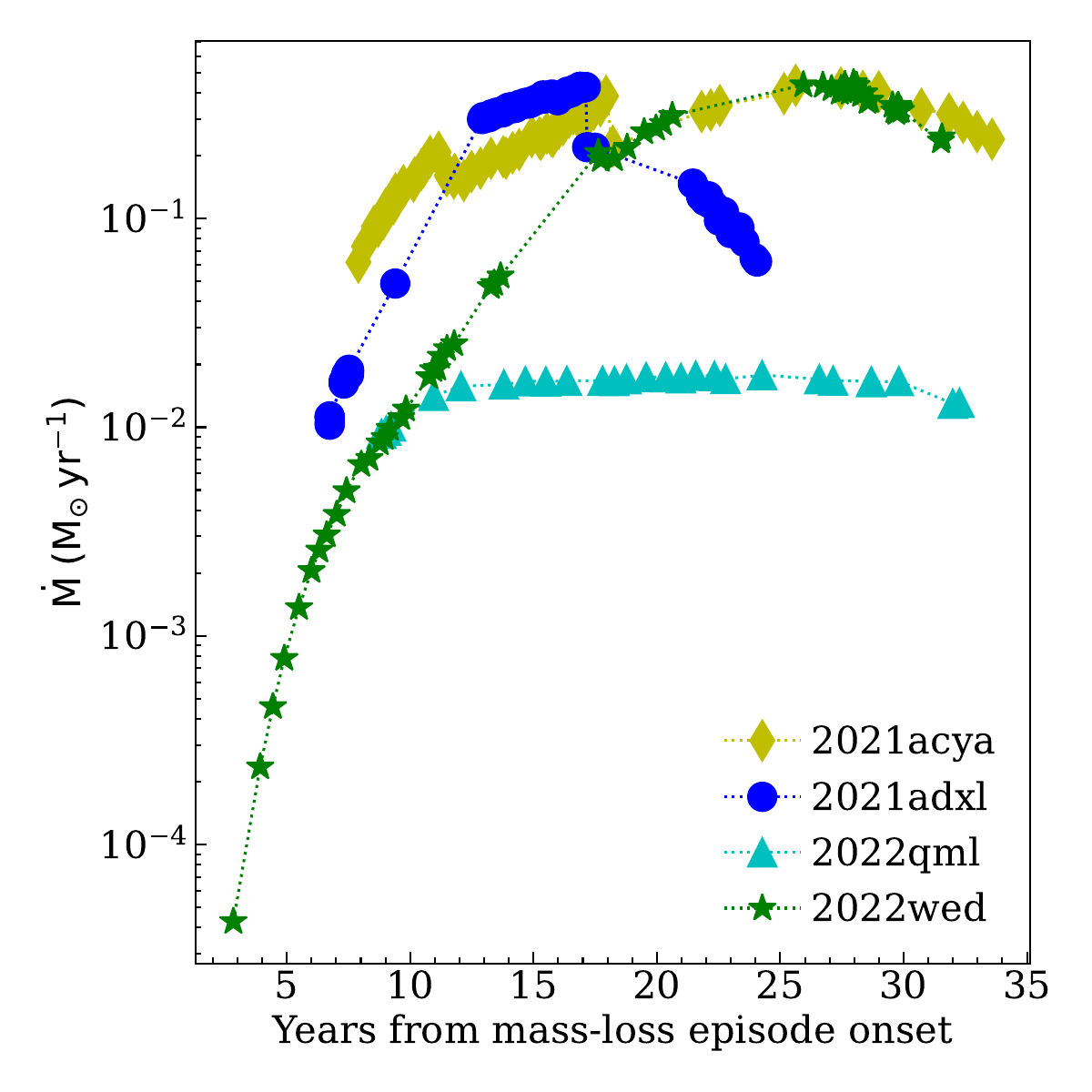}
    \caption{Mass-loss rate as a function of the time from the onset of the mass-loss episode, assuming a steady wind and shock velocity as derived in Sect.~\ref{subsec:multifit}.}
    \label{fig:tw_mloss}
\end{figure}

It is also interesting to explore the mass-loss rate as a function of time. 
Regardless of the CSM origin (wind or eruption), we assume that it is distributed in a shell within which the velocity is constant. Assuming a steady wind for simplicity, it is possible to derive the value of $\dot{M}$ on the timescale of the wind. The results for each SN are shown in Fig.~\ref{fig:tw_mloss}. Here, there is a strong dependence especially on the ratio between the shock and the wind velocity, which gives the overall shape to the trend, as well as the duration of the bolometric light curve. As shock velocity, we considered the trend derived on the H$\alpha$ blue shoulder fit for SNe~2021adxl and 2022qml, while for SNe~2021acya and 2022wed we take the same trend of SN~2021adxl rescaled to the velocity derived by \citet{fransson_2010jl_2014} at 320~days for SN~2010jl, as explained in Sect.~\ref{subsec:multifit}.
From Fig.~\ref{fig:tw_mloss}, it appears that the mass-loss rate of SN~2022qml was more constant, compared to the other SNe, and overall lower, once again in agreement with a different origin for this SN. The trends of SNe~2021adxl and 2022wed are similar but shifted, with SN~2022wed starting to significantly increase later on, around 20~years after the onset, while SN~2021adxl shows a high mass-loss rate already 10~years in. 
SN 2021acya, on the other hand, has a peculiar behaviour with possibly three distinct peaks that could indicate the onset of different mass-loss episodes. Interestingly, the timeline of the peaks seems to align with the other SNe: the second one happens around the same time as the one of SN~2021adxl and the third is almost in coincidence with the peak of SN~2022wed.
With these considerations, it appears even more evident that the progenitor stars of these SNe had to be massive, peculiar objects to be able to expel such a huge amount of mass within such a limited time period, possibly, in the case of SN~2021acya, in two separate episodes.
A mass-loss rate that 
starts high and increases for decades after the onset of the episode is also in line with what is found by \citet{moriya_massloss_2014},
although it may decline closer to the explosion.

In Table~\ref{tab:par_mis}, the total radiated energy obtained by integrating the bolometric light curve is reported. It can be converted to the kinetic energy $E_K$ produced in the explosion by assuming a conversion factor $\epsilon$, which is not well constrained and can vary from 0.1 to 1 \citep{moriya2014}. As before, we assumed $\epsilon=0.3$ as in \cite{kokubo_kiss15s_2019}, while the result is reported in Table~\ref{tab:csm}.

The CSM radius, its mass, and the kinetic energy are useful to compare our findings to models of neutrino production in shocked regions of interacting SNe. 
In particular, the semi-analytical model in \cite{pitik2023} shows that the largest probability to produce HE neutrinos is found for SNe with $E_K \gtrsim 10^{51}\; \mathrm{erg}$, $1 \lesssim M_{CSM} \lesssim 30\mathrm{\;M_{\odot}}$, and $R_{CSM}\gtrsim 10^{16}\mathrm{\;cm}$. These parameters are favoured in SNe that display luminous light curves ($L_{peak}\gtrsim 10^{43}-10^{44}\;\mathrm{erg\,s^{-1}}$) but average rise time ($10\,\mathrm{d} \lesssim t_{rise}\lesssim 40\,\mathrm{d}$).
All our SNe fulfill these criteria but for SN~2022qml (see Table~\ref{tab:par_mis}), whose total energy is smaller than the rest due to the shorter duration of the light curve, as well as its overall dimness compared to the others. 

\begin{figure}[htbp]
    \centering
    \includegraphics[width=\columnwidth]{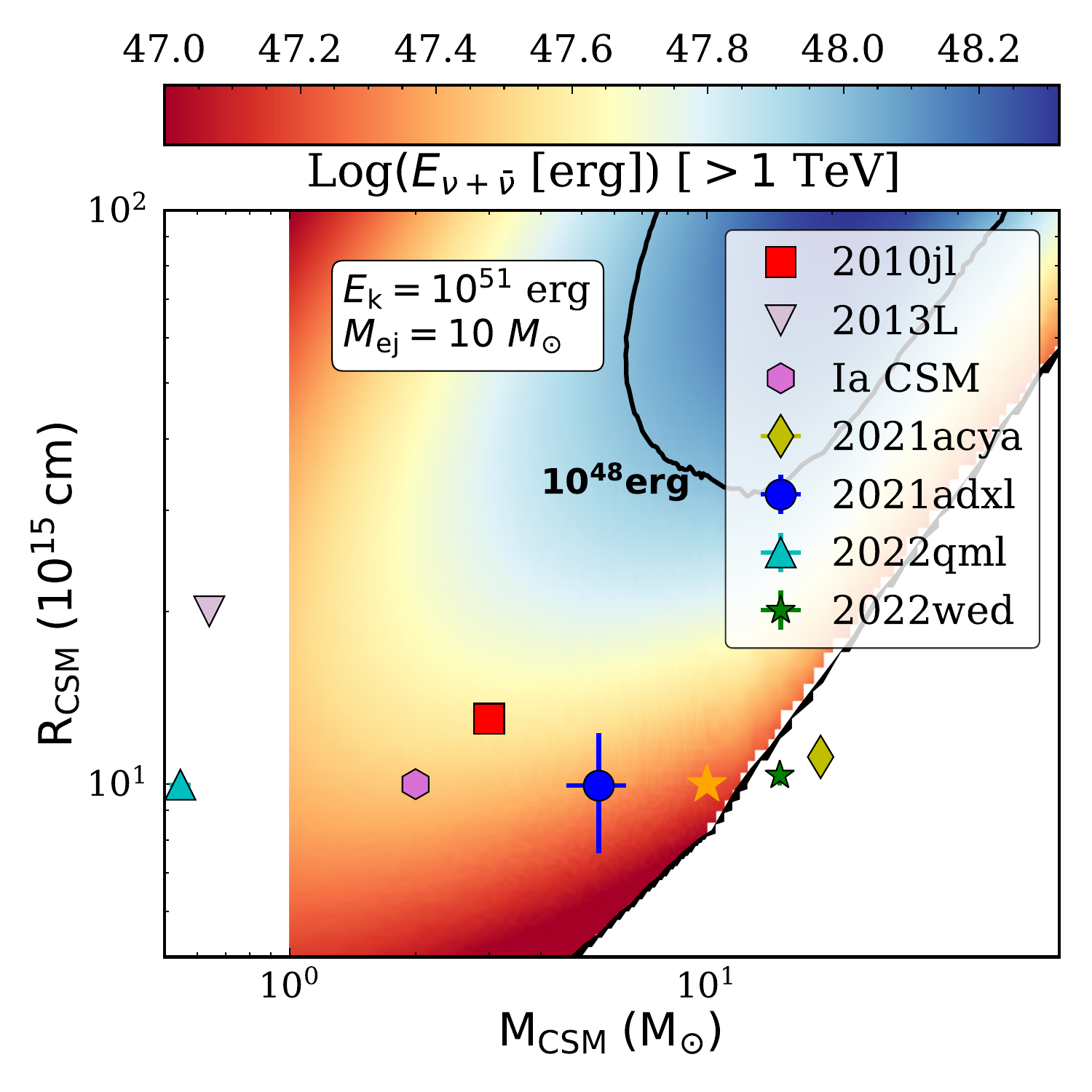}
    \includegraphics[width=\columnwidth]{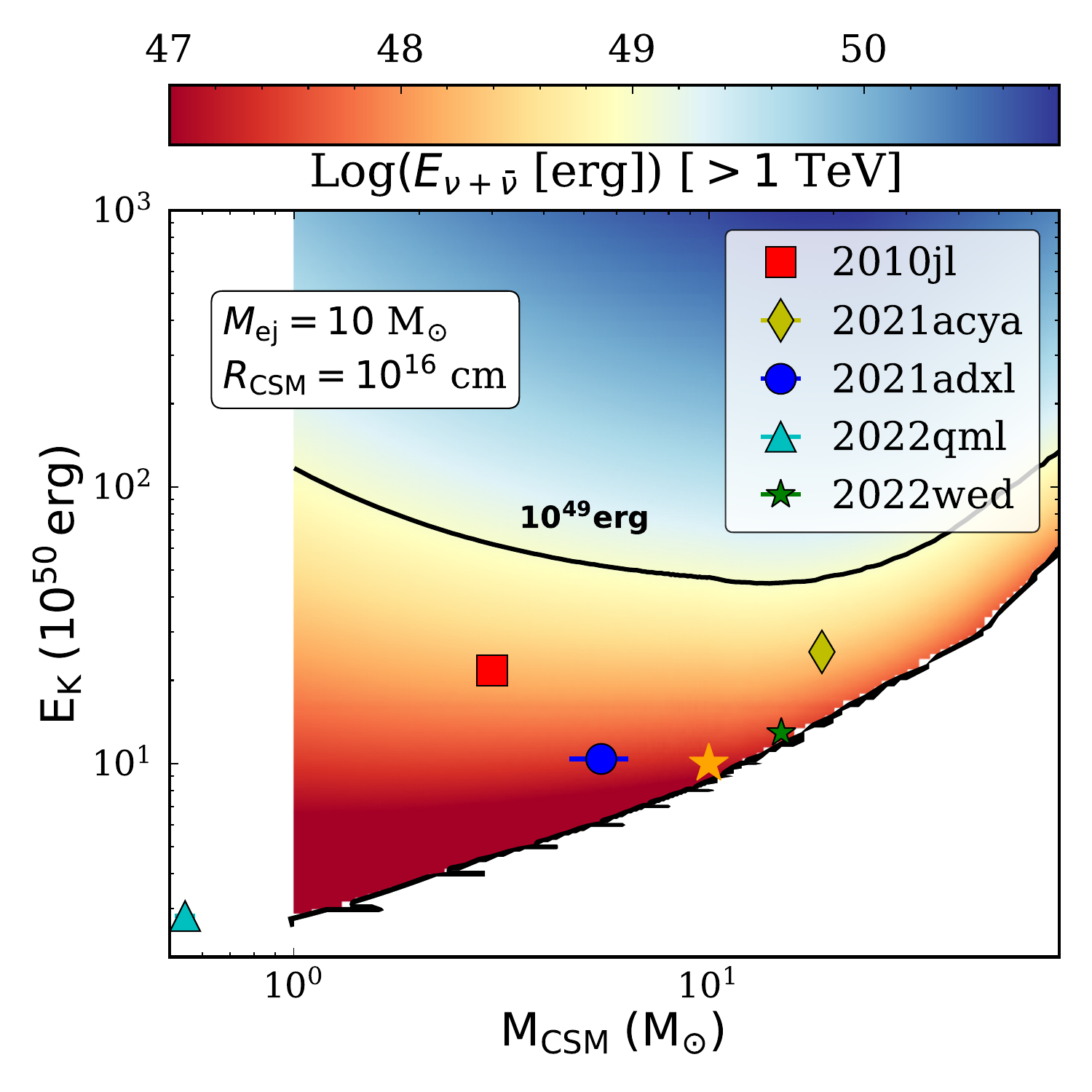}
    \caption{\textit{Top: }Contour plots of the total neutrino energy $\varepsilon_{\nu+\bar{\nu}}$ integrated for $E_{\nu}\geq1$ TeV from \cite{pitik2023} in function of the CSM radius and mass, over which the same parameters for our SNe are plotted. The orange star indicates the benchmark of the model. \textit{Bottom: } Same as above but in function of kinetic energy and CSM mass.}
    \label{fig:pitik}
\end{figure}

The parameters of the SNe in our sample are plotted over the contour plots of the integrated neutrino energy \citet{pitik2023} calculated for their model in the upper panel in Fig.~\ref{fig:pitik}, where it is assumed that a fraction of 10\% of the kinetic energy of the shock is used to accelerate protons. The plot shows the integrated neutrino energy $\varepsilon_{\nu+\bar{\nu}}$ for energy above 1~TeV as a function of $M_{CSM}$ and $R_{CSM}$ at fixed kinetic energy ($E_K=10^{51}\; \mathrm{erg}$) and ejecta mass ($M_{ej}=10\;\mathrm{M_{\odot}}$). To our measurements are also added those of SN~2010jl \citep{fransson_2010jl_2014}, SN~2013L \citep{taddia_2013L_2020}, and a prototypical Type~Ia-CSM whose parameters were inferred from an average on the sample presented in \cite{sharma_iacsm_2023}. While SNe~2022qml and 2013L are outside the calculated range of the model (which could be due to a different progenitor for SN~2022qml), SNe~2021acya and 2022wed have a CSM that is too massive, while the others fall within the limits for HE neutrino production. 
However, we caution that there is some degeneracy in the parameters. 
The same exercise is repeated in the bottom panel of Fig.~\ref{fig:pitik}, but this time plotting the kinetic energy with respect to the CSM mass at fixed ejecta mass ($M_{ej}=10\;\mathrm{M_{\odot}}$) and CSM radius ($R_{CSM}= 10^{16}\mathrm{\;cm}$). In this case, SNe~2021acya and 2022wed are within the model margins and actually SN~2021acya is the most favoured. Moreover, we remind that the integrated luminosity is a lower limit, since there is no full bolometric coverage at all epochs. In principle, this could push some SNe upward in the plot.
 
This test shows that the SNe in our sample indeed have the characteristics to produce HE neutrinos. We should stress that the parameters have some degeneracy and, in fact, multiple SN models could lead to the same neutrino flux. For example, a higher mass of the CSM disfavours neutrino production, but it can be compensated by a higher kinetic energy \citep{pitik2023}.

For reference, we can apply the same analysis also to SN~2020faa,
a SN powered by hidden interaction, which has a derived kinetic energy and CSM mass and radius: $E_K=4.8\times10^{51}\;\mathrm{erg}$, $M_{CSM}=1\;M_{\odot}$, and $R_{CSM}=\mathrm{10^{14}-10^{15}\;cm}$ \citep{salmaso_faa_2023}. The small CSM radius disfavours SN~2020faa as a likely HE neutrino source, however, the kinetic energy and the CSM mass are in agreement with what is found for the other SNe in our sample. As mentioned, degeneracy in the parameters implies uncertainties that are difficult to disentangle but it is clear that high values of CSM mass and kinetic energy, hence requiring a massive progenitor, seem to be needed.

\section{Conclusions}
\label{sec:disc_altre}

This paper shows that strongly-interacting SNe can display a variety of light curves and spectra. These differences mainly depend on the progenitor and its pre-explosion history, since this will determine the mass, radius, density and distribution of the CSM, other than the final mass at the explosion.  Some interesting clues are also inferred from the comparison among the events and with prototypical events from literature. The analysis of similarities and differences shows the range of diversity in the density and profile of the CSM and in the properties of the nuclear engine. In particular, there is circumstantial evidence that SN~2022qml could hide a Type~Ia SN explosion rather than a CC as for the other SNe.

The hosts of these SNe are dwarf galaxies consistent with high SFR. Top heavy IMFs, which favour the production of massive stars, could cause these strongly-interacting SNe from massive progenitors to be more frequent in smaller galaxies. This would be important for the chemical enrichment of such environments. Also, in a multimessenger context, massive SN progenitors are interesting because the compact remnant after the explosion could be a massive black hole in the range of those observed by LIGO-Virgo \citep{abbott_bbhmass_2020}.

Despite their differences, all the SNe analysed here are consistent with energetic explosions and strong interaction with the surrounding CSM. The CSM masses are high (extremely so in the case of SNe~2021acya and 2022wed) and the mass-loss rates are similar, around 30~years before the explosion. Instead, the trend of the mass-loss rate differs, since during this time SN~2022qml had a more constant mass-loss, while SNe~2021adxl and 2022wed underwent a significant acceleration and SN~2021acya possibly even three separate mass-loss episodes.

The energies and CSM mass and radius derived from our analysis are also mostly consistent with HE neutrino production. \cite{murase_neutrini_sne_2023} shows that, for SNe~IIn with parameters similar to those analysed here, the ideal timescale of neutrino detection is between $10^6$ and $10^8$~s, that is, between 10$\,\rm{days}$ and 3~years, in line with other models \citep{sarmah_neutrini_2022}.
Eventually, these transients appear to be located in regions of the parameter space where the production of HE neutrinos is marginal but not excluded.

\section*{Data availability}
Tables with photometric measurements, spectral log of observations, and fit on the H$\alpha$ and H$\beta$ are only available in electronic form at the CDS via anonymous ftp to cdsarc.u-strasbg.fr (130.79.128.5) or via http://cdsweb.u-strasbg.fr/cgi-bin/qcat?J/A+A/.

\begin{acknowledgements}
We thank the anonymous referee for their helpful insights.
The author acknowledges the help of Dr. T. Pitik in recreating Figure~\ref{fig:pitik}.
This paper is supported by fundings from MIUR, PRIN 2017 (grant 20179ZF5KS). IS, AR, and NER are partially supported by the PRIN-INAF 2022 project “Shedding light on the nature of gap transients: from the observations to the models”.
AR acknowledges the support GRAWITA Large Program Grant (PI P. D'Avanzo).
YZC is supported by the National Natural Science Foundation of China (NSFC, Grant No. 12303054), the National Key Research and Development Program of China (Grant No. 2024YFA1611603), the Yunnan Fundamental Research Projects (Grant No. 202401AU070063), and the International Centre of Supernovae, Yunnan Key Laboratory (No. 202302AN360001). 
JPA was funded by ANID, Millennium Science Initiative, ICN12\_009.
TP acknowledges the support by ANID through the Beca Doctorado Nacional 202221222222.
MN is supported by the European Research Council (ERC) under the European Union’s Horizon 2020 research and innovation programme (grant agreement No.~948381) and by UK Space Agency Grant No.~ST/Y000692/1.
LG, CPG and TEMB acknowledge financial support from the Spanish Ministerio de Ciencia e Innovaci\'on (MCIN), the Agencia Estatal de Investigaci\'on (AEI) 10.13039/501100011033, the European Social Fund (ESF) ``Investing in your future'', the European Union Next Generation EU/PRTR funds, the Horizon 2020 Research and Innovation Programme of the European Union, and by the Secretary of Universities and Research (Government of Catalonia), under the PID2020-115253GA-I00 HOSTFLOWS project, 2021-SGR-01270, the 2019 Ram\'on y Cajal program RYC2019-027683-I, the 2021 Juan de la Cierva program FJC2021-047124-I, the Marie Sk\l{}odowska-Curie and the Beatriu de Pin\'os 2021 BP 00168 programme, and from Centro Superior de Investigaciones Cient\'ificas (CSIC) under the PIE project 20215AT016, and the program Unidad de Excelencia Mar\'ia de Maeztu CEX2020-001058-M.
TWC acknowledges the Yushan Young Fellow Program by the Ministry of Education, Taiwan for the financial support.
MB, EC and TP acknowledge the financial support from the Slovenian Research Agency (grants I0-0033, P1-0031, J1-8136, J1-2460 and Z1-1853) and the Young Researchers program.
PC acknowledges support via Research Council of Finland (grant 340613).
MS is funded by the Independent Research Fund Denmark (IRFD, grant number  10.46540/2032-00022B).
Based on observations collected at Copernico 1.82m telescope and Schmidt 67/92 telescope (Asiago Mount Ekar, Italy) INAF - Osservatorio Astronomico di Padova.
This project has received funding from the European Union's Horizon 2020 research and innovation programme under grant agreement No 101004719. This material reflects only the authors views and the Commission is not liable for any use that may be made of the information contained therein.
Based on observations made with the Nordic Optical Telescope, owned in collaboration by the University of Turku and Aarhus University, and operated jointly by Aarhus University, the University of Turku and the University of Oslo, representing Denmark, Finland and Norway, the University of Iceland and Stockholm University at the Observatorio del Roque de los Muchachos, La Palma, Spain, of the Instituto de Astrofisica de Canarias. Observations from the Nordic Optical Telescope were obtained through the NUTS2 collaboration, which are supported in part by the Instrument Centre for Danish Astrophysics (IDA). 
The data presented here were obtained in part with ALFOSC, which is provided by the Instituto de Astrofisica de Andalucia (IAA). 
This work makes use of data from the Las Cumbres Observatory network. The LCO team is supported by NSF grants AST–1911225 and AST– 1911151, and NASA SWIFT grant 80NSSC19K1639.
The Liverpool Telescope is operated on the island of La Palma by Liverpool John Moores University in the Spanish Observatorio del Roque de los Muchachos of the Instituto de Astrofisica de Canarias with financial support from the UK Science and Technology Facilities Council.
This work was based in part on observations made with the Italian Telescopio Nazionale Galileo (TNG), operated on the island of La Palma by the Fundaci\'on Galileo Galilei of the INAF (Istituto Nazionale di Astrofisica) at the Spanish Observatorio del Roque de los Muchachos of the Instituto de Astrofisica de Canarias.
The Gran Telescopio CANARIAS (GTC) is a 10.4m telescope with a segmented primary mirror. It is located in one of the top astronomical sites in the Northern Hemisphere: the Observatorio del Roque de los Muchachos (ORM, La Palma, Canary Islands). The GTC is a Spanish initiative led by the Instituto de Astrofísica de Canarias (IAC). The project is actively supported by the Spanish Government and the Local Government from the Canary Islands through the European Funds for Regional Development (FEDER) provided by the European Union. The project also includes the participation of Mexico (Instituto de Astronomía de la Universidad Nacional Autónoma de México (IA-UNAM) and Instituto Nacional de Astrofísica, Óptica y Electrónica (INAOE)), and the US University of Florida.
Based on observations collected at the European Organisation for Astronomical Research in the Southern Hemisphere under ESO programmes 1103.D-0328 and 105.20TF.001.
We acknowledge the use of public data from the \textit{Swift} data archive.
Based on observations obtained with the Samuel Oschin Telescope 48-inch and the 60-inch Telescope at the Palomar Observatory as part of the Zwicky Transient Facility project. ZTF is supported by the National Science Foundation under Grant No. AST-2034437 and a collaboration including Caltech, IPAC, the Weizmann Institute for Science, the Oskar Klein Center at Stockholm University, the University of Maryland, Deutsches Elektronen-Synchrotron and Humboldt University, the TANGO Consortium of Taiwan, the University of Wisconsin at Milwaukee, Trinity College Dublin, Lawrence Livermore National Laboratories, and IN2P3, France. Operations are conducted by COO, IPAC, and UW.
The Pan-STARRS1 Surveys (PS1) and the PS1 public science archive have been made possible through contributions by the Institute for Astronomy, the University of Hawaii, the Pan-STARRS Project Office, the Max-Planck Society and its participating institutes, the Max Planck Institute for Astronomy, Heidelberg and the Max Planck Institute for Extraterrestrial Physics, Garching, The Johns Hopkins University, Durham University, the University of Edinburgh, the Queen's University Belfast, the Harvard-Smithsonian Center for Astrophysics, the Las Cumbres Observatory Global Telescope Network Incorporated, the National Central University of Taiwan, the Space Telescope Science Institute, the National Aeronautics and Space Administration under Grant No. NNX08AR22G issued through the Planetary Science Division of the NASA Science Mission Directorate, the National Science Foundation Grant No. AST–1238877, the University of Maryland, Eotvos Lorand University (ELTE), the Los Alamos National Laboratory, and the Gordon and Betty Moore Foundation.
This work presents results from the European Space Agency (ESA) space mission Gaia. Gaia data are being processed by the Gaia Data Processing and Analysis Consortium (DPAC). Funding for the DPAC is provided by national institutions, in particular the institutions participating in the Gaia MultiLateral Agreement (MLA). The Gaia mission website is https://www.cosmos.esa.int/gaia. The Gaia archive website is https://archives.esac.esa.int/gaia.
This work has made use of data from the Asteroid Terrestrial-impact Last Alert System (ATLAS) project. The Asteroid Terrestrial-impact Last Alert System (ATLAS) project is primarily funded to search for near earth asteroids through NASA grants NN12AR55G, 80NSSC18K0284, and 80NSSC18K1575; byproducts of the NEO search include images and catalogs from the survey area. This work was partially funded by Kepler/K2 grant J1944/80NSSC19K0112 and HST GO-15889, and STFC grants ST/T000198/1 and ST/S006109/1. The ATLAS science products have been made possible through the contributions of the University of Hawaii Institute for Astronomy, the Queen’s University Belfast, the Space Telescope Science Institute, the South African Astronomical Observatory, and The Millennium Institute of Astrophysics (MAS), Chile.
     
\end{acknowledgements}

\bibliographystyle{aa}
\bibliography{biblio}

\onecolumn
\begin{appendix}

\section{Observations}
\label{sec:obs}

Our targets were monitored using the Schmidt and Copernico telescopes of the Asiago Observatory\footnote{\url{https://www.oapd.inaf.it/sede-di-asiago/telescopes-and-instrumentations/}}, INAF (Italy); the Nordic Optical Telescope (NOT)\footnote{\url{https://www.not.iac.es}}, the Telescopio Nazionale Galileo (TNG)\footnote{\url{https://www.tng.iac.es}}, the Liverpool Telescope (LT)\footnote{\url{https://telescope.livjm.ac.uk}}, and the Gran Telescopio CANARIAS (GTC)\footnote{\url{https://www.gtc.iac.es}}, all located in La Palma (Spain); the Rapid Eye Mount (REM)\footnote{\url{http://www.rem.inaf.it}} of INAF at La Silla and the Very Large Telescope (VLT)\footnote{\url{https://www.eso.org/public/teles-instr/paranal-observatory/vlt}} of ESO at Paranal (Chile).
We also exploited the time allocated to other facilities to international collaborations, namely the NOT via NUTS2 (Nordic-optical-telescope Un-biased Transient Survey)\footnote{\url{https://nuts2.sn.ie/}} and the ESO NTT\footnote{\url{https://www.eso.org/public/teles-instr/lasilla/ntt}} via ePESSTO+ (advanced Public ESO Spectroscopic Survey for Transient Objects, \citealt{smartt_pessto_2015}).
These observations were complemented with data from transient surveys available from public archives, in particular, the Zwicky Transient Facility (ZTF, \citealt{bellm_ztf_2019}), the Las Cumbres Observatory (LCO, \citealt{brown_lcogt_2013}), the Panoramic Survey Telescope and Rapid Response System (Pan-STARRS, PS1, \citealt{ps1}), the Asteroid Terrestrial-impact Last Alert System (ATLAS, \citealt{tonry_atlas_2018}). 
Space observations obtained by the Gaia mission \citep{gaia_dr2}, and from NASA's Neil Gehrels Swift Observatory (SWIFT, \citealt{swift_uvot}) were also retrieved.

\section{Data reduction}
\label{sec:redu}
\subsection{Photometry}
\label{subsec:phot}
Photometric observations were reduced with standard techniques using \texttt{IRAF}\footnote{\url{https://iraf-community.github.io/}} recipes and a number of different \texttt{python} packages, in particular \texttt{astropy} \citep{astropy:2022} and affiliated packages (\texttt{astroquery}, \texttt{ccdproc}, \texttt{photutils}). At first, the detector signature was removed with bias and flat-field corrections. Then, an algorithm for cosmic-ray rejection\footnote{The algorithm is an implementation of the code described in \cite{vanDokkum_cosmici_2001} as implemented by \cite{mccully_cosmici_2018}.} was applied. For astrometric and photometric calibration and for the measurement of the SN magnitudes the required recipes in the \texttt{ecsnoopy} package\footnote{ecsnoopy is a python package for SN photometry using PSF fitting and/or template subtraction developed by E. Cappellaro. A package description can be found at \url{http://sngroup.oapd.inaf.it/ecsnoopy.html}} were implemented.
In most cases (when the SN magnitude was below $\sim16-17$~mag), the SN magnitude was measured after subtracting a template image from the observed frame, to better remove the contamination from the host galaxy. To this aim, reference archival images from public surveys such as PS1, SDSS (Sloan Digital Sky Survey, \citealt{kollmeier_sdssv_2019}), or Skymapper \citep{keller_skymapper_2007} were used. With \texttt{ecsnoopy}, the registration of the template image was secured to the same pixel grid of the science image and then the code \texttt{hotpants} \citep{Becker2015} was used for the convolution of the two images to the same PSF and photometric scale. 
Finally, the instrumental magnitudes (or upper limits) were calibrated using photometric zero points measured from local stars with photometry retrieved from the public surveys mentioned above.

For the SWIFT observations of SN~2021acya, the magnitudes were measured with aperture photometry using \texttt{uvotsource} from \texttt{heasoft} \citep{heasoft} adopting a circular radius of 5~arcsec and subtracting the sky background measured in an offset empty region.

Finally, for the near-infrared (NIR) observations, the same procedure as for the optical observations was used but including the preliminary subtraction of the sky background, which was obtained from the median combination of the dithered images for each filter. 
In this case, the nightly zero-points were computed with respect to the Two Micron All-Sky Survey (2MASS, \citealt{skrutskie_2mass_2003}) source catalogue photometry.
The magnitudes of the targets are available online only (see the Data availability section), together with the magnitudes acquired from the public surveys mentioned above.

The available pre-supernova imaging was also checked for possible eruptive episodes in our SNe. In particular, through the available web tool\footnote{\url{https://fallingstar-data.com/forcedphot/}},  ATLAS forced photometry at the SN locations was obtained. In all cases, we concluded that there is no evidence of variation in the observed flux in the last decade.
However, we stress that the limiting magnitude of ATLAS (20.0~mag in cyan and 19.5~mag in orange), corresponds to an absolute magnitude that ranges from $-14.7$ to $-18.8$ mag for the distance of our SNe, so the limit is not stringent and it is possible that any low-intensity precursor activity went undetected.

\subsection{Spectroscopy}
\label{subsec:spec}

The spectroscopic observations were in general reduced using standard prescriptions with the package \texttt{foscgui}\footnote{foscgui is a python/pyraf-based graphic user interface aimed at extracting SN spectroscopy and photometry obtained with FOSC-like instruments. It was developed by E. Cappellaro. A package description can be found at \url{http://sngroup.oapd.inaf.it/foscgui.html}}.
The spectra were corrected for bias, flat-field, and cosmics rejection, calibrated in wavelength in the 2D frame, and extracted to obtain the 1D spectrum. This was then calibrated in flux and corrected for second-order contamination (if required) and telluric absorptions, the latter with the aid of the spectrum of a hot spectrophotometric standard star.
For the spectra taken using X-shooter, the publicly available data reduction pipeline \texttt{EsoReflex}\footnote{\url{https://www.eso.org/sci/software/esoreflex/}} was used, following the same steps described above.
The logs of spectroscopic observations are available online only (see the Data availability section). The spectral evolution of the targets through the most significant spectra is shown in Figs.~\ref{fig:spec_evol_acya},~\ref{fig:spec_evol_adxl},~\ref{fig:spec_evol_qml}, and \ref{fig:spec_evol_wed}.
 
\subsection{Extinction and redshift corrections}
\label{subsec:reddening}
\begin{figure}[h!]
    \centering
    \includegraphics[width=\columnwidth]{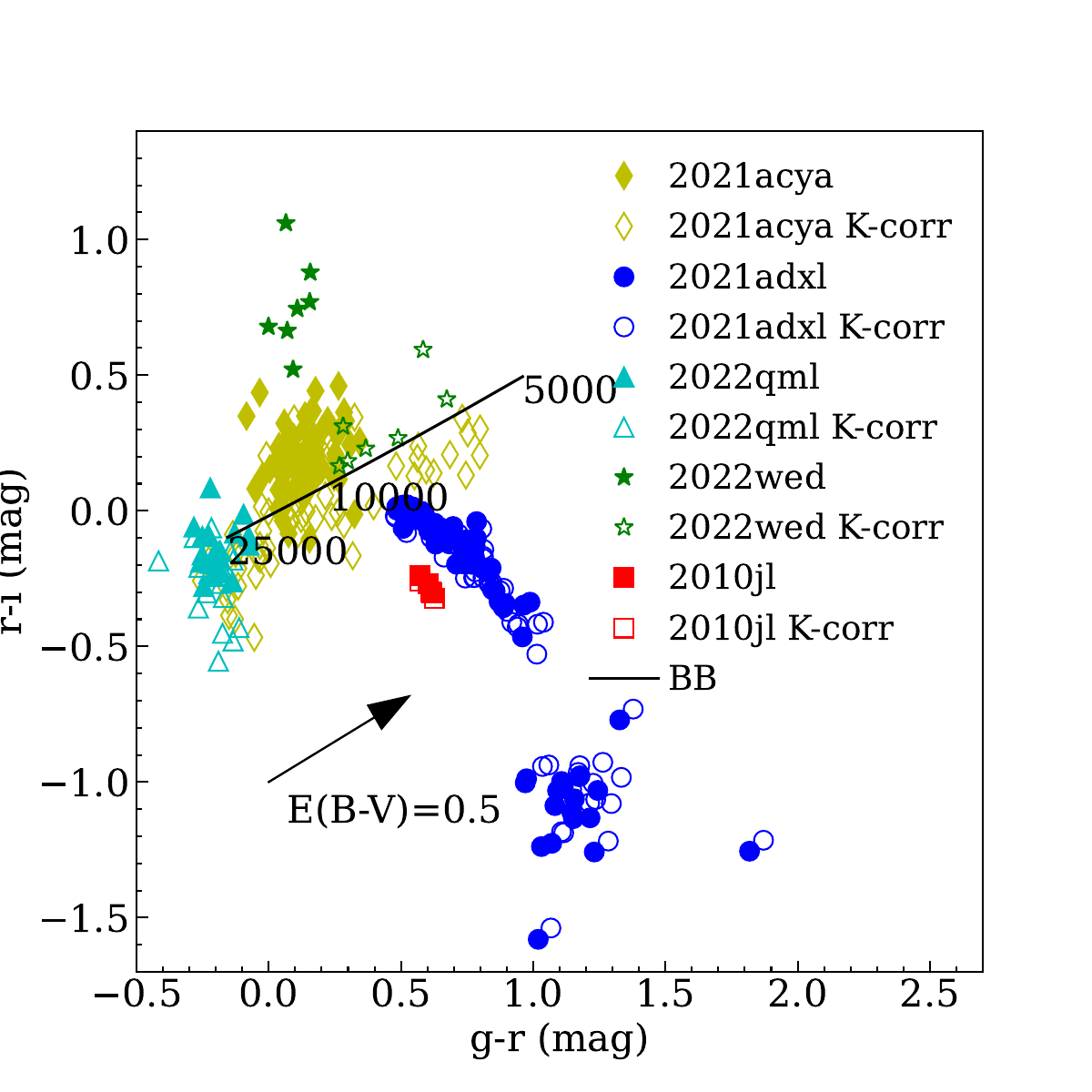}
    \caption{Colour-colour $r-i$ vs. $g-r$ diagram of our SNe, with and without the K correction. The extinction vector E(B-V) and a BB with temperature between 25000 and 5000~K are also added.}
    \label{fig:2cd_kcorr}
\end{figure}

Galactic extinction values were obtained from NED\footnote{\url{https://ned.ipac.caltech.edu}}, assuming $R_V=3.1$ \citep{schlafly_reddening_2011}. Moreover, we examined the spectra of all transients to search for evidence of \ion{Na}{I}~D~$\lambda\lambda5890,5896$ interstellar gas in the host galaxy. This was only identified in the X-shooter spectrum of SN~2022qml at +49$\,\rm{days}$. The lines have an equivalent width (\textit{EW}) of 0.1168~\AA~ and 0.1198~\AA~ for D1 and D2, respectively. Given the line \textit{EW} vs. extinction relation in \citet{poznanski_reddening_2012}, these values correspond to a total reddening $E(B-V)=0.028\pm0.2$~mag, where the error comes from the uncertainties in the relation, and give an absorption $A_V=0.09\pm{0.6}$ if $R_V=3.1$ also in this case, which is much lower than the extinction due to the MW (see Table~\ref{tab:tutte}).
However, \citet{phillips_na_ia_2013} found that the use of this method is not reliable for dust reddening estimation in SNe~Ia,
while \citet{rodriguez_reddening_2023} argue that it underestimates the dust contribution in SNe~II.
Another estimate of intrinsic reddening can be obtained through the relation proposed in \citet{turatto_reddening_2003}, which gives $E(B-V)=0.038$~mag, still small compared to the MW contribution.
Therefore, we conclude that in all cases the extinction inside the host galaxy is likely negligible.

The presence of strong emission lines combined with a significant redshift for some SNe in the sample suggested checking for the effect of the  K-correction on the photometry. 
To calculate the magnitude of the K-correction, we examined all the spectra and extrapolate the flux at $g,r,i$ bands. The measurement was performed twice, once on the original spectra and once on the redshift-corrected spectra. The difference is the K-correction that was then linearly interpolated and applied to the light curve at all epochs.
The magnitude of the K-correction is shown in Fig.~\ref{fig:2cd_kcorr}, where the location of each transient at different phases both with and without K-correction (empty and filled symbols, respectively) is plotted in a colour-colour diagram ($r-i$ vs. $g-r$). The strongly-interacting SN~2010jl \citep{fransson_2010jl_2014} is also added as reference. The shift after the K-correction appears more significant for SN~2022wed, as expected being the transient at the highest redshift, and also for SN~2021acya. In the case of SN~2022qml, the transient does not show a significant evolution and is always in agreement with an extremely hot black body (BB). This is probably due to the blue bump excess, a feature that is discussed in Sect.~\ref{subsec:indice}. On the other hand, SNe~2010jl and 2021adxl are in countertrend, showing redder colours at all phases. This is interpreted as the effect of a luminous H$\alpha$ emission line along with a small blue bump excess compared to the other SNe in our sample, although a contribution from higher host extinction cannot be completely ruled out.
The K-correction to all the SNe in our sample for the rest of the analysis.

The redshift is measured from the narrow H$\alpha$ emission line. To calculate the distance modulus for the absolute magnitude computation it is assumed $H_0=70 \; \rm{km\,s^{-1}\,Mpc^{-1}}$, while the measured redshifts are corrected to the V3K reference frame.

\end{appendix}

\end{document}